%% file: whitepaper.tex
\documentclass[a4paper,eqsecnum,nofootinbib,twocolumn,secnumarabic,superscriptaddress]{revtex4}

\pdfoutput=1

\usepackage{graphicx}
\usepackage{dcolumn}
\usepackage{bm}
\usepackage{textcomp}
\usepackage{amsmath}
\usepackage{multirow}
\usepackage{subfigure} 
\usepackage{setspace}

\usepackage{hhline}
\usepackage{array}
\usepackage{verbatim} 
\usepackage{mhchem} 
\usepackage{eurosym}
\usepackage{nomencl}

\makenomenclature 

\usepackage{titlesec}
\titleformat{\chapter}[hang]{\large\bfseries}{\thechapter\quad}{0pt}{}
\titleformat{\section}[hang]{\Large\bfseries}{\thesection\quad}{0pt}{}
\titleformat{\subsection}[hang]{\large\bfseries}{\thesubsection\quad}{0pt}{}
\titleformat{\subsubsection}[hang]{\bfseries}{\thesubsubsection\quad}{0pt}{}
\titleformat{\paragraph}[hang]{\bfseries}{\theparagraph)~}{0pt}{}

\titlespacing{\chapter}{0pt}{-3em}{6pt}
\titlespacing{\section}{0pt}{6pt}{6pt}
\titlespacing{\subsection}{0pt}{18pt}{6pt}
\titlespacing{\subsubsection}{0pt}{12pt}{6pt}
\titlespacing{\paragraph}{0pt}{9pt}{3pt}
\begin{document}


\title{The next-generation liquid-scintillator neutrino observatory LENA}

\date{\today}

\input{abstract}

\author{Michael Wurm}
\email[Corresponding author. e-mail: ]{michael.wurm@desy.de}
\affiliation{Physik-Department, Technische Universit\"at M\"unchen, Germany}
\affiliation{Institut f\"ur Experimentalphysik, Universit\"at Hamburg, Germany}
\author{John F. Beacom}
\affiliation{Department of Physics, Ohio State University, Columbus, OH, USA}
\author{Leonid B. Bez\-ru\-kov}
\affiliation{Institute for Nuclear Research, Russian Academy of Sciences, Moscow, Russia}
\author{Daniel Bick}
\affiliation{Institut f\"ur Experimentalphysik, Universit\"at Hamburg, Germany}
\author{Johannes Bl\"umer}
\affiliation{Institut f\"ur Kernphysik, Karlsruhe Institute of Technology KIT, Germany}
\author{Sandhya Choubey}
\affiliation{Harish-Chandra Research Institute, Allahabad, India}
\author{Christian Ciemniak}
\affiliation{Physik-Department, Technische Universit\"at M\"unchen, Germany}
\author{Davide D'Angelo}
\affiliation{Dipartimento di Fisica, Universit\`a degli Studi e INFN, Milano, Italy}
\author{Basudeb Dasgupta}
\affiliation{Department of Physics, Ohio State University, Columbus, OH, USA}
\author{Alexander Derbin}
\affiliation{Petersburg Nuclear Physics Institute, St. Petersburg, Russia}
\author{Amol Dighe}
\affiliation{Department of Theoretical Physics, Tata Institute of Fundamental Research, Mumbai, India}
\author{Grigorij Domogatsky}
\affiliation{Institute for Nuclear Research, Russian Academy of Sciences, Moscow, Russia}
\author{Steve Dye}
\affiliation{Hawaii Pacific University, Kaneohe, HI, USA}
\author{Sergey Eliseev}
\affiliation{Petersburg Nuclear Physics Institute, St. Petersburg, Russia}
\author{Timo Enqvist}
\affiliation{Oulu Southern Institute and Department of Physics, University of Oulu, Finland}
\author{Alexey Erykalov}
\affiliation{Petersburg Nuclear Physics Institute, St. Petersburg, Russia}
\author{Franz von Feilitzsch}
\affiliation{Physik-Department, Technische Universit\"at M\"unchen, Germany}
\author{Gianni Fiorentini}
\affiliation{Dipartimento di Fisica, Universit\`a e INFN, Ferrara, Italy}
\author{Tobias Fischer}
\affiliation{GSI, Helmholtzzentrum f\"ur Schwerionenforschung, Darmstadt, Germany}
\author{Marianne G\"oger-Neff}
\affiliation{Physik-Department, Technische Universit\"at M\"unchen, Germany}
\author{Peter Grabmayr}
\affiliation{Kepler Center f\"ur Astro- und Teilchenphysik, Eberhard Karls Universit\"at T\"ubingen}
\author{Caren Hagner}
\affiliation{Institut f\"ur Experimentalphysik, Universit\"at Hamburg, Germany}
\author{Dominikus Hell\-gart\-ner}
\affiliation{Physik-Department, Technische Universit\"at M\"unchen, Germany}
\author{Johannes Hissa}
\affiliation{Oulu Southern Institute and Department of Physics, University of Oulu, Finland}
\author{Shunsaku Horiuchi}
\affiliation{Department of Physics, Ohio State University, Columbus, OH, USA}
\author{Hans-Thomas Janka}
\affiliation{Max-Planck-Institut f\"ur Astrophysik, Garching, Germany}
\author{Claude Jaupart}
\affiliation{Institut de Physique du Globe de Paris, France}
\author{Josef Jochum}
\affiliation{Kepler Center f\"ur Astro- und Teilchenphysik,
             Eberhard Karls Universit\"at T\"ubingen}
\author{Tuomo Kalliokoski}
\affiliation{Department of Physics, University of Jyv\"askyl\"a, Finland}
\author{Alexei Kayunov}
\affiliation{Petersburg Nuclear Physics Institute, St. Petersburg, Russia}
\author{Pasi Kuusiniemi}
\affiliation{Oulu Southern Institute and Department of Physics, University of Oulu, Finland}
\author{Tobias Lachenmaier}
\affiliation{Kepler Center f\"ur Astro- und Teilchenphysik, Eberhard Karls Universit\"at T\"ubingen}
\author{Ionel Lazanu}
\affiliation{Faculty of Physics, University of Bucharest, Romania}
\author{John G. Learned}
\affiliation{Department of Physics and Astronomy, University of HawaiÕi, Honolulu, HI, USA}
\author{Timo Lewke}
\affiliation{Physik-Department, Technische Universit\"at M\"unchen, Germany}
\author{Paolo Lombardi}
\affiliation{Dipartimento di Fisica, Universit\`a degli Studi e INFN, Milano, Italy}
\author{Sebastian Lorenz}
\affiliation{Institut f\"ur Experimentalphysik, Universit\"at Hamburg, Germany}
\author{Bayarto Lubsandorzhiev}
\affiliation{Institute for Nuclear Research, Russian Academy of Sciences, Moscow, Russia}
\affiliation{Kepler Center f\"ur Astro- und Teilchenphysik, Eberhard Karls Universit\"at T\"ubingen}
\author{Livia Ludhova}
\affiliation{Dipartimento di Fisica, Universit\`a degli Studi e INFN, Milano, Italy}
\author{Kai Loo}
\affiliation{Department of Physics, University of Jyv\"askyl\"a, Finland}
\author{Jukka Maalampi}
\affiliation{Department of Physics, University of Jyv\"askyl\"a, Finland}
\author{Fabio Mantovani}
\affiliation{Dipartimento di Fisica, Universit\`a e INFN, Ferrara, Italy}
\author{Michela Marafini}
\affiliation{Laboratoire Astroparticule et Cosmologie, Universit\'e Paris 7 (Diderot), France}
\author{Jelena Maricic}
\affiliation{Department of Physics, Drexel University, Philadelphia, PA, USA}
\author{Teresa Marrod\'an Undagoitia}
\affiliation{Physik-Institut, Universit\"at Z\"urich, Switzerland}
\author{William F.\,McDonough}
\affiliation{Department of Geology, University of Maryland, MD, USA}
\author{Lino Miramonti}
\affiliation{Dipartimento di Fisica, Universit\`a degli Studi e INFN, Milano, Italy}
\author{Alessandro Mirizzi}
\affiliation{II Institut f\"ur Theoretische Physik, Universit\"at Hamburg, Germany}
\author{Quirin Meindl}
\affiliation{Physik-Department, Technische Universit\"at M\"unchen, Germany}
\author{Olga Mena}
\affiliation{Instituto de F\'isica Corpuscular, University of Valencia, Spain}
\author{Randolph M\"ollenberg}
\affiliation{Physik-Department, Technische Universit\"at M\"unchen, Germany}
\author{Valentina Muratova}
\affiliation{Petersburg Nuclear Physics Institute, St. Petersburg, Russia}
\author{Rolf Nahnhauer}
\affiliation{DESY, Zeuthen, Germany}
\author{Dmitry Nesterenko}
\affiliation{Petersburg Nuclear Physics Institute, St. Petersburg, Russia}
\author{Yuri N. Novikov}
\affiliation{Petersburg Nuclear Physics Institute, St. Petersburg, Russia}
\author{Guido Nuijten}
\affiliation{Rockplan Ltd., Helsinki, Finland}
\author{Lothar Oberauer}
\affiliation{Physik-Department, Technische Universit\"at M\"unchen, Germany}
\author{Sandip Pakvasa}
\affiliation{Department of Physics and Astronomy, University of HawaiÕi, Honolulu, HI, USA}
\author{Sergio Palomares-Ruiz}
\affiliation{Centro de F\'isica Te\'orica de Part\'iculas, Instituto Superior T\'ecnico, Lisboa, Portugal}
\author{Marco Pallavicini}
\affiliation{Dipartimento di Fisica, Universit\`a e INFN, Genova, Italy}
\author{Silvia Pascoli}
\affiliation{IPPP, Department of Physics, Durham University, Durham, UK}
\author{Thomas Patzak}
\affiliation{Laboratoire Astroparticule et Cosmologie, Universit\'e Paris 7 (Diderot), France}
\author{Juha Pel\-to\-nie\-mi}
\affiliation{Neutrinica Oy, Oulu, Finland}
\author{Walter Potzel}
\affiliation{Physik-Department, Technische Universit\"at M\"unchen, Germany}
\author{Tomi R\"aih\"a}
\affiliation{Oulu Southern Institute and Department of Physics, University of Oulu, Finland}
\author{Georg G. Raffelt}
\affiliation{Max-Planck-Institut f\"ur Physik, M\"unchen, Germany}
\author{Gioacchino Ranucci}
\affiliation{Dipartimento di Fisica, Universit\`a degli Studi e INFN, Milano, Italy}
\author{Soebur Razzaque}
\affiliation{George Mason University, Fairfax, VA, USA}
\author{Kari Rummukainen}
\affiliation{University of Helsinki and Helsinki Institute of Physics, Finland}
\author{Juho Sarkamo}
\affiliation{Oulu Southern Institute and Department of Physics, University of Oulu, Finland}
\author{Valerij Sinev}
\affiliation{Institute for Nuclear Research, Russian Academy of Sciences, Moscow, Russia}
\author{Christian Spiering}
\affiliation{DESY, Zeuthen, Germany}
\author{Achim Stahl}
\affiliation{III. Physikalisches Institut, RWTH Aachen University, Germany}
\author{Felicitas Thorne}
\affiliation{Physik-Department, Technische Universit\"at M\"unchen, Germany}
\author{Marc Tippmann}
\affiliation{Physik-Department, Technische Universit\"at M\"unchen, Germany}
\author{Alessandra Tonazzo}
\affiliation{Laboratoire Astroparticule et Cosmologie, Universit\'e Paris 7 (Diderot), France}
\author{Wladyslaw H. Trzaska}
\affiliation{Department of Physics, University of Jyv\"askyl\"a, Finland}
\author{John D. Vergados}
\affiliation{Physics Department, University of Ioannina, Greece}
\author{Christopher Wiebusch}
\affiliation{III. Physikalisches Institut, RWTH Aachen University, Germany}
\author{J\"urgen Winter}
\affiliation{Physik-Department, Technische Universit\"at M\"unchen, Germany}

\maketitle

\tableofcontents

\cleardoublepage

\onecolumngrid

\section{Introduction}
\label{sec::introduction}

\twocolumngrid

\input{introduction}

\onecolumngrid
\newpage

\section{Detector design}
\label{sec::hardware}

\twocolumngrid

\input{detector}

\subsection{Liquid scintillator}
\label{subsec::scintillator}

\input{scintillator}

\subsection{Light detection}
\label{subsec::pmts}

\input{pmts}

\subsection{Read-out electronics}
\label{subsec::readout}

\input{readout}

\onecolumngrid
\newpage

\section{Detector performance}
\label{sec::performance}

\twocolumngrid

\noindent The expected performance of the final LENA detector can be extrapolated to a large extent from the currently running large-volume liquid-scintillator detectors, first of all of course Borexino and KamLAND. However, a precise assessment of the accuracy achieved in the reconstruction of neutrino events and background levels present in the detector requires detailed Monte Carlo simulations. The necessary input on the properties of scintillator, PMTs and detector electronics is presented in Sec.~\ref{sec::hardware}. 

We start out with a set of baseline parameters for the detector components in Sec.~\ref{subsec::baspar} which are later on implemented in the phenomenological studies of Secs.\ref{sec::le}-\ref{sec::he}. The precision of vertex and energy reconstruction is discussed in Sec.~\ref{subsec::vertex}. A special emphasis is given to the reconstruction of extended particle tracks at higher energies in Sec.~\ref{subsec::tracking}, motivated by its importance for long-baseline oscillation physics at GeV energies. A discussion on the backgrounds affecting the neutrino and rare event searches is presented in the corresponding phenomenological sections.

\subsection{Baseline parameters}
\label{subsec::baspar}

\input{baspar}

\subsection{Low-energy vertex reconstruction}
\label{subsec::vertex}

\input{vertex}

\subsection{GeV event reconstruction}
\label{subsec::tracking}

\input{tracking}

\onecolumngrid
\newpage

\section{Astrophysical neutrino sources}
\label{sec::le}

\twocolumngrid

\noindent LENA's \nomenclature{LENA}{Low Energy Neutrino Astronomy}  
core science program is in the low-energy range with
neutrino energies up to a few tens of MeV. Most of the relevant
sources are based on nuclear reactions defining this energy scale, in
particular the Sun, Earth, and power reactors. The same energy range
is covered by the quasi-thermal emission of neutrinos by collapsing
stars. The science goals reach from a better understanding of
astrophysical sources and the Earth to the investigation of neutrino
properties based on flavor oscillations. We begin with neutrinos emitted from extraterrestrial sources:
Core-collapse supernovae (SNe) create intense bursts of neutrinos, and in particular the next galactic SN will provide more than ten thousand neutrino interactions in LENA (Sec.~\ref{subsec::sn}). In addition, the detector will be sensitive to the faint signal generated by the diffuse flux from all past SNe (Sec.~\ref{subsec::dsnb}). Also neutrinos from thermonuclear fusion reactions within the Sun will provide a high-statistics signal (Sec.~\ref{subsec::solar}). Finally, the annihilation signature of dark-matter particles is studied in Sec.~\ref{subsec::idms}.

\subsection{Galactic Supernova neutrinos}
\label{subsec::sn}

\input{sn}

\subsection{Diffuse Supernova neutrinos}
\label{subsec::dsnb}

\input{dsnb}

\subsection{Solar neutrinos}
\label{subsec::solar}

\input{solar}

\subsection{Indirect dark matter search}
\label{subsec::idms}

\input{idms}

\onecolumngrid
\newpage

\section{Terrestrial neutrino sources}

\twocolumngrid

\noindent Not only astrophysical bodies, but also our own planet is a source of low-energetic neutrinos (Sec.~\ref{subsec::geo}). Moreover, there is a number of anthropogenic neutrino sources intense enough at low energies to allow for neutrino oscillation physics: Antineutrinos from nuclear reactors (Sec.~\ref{subsec::reactor}) may allow for a high-precision measurement of the ``solar'' neutrino mixing parameters. Neutrinos from strong radioactive electron capture sources may provide a unique opportunity to investigate flavor oscillations on a very short
baseline, providing sensitivity to the mixing angle $\theta_{13}$ and especially $\nu_{e}$ disappearance into sterile neutrinos (Sec.~\ref{subsec::oscillometry}). Alternatively, low-energy neutrinos generated by a pion decay at-rest beam might offer sensitivity to $\theta_{13}$ and especially the CP-violating phase $\delta_\mathrm{CP}$
(Sec.~\ref{subsec::daedalus}). Doting the scintillator target might allow to search for neutrinoless double-beta decay (Sec.~\ref{subsec::doublebeta}). Finally, a large low-energy detector is sensitive to many other rare processes connected to non-standard interaction of neutrinos or exotic decay processes (Sec.~\ref{subsec::rareproc}).

\subsection{Geoneutrinos}
\label{subsec::geo}

\input{geo}

\subsection{Reactor neutrinos}
\label{subsec::reactor}

\input{reactor}

\subsection{Neutrino oscillometry}
\label{subsec::oscillometry}

\input{oscillometry}

\subsection{Pion decay at-rest experiment}
\nomenclature{DAE$\delta$ALUS}{Decay At-rest Experiment for $\delta_{CP}$ studies At the Laboratory for Underground Sciences}
\label{subsec::daedalus}
%
\input{daedalus}

\subsection{Neutrinoless double-beta decay}
\label{subsec::doublebeta}

\input{doublebeta}

\subsection{Search for other rare processes}
\label{subsec::rareproc}

\input{rareproc}

\onecolumngrid
\newpage

\section{GeV physics}
\label{sec::he}
\twocolumngrid

\noindent While the emphasis of the LENA physics program is on low-energy neutrinos ($E<100$\,MeV), the experiment can also contribute to several aspects of neutrino and particle physics associated to GeV energies. Actually, the search for proton decay into kaon and antineutrino was one of the first items considered to play an integral part in the LENA concept, since the visibility of the kaon's energy deposition in the scintillator highly increases the detection efficiency  substantially in comparison to water Cherenkov detectors (Sec.~\ref{subsec::pdecay}).

In the last years, it also became evident that liquid-scintillator detectors will be a serious option for the use as a far detector in a long-baseline neutrino beam experiment, and for the investigation of atmospheric neutrino oscillations. An overview of possible neutrino beam experiments and the information that could be won from atmospheric neutrinos are outlined in Secs.~\ref{subsec::beam} and \ref{subsec::atmospherics}, respectively.

\subsection{Nucleon decay search}
\label{subsec::pdecay}

\input{pdecay}

\subsection{Long-baseline neutrino beams}
\label{subsec::beam}

\input{beam}

\subsection{Atmospheric neutrinos}
\label{subsec::atmospherics}

\input{atmospherics}


\onecolumngrid
\newpage


\section{Conclusions}
\twocolumngrid
\label{sec::conclusions}

\input{conclusions}


\vspace{1.5cm}

\begin{acknowledgments} 
\noindent This work was supported by the Maier-Leibnitz-Laboratorium (Garching), by the Deutsche Forschungsgemeinschaft DFG (Transregio 27: ``Neutrinos and Beyond", SFB 676: ``Particles, Strings and Early Universe", and the Munich Cluster of Excellence ``Origin and Structure of the Universe"), by the German BMBF (WTZ Project Rus 07/015), by the Russian Minobrnauky Project No.\,2.2.1, by the Council of Oulu Region, by the European Union Regional Development Funds, by the Portuguese FCT through CERN/FP/109305/2009 and CFTP-FCT UNIT 777, which are partially funded through POCTI (FEDER), by the Spanish Grant FPA2008-02878 of the MICINN, and by the European Community via the ``LAGUNA - Design of a pan-European Infrastructure for Large Apparatus studying Grand Unification and Neutrino Astrophysics'' FP7 Grant (GA 212343).
\end{acknowledgments}

\onecolumngrid
\newpage

\section*{List of Abbreviations}
\twocolumngrid

\begin{spacing}{1.2}
\begin{tabbing}
LALALALal \= Links \kill
0$\nu$2$\beta$ \> Neutrino-less Double Beta decay \\
ADC \> Analog-to-Digital Converter \\
AP \> AfterPulses \\
ASIC \> Application-Specific Integrated Circuit \\
Bis-MSB \> BIS-o-Methyl-Styryl-Benzene, a fluor \\
Borexino \> derived from BORon EXperiment \\
BSE \> Bulk Silicate Earth model \\
C12 \> Dodecane, a hydrocarbon \\
CMB \> Core-Mantle Boundary \\
CNO \> Carbon-Nitrogen-Oxygen fusion cycle \\
CTF \> Counting Test Facility of Borexino \\
CUPP \> Center for Underground Physics in \\
	\> Pyh\"asalmi \\
DAE$\delta$ALUS \> Decay At-rest Experiment for $\delta_\mathrm{CP}$\\
	\> studies At the Laboratory for \\
	\> Underground Sciences \\ 
DC \> Dark Count \\
DIN \> DiIsopropyl-Naphtalene \\
DM \> Dark Matter \\
DSNB \> Diffuse Supernova Neutrino Background \\
EC \> Electron Capture \\
ECR \> Electron Cyclotron Resonance source \\
EURISOL \> European Isotope Separation On-Line \\
	\> radioactive ion beam facility \\
FADC \> Fast Analog-to-Digital Converter \\
FPGA \> Field-Programmable Gate Array \\
GALLEX \> GALLium EXperiment \\
GEANT4 \> GEometry ANd Tracking MC platform \\
GLACIER \> Giant Liquid Argon Charge Imaging\\
	\> ExpeRriment \\
GLoBES \> General LOng-Baseline Experiment \\
	\> Simulator \\
GNO \> Gallium Neutrino Observatory \\
GUT \> Grand Unified Theories \\
HPT \> Hybrid Photo Tube \\
HQE \> High Quantum Efficiency \\
K2K \> KEK-to-Kamiokande neutrino beam \\
Kamiokande \> KAMIOKA Nucleon Decay Experiment \\
KamLAND \> KAMioka Liquid-scintillator \\
	\> Anti-Neutrino Detector \\
	\> \\	
LAGUNA \> Large Apparatus for Grand Unification \\
	\> and Neutrino Astrophyics \\
LENA	 \> Low Energy Neutrino Astronomy \\
LETA \> Low Energy Threshold Analysis of SNO \\
LINAC \> LINear ACcelerator \\
LMA \> Large Mixing Angle oscillation scenario \\
LSM \> Laboratoire Souterrain de Modane \\
MC \> Monte Carlo simulation \\
MEMPHYS \> MEgaton Mass PHYSics \\
OC \> Optical Coverage \\
PDE \> Photo Detection Efficiency \\
pe \> PhotoElectron \\
pep \> Proton-Electron-Proton fusion \\
PMP \> Phenyl-Mesityl-Pyrazoline, a fluor \\
PMT \> PhotoMultiplier Tube \\
pot \> Protons On Target \\
pp \> Proton-Proton fusion \\
PPO \> diPhenyl-Oxazole, a fluor \\
PREM \> Preliminary Reference Earth Model \\
PXE	 \> Phenyl-Xylyl-Ethane, organic solvent \\
QE \> Quantum Efficiency \\
RAA \> Reactor Antineutrino Anomaly \\
RPC \> Resistive Plate Chamber \\
SAGE \> Sowjet-American Gallium Experiment \\
SER \> Single Electron Resolution \\
SiPM \> SIlicon Photo Multiplier \\
SM \> the Standard Model of particle physics \\
SN \> Supernova \\
SNO \> Sudbury Neutrino Observatory \\
SNP \> Solar Neutrino Problem \\
SPS \> 	Super Proton Synchrotron \\
SSM \> Standard Solar Model \\
SUSY \> SUper SYmmetry \\
T2K \> Tokai-to-Kamiokande neutrino beam \\
TDC \> Time-to-Digital Converter \\
TNU \> Terrestrial Neutrino Unit \\
TOF \> Time Of Flight \\
TTS \> Transit Time Spread \\
\end{tabbing}
\end{spacing}

\onecolumngrid
\newpage
\twocolumngrid

\bibliographystyle{h-physrev}
\bibliography{whitepaper}

\end{document}

%% file: abstract.tex

\begin{abstract}

\vspace{2.5cm}

\noindent As part of the European LAGUNA design study on a next-generation neutrino detector, we propose the liquid-scintillator detector LENA (Low Energy Neutrino Astronomy) as a multipurpose neutrino observatory. The outstanding successes of the Borexino and KamLAND experiments demonstrate the large potential of liquid-scintillator detectors in low-energy neutrino physics. Low energy threshold, good energy resolution and efficient background discrimination are inherent to the liquid-scintillator technique. A target mass of 50\,kt will offer a substantial increase in detection sensitivity. 

At low energies, the variety of detection channels available in liquid scintillator will allow for an energy- and flavor-resolved analysis of the neutrino burst emitted by a galactic Supernova. Due to target mass and background conditions, LENA will also be sensitive to the faint signal of the Diffuse Supernova Neutrino Background. Solar metallicity, time-variation in the solar neutrino flux and deviations from MSW-LMA survival probabilities can be investigated based on unprecedented statistics. Low background conditions allow to search for dark matter by observing rare annihilation neutrinos. The large number of events expected for geoneutrinos will give valuable information on the abundances of Uranium and Thorium and their relative ratio in the Earth's crust and mantle. Reactor neutrinos enable a high-precision measurement of solar mixing parameters. A strong radioactive or pion decay-at-rest neutrino source can be placed close to the detector to investigate neutrino oscillations for short distances and sub-MeV to MeV energies. 

At high energies, LENA will provide a new lifetime limit for the SUSY-favored proton decay mode into kaon and antineutrino, surpassing current experimental limits by about one order of magnitude. Recent studies have demonstrated that a reconstruction of momentum and energy of GeV particles is well feasible in liquid scintillator. Monte Carlo studies on the reconstruction of the complex event topologies found for neutrino interactions at multi-GeV energies have shown promising results. If this is confirmed, LENA might serve as far detector in a long-baseline neutrino oscillation experiment currently investigated in LAGUNA-LBNO.



\vspace{20cm}

\end{abstract}

%% file: introduction.tex
\noindent Over the past decades, neutrinos have been firmly established as
astronomical messengers. The feeble interaction strength of these
elusive particles requires unusually large detectors, but on the
other hand allows us to investigate processes in the deep interior
of stars that are shrouded from view in other forms of radiation.
Neutrino astronomy therefore complements observations in the
electromagnetic spectrum, charged cosmic rays, and gravitational
waves. In recognition of this importance, the pioneering first
observations of solar and supernova (SN\nomenclature{SN}{Supernova}) neutrinos were honored with
the Physics Nobel Prize in 2002.

Even the first solar neutrino observations about 40 years ago showed
an apparent deficit that today is unambiguously explained by flavor
oscillations. In this way neutrino astronomy has triggered an
avalanche of fundamental discoveries, shedding completely new light
on the inner properties of neutrinos with ramifications both for the
fundamental theory of elementary particles and the universe at
large.

With the standard three-flavor oscillation scenario established,
neutrinos can be used as new messengers from astrophysical sources.
In this context, a variety of far-reaching questions can be
addressed, notably
\begin{itemize}
\item Is the core-collapse SN paradigm correct? Are there
    substructures in the neutrino signal?
\item How large is the flux of the diffuse SN neutrino
    background (DSNB\nomenclature{DSNB}{Diffuse Supernova Neutrino Background}) and what is its spectrum?
\item What is the Sun's metal content? Are there time-variations
    in the solar fusion rate?
\item How large is the concentration of radioactive elements in
    the Earth and what is their contribution to its heat flow?
\item What is the dark matter of the universe?
\end{itemize}
On the other hand, neutrinos remain fundamental particle-physics
messengers and large-scale detectors will shed new light on topics
like
\begin{itemize}
\item What is the value of $\theta_{13}$?
\item Is the CP symmetry violated among leptons?
\item What is the neutrino mass hierarchy?
\item Do sterile neutrinos exist?
\item Are there non-standard neutrino interactions and how do
    they affect flavor oscillations?
\item Is baryon number conserved?
\end{itemize}
These question will be addressed by observing low-energy neutrinos
from SNe, the Sun, Earth, reactors, and radioactive sources, by neutrino beams and
atmospheric neutrinos in the GeV range, and finally by signatures of
possible nucleon decays in the detector material itself.

\begin{figure}[b!]
\centering
\includegraphics[width=0.3\textwidth]{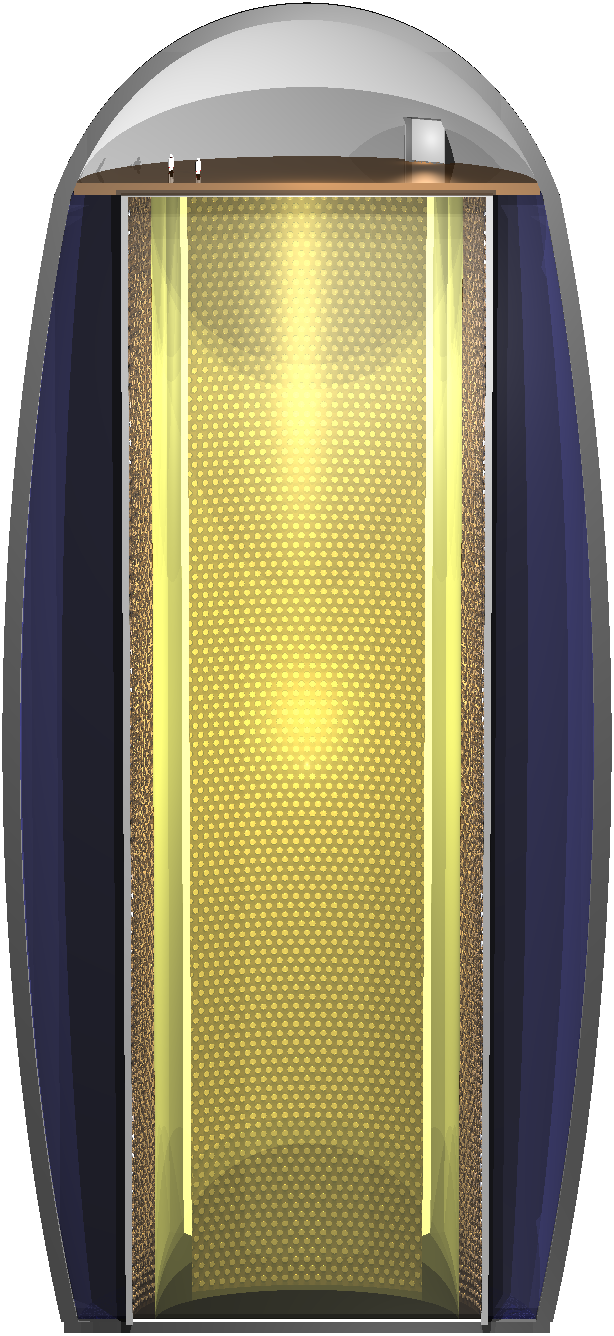}
\caption{Artist's view of the LENA detector: The detector tank is 100\,m in height and 30\,m in diameter. See Fig.\,\ref{fig:lena} for details.}
\label{fig::lenascience}
\end{figure}

The small event rate of neutrino interactions or the search for
extremely rare processes requires a large target mass. In the past,
large-volume unsegmented detectors have played a dominant role in
this field. Originally triggered by the search for nucleon decay,
the Kamiokande and later Super-Kamiokande water Cherenkov detectors
provided crucial measurements of solar, atmospheric, SN and beam
neutrinos. It was only the huge target mass of 50~kt that allowed
Super-Kamiokande to accrue enough statistics to measure precisely
the deformation of atmospheric neutrino spectra caused by flavor
oscillations. In a parallel development, liquid-scintillator
detectors on the kiloton scale explored neutrino fluxes at energies
below 5\,MeV. In particular, the KamLAND measurements of reactor
neutrino oscillations tightly confine the mass-squared difference of
solar neutrino mixing, while Borexino confirmed solar neutrino
oscillations at sub-MeV energies. Both detectors provided first
evidence for the faint geoneutrino signal originating from
radioactive elements embedded in the Earth's crust.

\medskip Based on this success, we propose a next-generation neutrino
observatory LENA (``low energy neutrino astronomy''). It is foreseen
as an unsegmented liquid-scintillator detector of 50\,kt target mass
(Fig.\,\ref{fig::lenascience}), combining the advantages of the
low-energy threshold and background discrimination capabilities of
Borexino and KamLAND with the size of Super-Kamiokande.

LENA will profit from the virtues of the scintillator technique that
were impressively demonstrated by KamLAND and Borexino.

\begin{itemize}
\item\textbf{Good energy resolution below 10~MeV.} The light
    yield is at least 200 photoelectrons per MeV, corresponding
    to about 3\% energy resolution at 5~MeV.
\item\textbf{Low detection threshold.} A neutrino energy of
    1.8~MeV is the threshold for inverse beta decay. For
    electron scattering, the threshold can be at a recoil energy
    as low as 200\,keV, the limit arising from the intrinsic
    background of radioactive {$^{14}$C} in the scintillator.
\item\textbf{Excellent background discrimination.} The
    final-state neutron of inverse beta decay provides a clear
    coincidence signature for $\bar\nu_e$ detection. Pulse shape
    analysis allows for an efficient discrimination against fast
    neutrons and, for detecting elastic neutrino-electron
    scattering, against alphas and even positrons.
\item\textbf{Radio purity.} Years of development and experience
    in Borexino have advanced the techniques for scintillator
    purification, identifying the most efficient methods.
\item\textbf{Self shielding.} A large monolithic detector
    shields its central detection volume against external
    backgrounds.
\end{itemize}
In most respects, the performance is competitive with a water
Cherenkov detector of several times its size.

LENA will be a true multi-purpose facility. High-statistics
measurements of strong neutrino sources like a galactic
core-collapse SN, the Sun or the Earth's interior will resolve
energy spectra and their time evolution in unprecedented detail.
Reactor neutrinos enable a high-precision measurement of the
``solar'' neutrino mixing parameters. In addition, a strong
radioactive neutrino source can be placed close to the detector to
investigate flavor oscillations at short distances and sub-MeV
energies. At the same time, the search for very rare events becomes
possible because the excellent background rejection allows to
identify a handful of events out of several years of data. Thus, the
faint flux of the predicted Diffuse Supernova Neutrino Background
(DSNB) is well within reach. Likewise, observation of rare
annihilation neutrinos allows for indirect dark matter search.

This rich low-energy program is complemented by several physics
objectives at GeV energies. LENA will further advance the search for
proton decay and thus baryon number violation. The new lifetime
sensitivity for the proton decay mode into kaon and antineutrino,
favored by supersymmetric theories, will surpass current
experimental limits by about one order of magnitude. Moreover,
recent studies indicate that a large-volume liquid-scintillator
detector can resolve both momentum and energy of GeV particles with
a precision of a few percent. Monte Carlo simulations of the complex
event topologies of charged-current neutrino interactions show
promising results for the reconstruction capabilities. These
techniques may offer the opportunity to use LENA as far detector in
long-baseline neutrino oscillation experiments, either for an
accelerator-produced neutrino beam or atmospheric neutrinos.

LENA is one of three options discussed within the
LAGUNA\nomenclature{LAGUNA}{Large Apparatus for Grand Unification and Neutrino Astrophyics} and the forthcoming LAGUNO-LBNO design studies that are sponsored by the European Union under the 7th Framework
Programme. This design study aims at the eventual construction of a
large-volume neutrino observatory in a European underground
laboratory based on scintillator, water Cherenkov (MEMPHYS), or liquid argon (GLACIER)
techniques. Due to the high level of expertise built up in several
European and international research groups and dedicated R\&D
activities over the past years, the liquid-scintillator technique
can be regarded as sufficiently mature to allow for an early start
of detector realization. Based on recent feasibility studies, the
LENA construction time is estimated to about eight years.

This paper lays out the science case for LENA and the current state
of R\&D and detector design activities. It is meant to support and
justify a proposal for the construction of a next-generation large
neutrino observatory based on the liquid-scintillator technique. It
provides a work of reference for future discussions and decision
making. The paper starts out with the state of R\&D and detector design
activities in Sec.\,\ref{sec::hardware}, followed by an outline of
the technical detector properties (Sec.\,\ref{sec::performance}).
The core physics objectives are in the low-energy domain
where a liquid-scintillator detector can play out its unique
capabilities particularly well. The main low-energy topics revolve
around solar, supernova, reactor and geo neutrinos that are
described in Sec.\,\ref{sec::le}. In addition, LENA has convincing
capabilities in the range of GeV energies, where the search for
nucleon decay and flavor oscillation physics with
accelerator-produced beams and atmospheric neutrinos form the main
topics that are discussed in Sec.\,\ref{sec::he}. 
The paper concludes with a brief summary and outlook in
Sec.\,\ref{sec::conclusions}.

%% file: detector.tex

%

\noindent Design, construction, and operation of the LENA detector will be a challenging endeavor. However, there are two neutrino detectors in operation that already anticipate scale and techniques of the LENA project: The Super-Kamiokande detector is of almost the same volume, featuring similar requirements concerning detector cavern, photocoverage and number of channels. On the other hand, the enormous amount of R\&D that led to the tremendous success of the Borexino experiment can be re-applied for LENA, covering questions concerning the liquid scintillator, the radiopurity of the used materials and their purification, requirements for photosensors and electronic read-out and so forth. Based on this foundation, but also on the laboratory and design activity carried out in the last few years especially for LENA, the following section describes the current design draft for LENA.

\medskip
\noindent Fig.\,\ref{fig:lena} shows a schematic overview of the current LENA design:

\begin{figure}[!h]
\centering\includegraphics[width=0.45\textwidth]{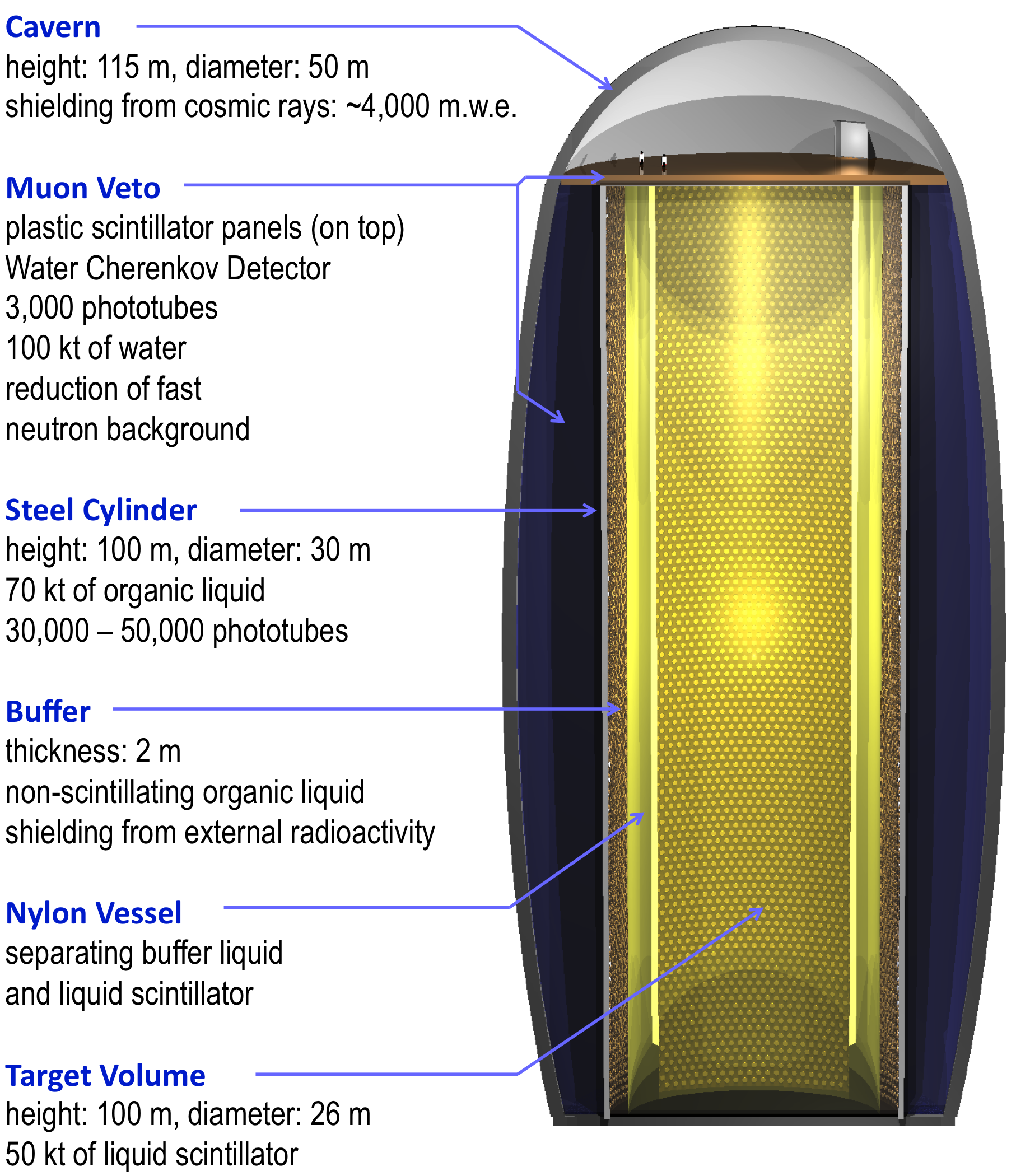}
\caption[Schematical view of the LENA detector]{Schematical view of the LENA detector \cite{lenastudy}.} 
\label{fig:lena}
\end{figure}

\medskip\noindent\textbf{Laboratory.}
The detector will be constructed in a dedicated cavern, about 115\,m in height. The shape will depend on the laboratory site: Pyh\"asalmi and Fr\'ejus will be described as exemplary sites in Sec.\,\ref{subsec::site}. The aspired rock shielding above the detector corresponds to 4\,000\,mwe, a requirement fulfilled by both sites.

\medskip\noindent\textbf{Tank.}
The liquid-scintillator will be contained in a cylindric steel or concrete tank of 100\,m height and 30\,m diameter. Several design options are discussed in Sec.\,\ref{subsec::tank}. Inside the tank, the volume is divided by a thin nylon vessel into the buffer volume shielding external radioactivity and the target volume. 

\medskip\noindent\textbf{Liquid scintillator.}
The target volume is 26$\,$m in diameter and 100\,m in height,
corresponding to 5.3$\times$10$^{4}$\,m$^{3}$. Depending on the exact composition of the liquid scintillator (Sec.\ref{subsec::scintillator}), the target mass ranges from 45 to 53$\,$kt. The buffer volume is filled with an inactive liquid, which should have a similar density as the scintillator in order to minimize buoyancy forces on the nylon vessel. The scintillator transparency will be monitored by a system of reference LEDs viewed by PMTs across the tank diameter. Light yield calibration at low energies will be performed by introducing radioactive sources at different positions inside the detection volume, while cosmic muons tracks reconstructed by the top muon veto provide information on the high-energy detector response.

\medskip\noindent\textbf{Nylon vessel.} A cylindrical nylon tube of 26\,m diameter and a surface of $\sim$10$^4$\,m$^2$ is needed to separate the active target from the inactive buffer volume. The vessel will be very thin ($\sim$100\,\textmu m), and fastened by two flanges to the tank lid and bottom. Intermediate support by nylon strings reaching out to the tank walls will help to carry the weight and to keep the cylindrical shape. Compared to Borexino, the construction of the vessel might prove easier as -- compared to a spheciral vessel -- a cylinder needs much less glued seams. Nevertheless, clean and radiopure fabrication of a vessel of these dimensions will be challenging. Currently, studies are on-going to integrate a small buffer volume within the encapsulation of each individual PMT which might allow to omit the vessel, filling the whole tank with active scintillator.

\medskip\noindent\textbf{Photomultipliers.} The intended photosensitive coverage is 30\,\% of the inner tanks walls. This requires e.\,g.\,$\sim$45\,000 eight-inch photomultiplier tubes (PMTs). Currently PMTs with a photocathode diameter between 5 and 10 inch are the most likely solution. Reflective light-concentrators mounted on the PMTs will be used to reduce the number of PMTs (Sec.\ref{subsec::pmts}). PMT response and timing can be calibrated by a system of optical fibers fed by ns-long LASER pulses.  

\medskip\noindent\textbf{Readout electronics.} A further option to reduce the large number of channels is to group several PMTs into a PMT array, featuring a common high voltage supply, signal digitization and readout channel. Possible solutions for the read-out electronics are discussed in Sec.\ref{subsec::readout}. 

\medskip\noindent\textbf{Muon veto.} Cosmic muons crossing the main detector will be identified by layers of plastic scintillator panels, Resistive Plate Channels (RPCs\nomenclature{RPC}{Resistive Plate Chamber}), or limited streamer tubes mounted above the upper lid of the detector. A dense instrumentation featuring several layers would offer the possibility to aid the reconstruction of muon tracks in the scintillator. On the outside of the tank, the interspace to the cavern walls is filled with water (at least 2$\,$m in width) shielding the inner detector from external radiation coming from the rock and from muon-induced neutrons. The outer tank walls can be equipped with PMTs to identify cosmic muons passing the detector by their Cherenkov light.

\subsection{Laboratory sites}
\label{subsec::site}

\noindent No underground laboratory existing today is sufficiently large to host the LENA detector. This implies that a cavern of appropriate size must be excavated, along with additional shafts and tunnels to house the auxiliary systems for filling and operation of the detector. To minimize the associated costs, it seems reasonable to construct the detector adjacent to an underground infrastructure already existent, either an underground science laboratory or a deep mine. In the following, two exemplary sites in Europe are presented that would suit the depth and infrastructure requirements of LENA. These places have been identified in the course of the FP7 LAGUNA design study which will publish its conclusive results in 2011.

\subsubsection{Pyh\"asalmi}

\noindent The Pyh\"asalmi mine is located close to the geographic center of Finland, near the town of Pyh\"aj\"arvi. The distance from CERN is 2288\,km. The mine is the deepest in Europe, the bottom level at $\sim$1450\,m. The products are copper, zinc and pyrite, and operation will last at least until 2018. The mine already hosts a small underground laboratory, the Finnish Center for Underground Physics in Pyh\"asalmi (CUPP\nomenclature{CUPP}{Center for Underground Physics in Pyh\"asalmi}). The feasibility study for LENA at Pyh\"asalmi was carried out by the Finnish company Rockplan Ltd.\footnote{Kalliosuunnittelu Oy Rockplan Ltd, Asemamiehenkatu 2, 00520 Helsinki (Finland).} \cite{rockplan-site}.

\medskip\noindent\textbf{Geology.} The characteristics of the rock surrounding the mine are well known due to the exploratory work performed by the mining company. The cavern will be constructed adjacent to the deepest level of the mine, about 500\,m from the central mine shaft. At this depth, the rock will be very hard, dry and at 23$^{\circ}$C. The seismic activity in the region is very low. 

\medskip\noindent\textbf{Background levels.} The air content of radon at the deepest level is 20\,Bq/m$^3$, the muon flux is 1.1$\times$10$^{-4}$\,/m$^{2}$s. The closest nuclear power plant is 350\,km from the mine. The expected reactor $\bar\nu_e$ background has been calculated to 1.9$\times$10$^{5}$\,/cm$^{2}$s \cite{Wurm:2007cy}. However, Finland plans to construct two additional reactors within the next decades.

\begin{figure} [htp]
\begin{center}
\includegraphics[width=0.47\textwidth]{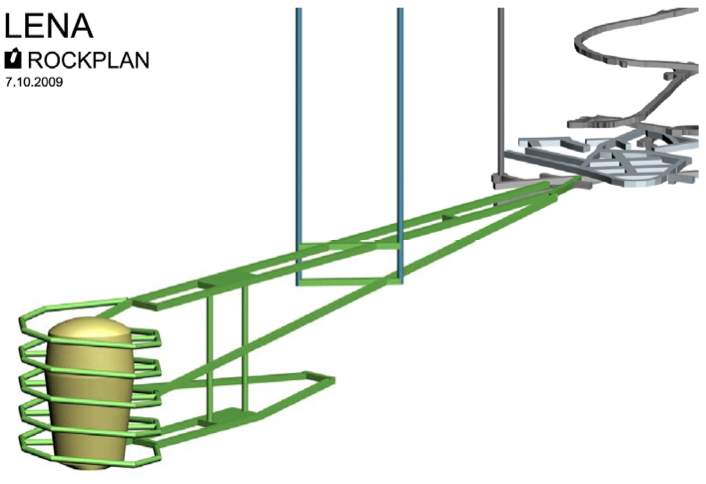}
\caption{LENA at Pyh\"asalmi (artistic impression by Rockplan Ltd.).}
\label{fig:lena_pyh}
\end{center}
\end{figure}

\medskip\noindent\textbf{Excavation.} The excavation will begin from the deepest mine level, creating two connecting tunnels from mine to cavern, a vertical shaft close to the laboratory, and the detector cavern itself (Fig.\,\ref{fig:lena_pyh}). To accommodate the high vertical and horizontal stresses acting on the final cavern, its walls will be curved and the ceiling domed, as depicted in Fig.\,\ref{fig:lena}. Moreover, the plan view will be elliptical, the semi-major axis aligned to the direction of main stress. Overall, a volume of 200\,000\,m$^3$ will be excavated, leaving room for an extensive water buffer.

\medskip\noindent\textbf{Infrastructure.} The laboratory can profit from the already available underground infrastructure of the working mine (power, ventilation, transport). In addition to the main shaft, a road tunnel spiraling from the surface to the deepest level of the mine will allow to bring large building elements to the detector cavern. There exists also the possibility to share to a certain extent the equipment and machines for underground excavation with the mining company. The transport of liquid scintillator to the laboratory to the mine will be possible both by road truck and freight trains as the mine is directly connected to the Finnish railway network.   

\subsubsection{Fr\'ejus}

\begin{figure*}[bht]
\begin{center}
\includegraphics[width=0.75\textwidth]{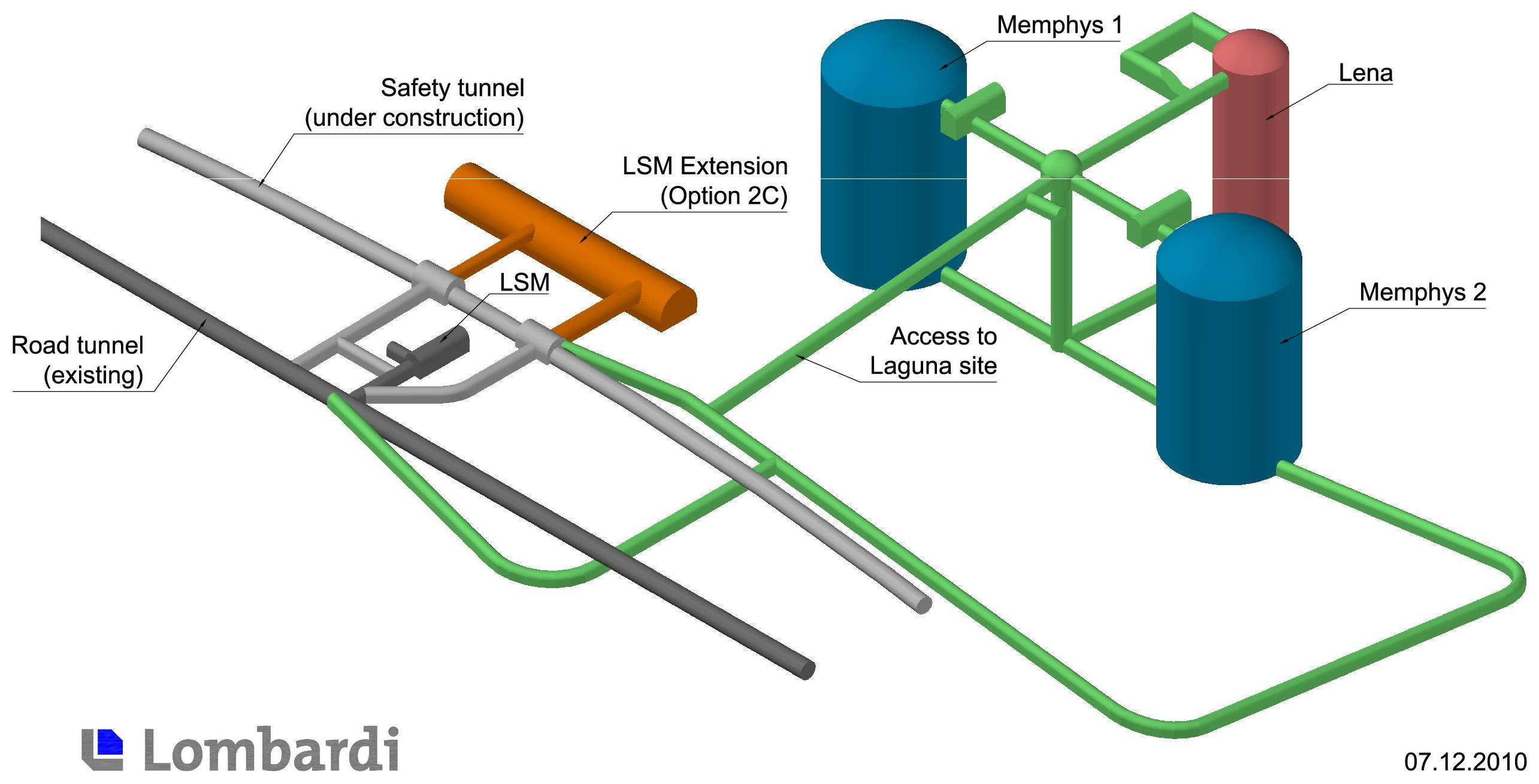}
\end{center}
\caption{LENA and MEMPHYS at Fr\'ejus (artistic impression by Lombardi Ltd.).}
\label{fig:lena_frejus}
\end{figure*}

\noindent The Laboratoire Souterrain de Modane (LSM\nomenclature{LSM}{Laboratoire Souterrain de Modane}) is located adjacent to the Fr\'ejus road tunnel in the French-Italian Alps, connecting Modane (F) and Bardonecchia (I). Originally, a new laboratory nearby has been discussed in the context of the MEMPHYS\nomenclature{MEMPHYS}{MEgaton Mass PHYSics}\nomenclature{GLACIER}{Giant Liquid Argon Charge Imaging ExpeRriment} detector. However, the FP7 LAGUNA design study has shown that Fr\'ejus will suit well the requirements of LENA. Lately, a laboratory hosting both detectors in a common infrastructure has been discussed (Fig.\,\ref{fig:lena_frejus}). The feasibility study was carried out by the Suisse company Lombardi Ltd.\footnote{Lombardi SA
Engineering Limited, Via R. Simen 19, 6648 Minusio (Switzerland).} \cite{lombardi-site}.

\medskip\noindent\textbf{Geology.} The characteristics of the rock surrounding the road tunnels have been investigated thoroughly during its excavation in the 1970s. In spite of its ductile behaviour, the calc-schist formation is of good quality for building, relatively dry and at a temperature of 30$^{\circ}$C. Seismic activity is present but not dangerous.

\medskip\noindent\textbf{Background levels.} The air content of radon was measured to 15\,Bq/m$^3$ in the LSM. Due to the large rock overburden of the Fr\'ejus mountain, corresponding to 4\,800\,mwe, the muon flux is very low, 5$\times$10$^{-5}$\,/m$^{2}$s. However, Fr\'ejus is close to the nuclear power plants of France, the closest at Bugey is merely 130\,km from the laboratory. The expected reactor $\bar\nu_e$ background has been calculated to 1.6$\times$10$^{6}$\,/cm$^{2}$s \cite{Wurm:2007cy}.

\medskip\noindent\textbf{Excavation.} The excavation of the large detector caverns will be made in various stages, using a preliminary support of anchors and shotcrete. Once excavated, the cavern walls will be sealed by a strong layer of concrete, more than 1\,m in width, in order to compensate for plasticity of the rock. The LENA cavern will be cylindrical with vertical walls, corresponding to an excavation volume of 100\,000\,m$^3$. Two additional caverns would hold the MEMPHYS detector (Fig.\,\ref{fig:lena_frejus}).

\medskip\noindent\textbf{Infrastructure.} The laboratory can profit from the already available underground infrastructure of the road tunnel (ventilation). Currently, a safety tunnel is being excavated close to the already existing road tunnel. Excavation works and transport of materials will mainly use this safety tunnel to minimize interference with road traffic. Liquid scintillator will be supplied by road trucks.   

\subsection{Detector tank}
\label{subsec::tank}

\noindent The Rockplan prefeasibility study on the LENA detector tank resulted in four options, two out of steel and two out of concrete \cite{rockplan-tank}. A concrete tank will be much more resistive to the compression generated by the water-scintillator density difference. However, it is also significantly more radioactive. To obtain the same fiducial volume for low-energy neutrinos, the diameter of the tank would have to be increased by 1-2\,m. The cost saving due to the low price of concrete will roughly compensate the additional expenses for organic solvent.

\begin{figure} [htp]
\centering
\includegraphics[width=0.3\textwidth,height=0.18\textheight]{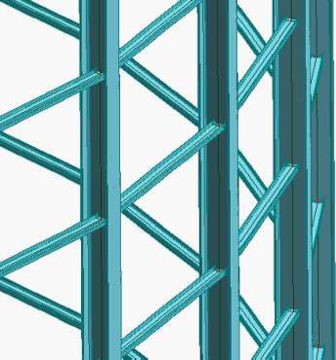}
\caption{Conventional steel tank (Rockplan Ltd.)}\label{fig:convsteeltank}
\end{figure}

\medskip\noindent\textbf{Conventional Steel Tank.}
A conventional tank requires a sizable amount of steel (driving the costs), and consists of many structural elements that would have to be brought separately into the laboratory and to be joined during underground construction. A basic steel mainframe will be erected first, followed by load-bearing plates and a final stainless steel sheet welded on (Fig.\,\ref{fig:convsteeltank}). There is also only one load-bearing layer separating scintillator and water. However, such a tank could be built straightforward and will be robust.

\begin{figure} [htp]
\centering
\includegraphics[width=0.3\textwidth,height=0.18\textheight]{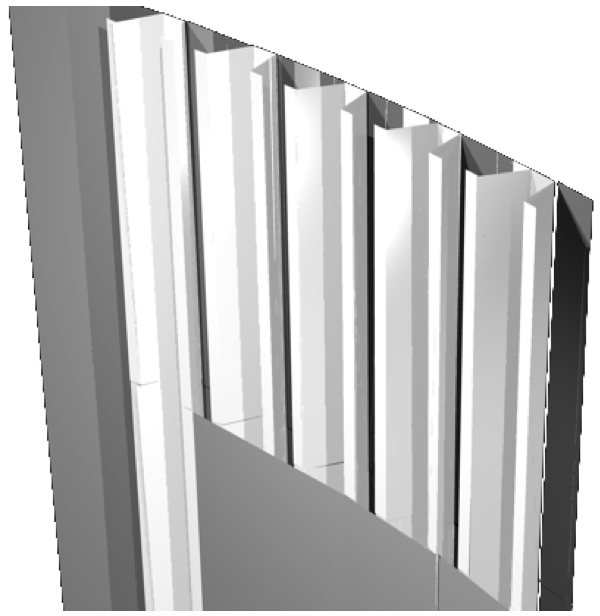}
\caption{Sandwich steel tank (Rockplan Ltd.)}\label{fig:sandwsteeltank}
\end{figure}

\medskip\noindent\textbf{Sandwich Steel Tank.}
This tank will consist of thin walled sandwich elements, featuring a very high strength-to-weight ratio and providing a multiple-layer defense against liquid leaks (Fig.\,\ref{fig:sandwsteeltank}). The elements can be prefabricated, reducing costs and allowing for extensive quality control. They will be lifted into place and welded together. There is also the opportunity to equip the interior of the elements with thermal insulation or cooling pipes, or to use it for active leak prevention. However, this tank will require a lot of welding (bearing the risk of radioactive contamination), and the mechanical design for tangential pressure will be challenging.

\medskip\noindent\textbf{Sandwich Concrete Tank.}
To assure water tightness, the concrete tank will be a steel-concrete-steel plate sandwich, about 30\,cm in width. The external steel plates are connected to each other with welded rebar. In construction, rings of steel plates will be lifted in place, the concrete being cast in between. Due to the slow curing of the concrete, construction will take a long time. Finally, an additional thin stainless steel sheet will be laser-welded on the inside for purity.

\medskip\noindent\textbf{Hollow Core Concrete Tank.}
Based on the Sandwich Concrete Tank, hollow cores are now added to the concrete layer of the tank. This increases mechanical strength, and allows to install a cooling system or active leak prevention. However, this option is up to now little used in tank construction.

%% file: scintillator.tex

%

\noindent The power of the liquid scintillator technology has been demonstrated in the past by successful neutrino experiments like Borexino \cite{Alimonti:2008gc} or KamLAND \cite{sue04kl}. Large target masses, high energy resolution and a low energy threshold are beneficial characteristics that enable real-time detection of rare low-energy events. 
As liquid scintillator is the central component of the detector, this chapter will cover the main properties of liquid scintillators as active material and their interplay with the detector hardware. Finally in Sec.\,\ref{sec::scintillatorcandidates}, the most promising scintillator mixtures are presented. At present, LAB as solvent with the admixture of PPO\nomenclature{PPO}{diPhenyl-Oxazole, a fluor} and Bis-MSB\nomenclature{Bis-MSB}{BIS-o-Methyl-Styryl-Benzene, a fluor} as solutes is favored.

\subsubsection{Scintillator properties}
\label{sec::scintillatorproperties}

\medskip\noindent\textbf{Light output and quenching.} Organic scintillators are excited by charged particle radiation or ultraviolet (UV) light. In the molecular deexcitation process, UV light is emitted. Charged particles which cross a scintillating medium ionize and excite molecules on their track. However, ionization and radiationless deexcitation processes lead to a loss of fluorescence efficiency. In general, processes reducing the efficiency of energy to light conversion are known as quenching.\\
Ionization and excitation densities are high for large energy deposition per unit length, which is the case for heavy particles, such as protons or $\alpha$s. This affects not only fluorescence efficiency, but also the scintillation pulse shapes and can thus be used for particle identification.

\medskip\noindent\textbf{Emission spectra.} The emission spectra of a single-component scintillator has a significant overlap with its own absorption spectra. This results in  multiple absorption and reemission processes where an important part of the information gets lost. In order to prevent additional losses in the energy conversion efficiency, usually one or multiple organic solutes are added. The solvent transfers its excitation energy mainly non-radiatively by dipole-dipole interaction to the solute (also called wavelength shifter or fluor) emitting a higher wavelength region (usually blue light) at which the solvent is transparent.

\medskip\noindent\textbf{Scintillation pulse shape.} For excited states of the scintillator molecules, there are several processes to decay: photon emission, radiationless electronic relaxation, inter-system crossing processes (i.e.\,transition between singlet and triplet states), and energy transfer by collision to other molecules.  
As several deexcitation modes are possible, scintillating pulse shapes commonly show more than one radiative decay constant. The shape of the scintillation pulse can be described by the sum of several exponential functions:\\
\begin{equation}
n(t) = \sum n_i e^{-\frac{t}{\tau_i}}
\end{equation}
In this case, $\tau_i$ denotes the decay constants and $n_i$ the amplitudes of the corresponding decay processes. These constants are typical parameters for each scintillator material; for most organic scintillator the fast decay component dominates the emission.

The amplitudes of the time components depend on the energy deposition per unit length. Consequently, the pulse shape can be used for particle discrimination of $\alpha$-particles or neutron-induced proton recoils from electron signals and thus provides a fundamental method for background rejection.

\medskip\noindent\textbf{Attenuation length.} As the scintillation photons propagate through the medium, absorption and scattering processes can occur. These processes strongly depend on the emitted wavelength; in general, the transparency of the medium increases with the wavelength. The main parameters for the description of the light propagation are the absorption length, the scattering length, and  the self-absorption length of the solute. For short wavelengths, the solute self-absorption dominates.
For large-volume particle detectors, long absorption and scattering lengths are required. As absorption processes decrease the total number of photons which arrive at the photo-sensors, the effective light yield of the detector is reduced. However, a high effective light yield is desirable, since it is directly connected to the energy resolution and energy threshold of the detector.

\medskip\noindent\textbf{Scattering length.}
Scattering processes change the direction of the scintillation photons, thereby elongating the photon path from the scintillation event to the photosensors. Due to the increase in photon time of flight and the corresponding smearing of the photon arrival time patterns, scattering has a potentially deteriorating effect on timing and pulse shape discrimination. Laboratory studies have shown that the scattering lengths of the investigated solvents are of the order of 20 to 30\,m and therefore of the same order as the envisaged detector diameter of LENA \cite{Wurm:2010ad}. The dominating processes are Rayleigh scattering off the solvent molecules and absorption-reemission  processes on organic impurities. The application of Winston cones reduces the fraction of scattered photons registered by the PMTs (Sec.\,\ref{subsec::pmts}).

While MC studies for low event energies show that scattering has to be taken into account for correct vertex reconstruction and pulse shape discrimination, it has only a subdominant effect on the time resolution of the detector (Sec.\,\ref{subsec::vertex}). The latter is mainly governed by the fast fluorescence decay constant of the initial scintillation light \cite{wur09phd}. The situation at GeV energies is comparable: As discussed in Sec.\,\ref{subsec::pdecay}, scattering has only a small effect on proton decay sensitivity \cite{mar08phd}, while it has to be included in track reconstruction algorithms \ref{subsec::tracking} to obtain correct results. 

\medskip\noindent\textbf{Radiopurity.} As solvents of organic liquid scintillators are hydrocarbons, intrinsic radioactivity of the scintillator originates mainly from the $^{14}$C $\beta$ decay. The $^{14}$C $\beta$ background rate by far surpasses all neutrino signals at energies below the endpoint of the $\beta$ spectrum at 156\,keV. While the original solvent produced in distillation plants is usually rather pure at the refinery, the surface contamination of the transportation and experimental containers may dissolve in the liquid scintillator. However, experiments like Borexino have demonstrated the feasibility of ultrahigh radiopurity levels in liquid scintillators, with contamination levels of $^{238}$U at the order of $10^{-17}$\,g/g. The achieved concentrations of $^{232}$Th and $^{40}$K are even at the level of $10^{-18}$\,g/g. The scintillator purification for LENA aims at the same level of radiopurity.

\subsubsection{Influence on detector design}

The properties of a liquid scintillator mixture directly sets constraints on the technical design of the detector and vice versa. In order to optimize the detector performance, one has to take a look at the impact of the scintillator properties on the detector geometry, its demands on the photosensors, on health and on handling issues.

\medskip\noindent\textbf{Geometry.} The absorption length of the liquid scintillator is the main defining parameter for the detector geometry. It strongly affects the effective light yield of the scintillator detector, which accounts for the energy resolution and the energy threshold. With the most foreseen liquid scintillator compounds having absorption lengths of 10-20\,m (see Sec \ref{sec::scintillatorcandidates}), it is unfeasible to build an unsegmented spherical 50\,kton detector. Still, the preferred geometry of the detector is a cylindrical shaped tank of 15\,m radius, which ensures a good effective light yield and volume to surface ratio. With a liquid-scintillator density close to 1\,g/cm$^{3}$, the corresponding height of the cylinder is about 100\,m in case of a 50\,kton detector. 

\medskip\noindent\textbf{Photosensors.} The photosensors are the link between scintillation light and data acquisition. Therefore, their properties play a crucial role for the performance of the detector. Especially the spectral sensitivity of the photosensors should match the emission spectra of the liquid scintillator, or at least be maximal between 320-450 nm. For conventional bialkali photomultiplier tubes this wavelength range is quite common (Sect.\,\ref{subsec::pmts}).

\medskip\noindent\textbf{Liquid handling.} Liquid handling comprises mainly the filling, pumping and storing of the liquid scintillator starting from its production. In this context radiopurity, purification, chemical compatibility, and safety issues have to be considered.

The purification of the scintillator is necessary for reaching a low-level radioactive contamination. For example, this can be done by water extraction and prevention of exposing the scintillator to cosmic rays, i.e.\,fast transport on surface and underground storage. For such a large detector as LENA, it seems most feasible to put purification plants on surface.
At all times, the detector has to be kept away from oxygen contamination, which would induce a degradation of the scintillator. Thus, ultra-clean nitrogen gas must be used to flush pipes and tanks. Moreover, any material that comes into contact with liquid scintillator has to be tested on its chemical compatibility. Possible materials are teflon or passivated stainless steel.

Chemically, the solvent of an organic liquid scintillator is a hydrocarbon. Thus, flammability is the most important concern. The Hazardous Materials Identification System (HMIS) classifies the danger in handling the liquid (from 0 - safe to 4 - dangerous) concerning flammability, reactivity and health. The HMIS ratings for scintillators in flammability range from 1-3, reactivity (almost all 0) and health (0-1) are in general not a big concern. The flash point\footnote{The flash point is the lowest temperature at which a liquid generates sufficient vapor to form a flammable mixture with air. Above this temperature, a spark is sufficient for ignition.} of the material is another characteristic parameter for its flammability.

\subsubsection{Candidate scintillator mixtures}
\label{sec::scintillatorcandidates}

All scintillator options consist of a mixture of a solvent and one or more solute powders, both scintillating organic compounds. This section is a compilation of properties of the most promising scintillation compounds and mixtures for LENA. Prospective scintillator mixtures have to provide an emission spectra with $\lambda > 400$\,nm in order to guarantee large absorption and scattering lengths (see Sect.\,\ref{sec::scintillatorproperties}). In addition, this is the region where the photomultipliers are most sensitive. Other critical parameters of the mixture are the scintillation pulse shape and light yield, which are important for the energy and time resolution of the detector. Detailed studies on these parameters for various organic liquid scintillation mixtures have been carried out in \cite{mar08phd}.

The mixtures under consideration are both \emph{linear-alkyl-benzene} (LAB) and \emph{phenyl-o-xylylethane} (PXE, with an admixture of non-scintillating dodecane) with PPO+Bis-MSB or PMP\nomenclature{PMP}{Phenyl-Mesityl-Pyrazoline, a fluor} as fluors. Tabs.\,\ref{tab::solventproperties} and \ref{tab::fluorproperties} summarize the main properties of solvents and fluors, while 
Tab.\,\ref{tab::scintillatorresults} provides an overview on the main properties of the resulting mixtures. Amongst this selection, LAB with PPO (2 g/$\ell$) and Bis-MSB (20 mg/$\ell$) seems most promising.

\begin{table}
\begin{tabular}{lccc}
\toprule
\textbf{Solvent} & \textbf{LAB}& \textbf{PXE}  & \textbf{C12}\\
\colrule
\multicolumn{2}{l}{\textit{Physical and Chemical Data}} &
\multicolumn{2}{r}{\cite{pxe-spec, lab-spec, c12-spec}}\\
Chemical formula & C$_{18}$H$_{30}$ & C$_{16}$H$_{18}$ & C$_{12}$H$_{26}$ \\
Molecular weight [g/mol] & 241 & 210 		& 170\\
Density $\rho$ [kg/$\ell$] & 0.863 & 0.986   & 0.749\\
Viscosity [cps]						&4.2		& 	& 1.3\\
Flash point [$^\circ$C]				& 140 & 167		& 83\\
Molecular density [10$^{27}$/m$^3$] & 2.2   & 2.8	 & 2.7\\
Free protons [10$^{28}$/m$^3$]		& 6.6	& 4.7    & 7.0\\
Carbon nuclei [10$^{28}$/m$^3$]		& 4.0   & 4.2	& 3.2\\
Total p/e$^-$ [10$^{29}$/m$^3$]		& 3.0 	& 3.2	& 2.6\\
\colrule
\multicolumn{2}{l}{\textit{HMIS Ratings}} & \multicolumn{2}{r}{\cite{pxe-spec, lab-spec, c12-spec}}\\
Health			& 1	& 1	& 1\\
Flammability		& 1	& 1	& 0\\
Reactivity		& 0	& 0	& 0\\
\colrule
\multicolumn{2}{l}{\textit{Optical Properties ($n$,$L$,$\ell_\mathrm{s}$@430\,nm)}} &\multicolumn{2}{r}{\cite{Back:2004zn, kay04, sow72, wah95, wur05dpl, win07dpl, mar08phd, SNO08phd, SK99phd}}\\
Refractive index $n$						& 1.49 & 1.57	& 1.42\\
Absorption maximum [nm]					& 260	& 270&	-	\\
Emission maximum [nm]						& 283	& 290&	-	\\
Attenuation length $L$ [m]				& $\sim$20& 12	& $>$12\\
Scattering length $\ell_\mathrm{s}$ [m]	& 25		& 22		& 35 \\
\botrule
\end{tabular}
\caption{Overview of the solvent parameters of PXE, LAB and Dodecane (C12). The information on physical parameters, HMIS (Hazardous Material Identification System) rating, and refractive index are cited from material safety and product specification sheets of the producers \cite{pxe-spec, lab-spec, c12-spec}. The HMIS rating quantifies the danger in handling the liquid (from 0 - save to 4 - dangerous). The molecular, proton and Carbon densities were computed using this information. The absorption and emission maxima of PXE and LAB are from \cite{mar08phd}. Attenuation and scatterings lengths at 430\,nm have been experimentally determined in \cite{Back:2004zn, SNO08phd,wur05dpl,Wurm:2010ad}.}
\label{tab::solventproperties}
\end{table} 

The first solvent under consideration, LAB, has first come to attention in the R\&D studies for SNO+ \cite{Kraus:2006qp, Ding:2008zzb, SpecLAB_Buck}. It is very appealing due to its high transparency, high light yield, and its low cost. Moreover, it is a non-hazardous liquid with relatively high flash point of 140$^{\circ}$C.
The second scintillator solvent option PXE has already been tested in the \emph{Counting Test Facility} (CTF\nomenclature{CTF}{Counting Test Facility of Borexino}) of Borexino \cite{Ranucci:1999az}. Results look very promising as it shows a high light yield, good transparency and a high flash point of 145$^{\circ}$ \cite{Back:2004zn}. Dodecane (C12\nomenclature{C12}{Dodecane, a hydrocarbon}) is a non-scintillating mineral oil which is a possible admixture to PXE. It is highly transparent and increases the total number of free protons in the mixture. This is a key issue, since free protons play a key role in the proton decay search and in the inverse $\beta$ decay detection channel.\\
~\\
The emission and absorption spectra of the scintillating solvents have a significant overlap. In order to shift the wavelength to a region where the solvent is transparent, one or more organic solutes are commonly added. There are two very promising candidates as wavelength shifters for LENA. First, a mixture of \emph{2,5-diphenyl\--oxazole} (PPO) and \emph{1,4-bis-(o-methyl\--styryl)\--benzene} (bis-MSB) or \emph{1-phenyl-3\--mesityl-2\--pyrazoline} (PMP). Their properties are summarized in Tab.\,\ref{tab::fluorproperties} \cite{Marrodan:2010qp, mar08phd}.

The fluor PMP\,\cite{Guesten1978} is a primary solute with a large \emph{Stokes shift} of about 120\,nm, resulting in a marginal overlap of the absorption and emission spectra and thus a small self-absorption.
The primary fluor PPO has a Stokes shift of $\sim$60\,nm and is therefore usually used in combination with a secondary fluor, Bis-MSB. It is added in small quantities (mg/$\ell$) to avoid self-absorption at large wavelengths.
The PPO absorption spectra significantly overlaps with the emission spectra of both the solvents PXE and LAB. Bis-MSB absorbs where the primary fluor PPO emits and produces a further shift of $\sim 60$\,nm.

\begin{table}
\begin{tabular}{l|ccc}
\toprule
\textbf{Solute} & \textbf{PMP} &  \textbf{PPO} & \textbf{Bis-MSB} \\
\colrule
Chemical formula & C$_{18}$H$_{20}$N$_{2}$ & C$_{15}$H$_{11}$NO &  C$_{24}$H$_{22}$\\
Absorption maximum  & 294\,nm & 303\,nm & 345\,nm\\
Emission maximum & 415\,nm & 365\,nm & 420\,nm\\
\botrule
\end{tabular}
\caption{Properties of the fluors PMP (1-phenyl-3-mesityle-2-pyrazolin), PPO (2,5-diphenyl-oxazole) and Bis-MSB (1,4-bis-(o-methylstyryl-benzene). Absorption and emission maxima are taken from \cite{Marrodan:2010qp, mar08phd}}
\label{tab::fluorproperties}
\end{table}

\begin{table*}
\begin{center}
\begin{tabular}{lccccccc}
\toprule
\multicolumn{2}{l}{Solvents} & \multicolumn{3}{c}{Wavelength shifters} & $Y$\,[\%]  & $n_1$\,[\%]& $\tau_1$\,[ns]\\
& & $1^\mathrm{st}$ & & $2^\mathrm{nd}$ & \cite{win07dpl}& \cite{MarrodanUndagoitia:2009kq, Lombardi:2010} & \cite{MarrodanUndagoitia:2009kq, Lombardi:2010}\\
\colrule

PXE & + & 2\,g/$\ell$ PMP & & - & 79.1 $\pm$ 3.1 & 95.9 $\pm$ 0.02 & 4.15 $\pm$ 0.02\\
   & + & 2\,g/$\ell$ PPO & + & 20\,mg/$\ell$ Bis-MSB & 102.0 $\pm$ 3.3 & 85.3 $\pm$ 1.4 & 2.61 $\pm$ 0.05\\

LAB & + & 2\,g/$\ell$ PMP & & - & 83.9 $\pm$ 3.0 & 85.1 $\pm$ 0.9 & 8.53 $\pm$ 0.15\\
   & + & 2\,g/$\ell$ PPO & + & 20\,mg/$\ell$ Bis-MSB & 99.7 $\pm$ 3.2 & 77.7 $\pm$ 0.8 & 5.21 $\pm$ 0.005\\
DIN & + & 1.5\,g/$\ell$ PPO & & - & - & 86.2 $\pm$ 0.2 & 6.95 $\pm$ 0.02\\
\botrule
\end{tabular}
\end{center}
\caption{Summary of the scintillation properties of different scintillation mixtures. The solvents are PXE and LAB, the dissolved wavelength shifters are PMP and a combination of the fluors PPO and Bis-MSB. The relative light yield $Y$ refers to the mixture of PXE with 2g/$\ell$ PPO \cite{Back:2004zn}, $n_1$ and $\tau_1$ are the weight and the decay constant of the fast pulse component, respectively, as measured in a small cylindrical cell of 2.5\,cm in diameter and 2\,cm in length.\\
Measurements on PXE with PPO show that adding small quantities of the secondary fluor Bis-MSB hardly affects the light yield and fast pulse shape component of the scintillation mixture. The values of the scintillation mixtures LAB + 2g/$\ell$ PPO + 20mg/$\ell$ Bis-MSB are expectations based on measurements of LAB + 2g/$\ell$ PPO \cite{mar08phd}. For DIN, the light yield measurement is presently carried out.}
\label{tab::scintillatorresults}
\end{table*}

\subsubsection{Summary and Outlook}

The final choice of a suited liquid-scintillator mixture for the LENA experiment must be made most carefully. Different physics goals put different constraints on the scintillator properties, making the optimization process difficult. All available solutes and solvents cannot be considered independently, but have to be tested in specific mixtures.Studies on possible scintillation mixtures are not yet concluded.

The most promising state-of-the-art mixtures are listed in Tab.\,\ref{tab::scintillatorresults}. The solutes under consideration have been LAB and PXE, as solvents PPO, PMP and Bis-MSB have been investigated. As can be seen, the combination of PPO and Bis-MSB provides a good light yield for both solutes LAB and PXE. Due to its high Stokes' shift, taking PMP as solvent makes the use of a secondary fluor needless, however the light yield is reduced distinctly. 
Recently, also the solvents \emph{2,5-diisopropyl\--naphtalene} (DIN\nomenclature{DIN}{DiIsopropyl-Naphtalene}) and n-paraffin drew attention and first results concerning decay time constant and weight look competitive.

Comparing the fastest decay processes of the different mixtures, one has to take into account both time constant and weight relative to the total light emitted. Obviously, using PMP as fluor, makes the mixture much slower. In the case of PPO and Bis-MSB as fluors, both LAB and PXE show good timing properties. PXE is slightly preferable since it is faster and the weight of the fastest decay is higher, too.

With respect to the scintillation light yields, LAB and PXE show comparable results. However, the light propagation properties of the liquid-scintillator mixture strongly influence the light collection efficiency of the detector. In terms of effective light yield, LAB (attenuation length $L$ of $\sim$20\,m) is preferable to PXE ($L = 12$\,m).

%% file: pmts.tex
%

%

\noindent The photosensors used for the detection of the scintillation light play an important role for the possible physics yield of LENA: The light detection efficiency affects energy resolution and detection threshold, while the timing influences event reconstruction and particle identification. In the following, we present a review of the most important sensor parameters, focussing on bialkali Photomultiplier Tubes (PMTs) and the existing possibilities to enhance their performance. The properties of alternative light sensors, Silicon Photomultipliers and Hybrid Phototubes, are shortly addressed.

\subsubsection{Photosensor requirements}

{\renewcommand{\thefootnote}{\alph{footnote}}
\begin{table*}[htb]
\begin{tabular}{lcccc}
\toprule
Property & ~~~~LENA~~~~ & ~~Borexino~~  & ~Double Chooz~ & ~~~~SNO+~~~~ \tabularnewline
\colrule
TTS\,(FWHM) [ns] & 3.0 & 3.1 & a few & $<$4 \tabularnewline
Early pulses \footnotemark[1]& $<$1\,\% & & & $<$1.5\,\% \tabularnewline
Late pulses \footnotemark[1]& $<$4\,\% & $<$4\,\% & & $<$1.5\,\% \tabularnewline
QE for $\lambda_\mathrm{peak}$ & $>$21\,\% & $>$21\,\% & $>$20\,\% & \tabularnewline
$\lambda_\mathrm{peak}$ [nm]& 420 & 420 & 400 & \tabularnewline
Optical coverage & 30\,\% & 30\,\% & 13.5\,\% & 54\,\% \tabularnewline
Winston cones & yes & yes & no & yes \tabularnewline
$\to$ effective area & $\times$1.75 & $\times$2.5 & $-$ & $\times$1.75 \\
Dynamic range\,\footnotemark[2] & \multicolumn{2}{l}{spe$\to$0.3\,pe/cm$^2$} & & \tabularnewline
Gain PMTs & 3$\cdot$10$^6$ & 1$\cdot$10$^7$ & 1$\cdot$10$^7$ & 1$\cdot$10$^7$ \tabularnewline
spe p/V & $>$2 & $>$1.5 & $>$2\,(typ.) & $>$1.2 \tabularnewline
DC per area [Hz/cm$^2$] & $<$15 & $<$62 & & $<25$ \tabularnewline
Ionic AP (0.2-200\,\textmu s)\footnotemark[1]& $<$5\,\% & $<$5\% & & $<1.5\%$ \tabularnewline
Fast AP (5-100\,ns)\footnotemark[1]& $<$5\,\% & & & \tabularnewline
Pressure resistance [bar] & $>$13 & & & $>$3\,\footnotemark[3]\tabularnewline
\ce{^{238}U} content [g/g] & $<$3$\cdot$10$^{-8}$ & $<$3$\cdot$10$^{-8}$ & $<$1.2$\cdot$10$^{-7}$ & \tabularnewline
\ce{^{232}Th} content [g/g] & $<$1$\cdot$10$^{-8}$ & $<$1$\cdot$10$^{-8}$ & $<$9$\cdot$10$^{-8}$ \tabularnewline
$^{\mathrm{nat}}K$ content [g/g] & $<$2$\cdot$10$^{-5}$ & $<$2$\cdot$10$^{-5}$ & $<$2$\cdot$10$^{-4}$ & \tabularnewline
Detector lifetime [yrs] & $>$30 & & & 10\,\footnotemark[3]\\
\botrule
\end{tabular}
\caption{Overview of the parameter limits for photosensors in LENA and other running or upcoming liquid-scintillator detectors \cite{Alimonti:2008gc, Arpesella:2001iz, Ardellier:2006mn, Boger:1999bb}. spe: single photoelectron, TTS: transit time spread for spe,  QE: quantum efficiency, PDE: photo detection efficiency, $\lambda_\mathrm{peak}$: peak wavelength of spectral response, pe: photoelectrons, p/V: peak-to-valley ratio, DC: dark count, AP: afterpulses.}\label{TabLimitsSensor}
\end{table*}
}

\noindent The crucial parameters of photosensors can be split up in four aspects: sensor performance, environmental properties, availability until start of construction and cost-performance ratio. The most important properties are:\\
~\\
\noindent\textbf{Photo detection efficiency (PDE).} The PDE\nomenclature{PDE}{Photo Detection Efficiency} is determined by the quantum efficiency (QE)\nomenclature{QE}{Quantum Efficiency} of the photocathode, the collection efficiency for photoelectrons (pe)\nomenclature{pe}{PhotoElectron} as well as backscattering losses of pe at the first dynode. In LENA, the baseline value for PDE is 20\,\% at 420nm.

\medskip
\noindent\textbf{Spectral response.} The sensitivity of the photosensors must match the spectrum of the scintillation light arriving at the detector walls. A bialkali photocathode is arguably the best choice in case of a PMT as the maximum PDE corresponds well to the effective scintillator emission spectrum.

\medskip\noindent\textbf{Optical coverage (OC)\nomenclature{OC}{Optical Coverage}} denotes the fraction of the detector walls covered with photosensors. Together with the detection efficiency of the PMTs, the OC determines the overall efficiency of light collection. Light-collecting reflective concentrators \cite{Oberauer:2003ac} might be used to increase this area beyond the active area, i.e. the photocathodes. The present design foresees an OC of 30\,\%. Primarily the energy resolution, but also both time and spatial resolution of LENA will depend on the product of PDE and OC.

\medskip\noindent\textbf{Time jitter.} The timing uncertainty for single photoelectrons (spe) for individual sensors is vital for the overall timing and position resolution of the detector. For PMTs, it is given by the transit time spread (TTS\nomenclature{TTS}{Transit Time Spread}).

\medskip\noindent\textbf{Afterpulses (AP\nomenclature{AP}{AfterPulses}).} AP are spurious pulses occurring in correlation to primary pulses. Fast afterpulses appearing within several tens of ns after the primary pulse might impede the resolution of fast double peak structures (e.g.\,for proton decay, Sect.\,\ref{subsec::pdecay}). APs induced by ions produced by the initial electron avalanche occur with delays of several $100\,$ns to \textmu s. These ionic APs affect the veto of cosmogenic backgrounds by lowering the detection efficiency of muon-induced neutrons \cite{Bellini:2011yd}. Usually, the probability of appearance is quoted as percentage of primary pulses.

\medskip\noindent\textbf{Dark count (DC\nomenclature{DC}{Dark Count}).} The number of dark counts (per photosensitive area) has to be low since it affects position and energy resolution by introducing fake hits. In extreme cases, it might cause triggers to random coincidences. 

\medskip\noindent\textbf{Dynamic range.} The dynamic range of the response of the whole detector has to extend from low energy events with mostly only a spe per sensor up to events depositing several GeV of energy, corresponding to hundreds of pe for the sensors closest to the event.

\medskip\noindent\textbf{Gain, single electron resolution (SER\nomenclature{SER}{Single Electron Resolution}).} The amplification gain must be sufficiently large and the single electron pulse height resolution sufficient (corresponding to a large peak-to-valley ratio) in order to distinguish reliably between noise and spe signals.

\medskip\noindent\textbf{Pulse shape, timing effects.} Short rise and fall times of the spe voltage pulses, in the order of a few ns, are advantageous for tracking and position reconstruction. Effects shifting the detection time, like early pulses, prepulses, and to some extent also late pulses, have a detrimental influence on the reconstruction.

\medskip\noindent\textbf{Photosensitive area.} For a given optical coverage, the granularity of the detector increases with lower sensor areas, improving the spatial resolution of the photon arrival pattern. Moreover, for smaller areas a smaller dynamic range of individual sensors is sufficient.

\medskip\noindent Environmental properties encompass \textbf{radioactive purity} of the materials used in the sensors, \textbf{pressure resistance} of the sensors placed near the tank bottom, their \textbf{long-term reliability} over 30+ years and \textbf{susceptibility to magnetic fields}. The availability until start of construction will define the candidate sensor types. To calculate the cost-performance ratio, it is necessary to know all sensor parameters and their effect on detector performance.

{\renewcommand{\thefootnote}{\alph{footnote}}
\footnotetext[1]{Probability of occurrence per primary pulse}
\footnotetext[2]{Assuming $\times1.75$ Winston cones; estimate valid for large sensor sizes only} 
\footnotetext[3]{In ultrapure water, at a water pressure of 3\,bar and exposed to earthquakes from mining activity} \stepcounter{footnote}
}

\medskip\noindent\textbf{Parameter constraints.} Minimum requirements for the photosensors to be used in LENA can be formulated based on the experimental experience gathered in the Borexino and Double Chooz experiments, and relying on MC studies carried out for LENA. Tab.\,\ref{TabLimitsSensor} lists the preliminary limits for LENA in comparison to the requirements of other liquid-scintillator detectors. MC simulations to further refine the values for LENA are ongoing.

\subsubsection{Bialkali photomultipliers}

A detailed description of the functionality and properties of photomultiplier tubes (PMTs) can be found in \cite{RCA, Leo}. In view of the requirements on photosensors, head-on hemispherical or plano-convex bialkali PMTs with low-background borosilicate glass are presumably the subtype best suited for LENA.

\newcounter{fnnumber}
\paragraph{Survey of available PMT series}

\begin{figure}
\centering{
\includegraphics[width=0.44\textwidth]{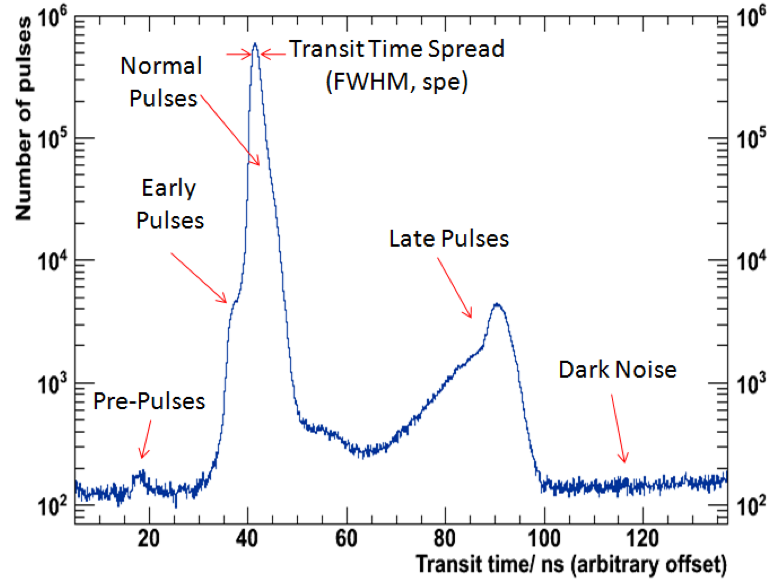}
\caption{Pulse timing effects in PMTs. Naming convention as used in Borexino publications. In this example a Hamamatsu R5912 (8") with +1425V applied was measured (threshold $0.2\,$pe) with the Borexino testing facility. A close description of the various features of the PMT response can be found in \cite{Smirnov:2004kz, Ianni:2004wj}.}
\label{PMTtiming}}
\end{figure}

\noindent PMTs are the most natural choice of photosensors for LENA, fulfilling all technical requirements (pricing, availability and environmental properties). A comprehensive study to identify the most promising commercially available PMT models is ongoing. Also, the next generation of PMTs featuring high quantum efficiency (HQE) photocathodes are being evaluated. So far, the characterization has been carried out based on three test setups: 

\medskip
\noindent\textbf{Borexino PMT testing facility.} This setup at the LNGS was originally used to screen $\sim$2\,300 PMTs for Borexino \cite{Brigatti:2004ix}. For LENA, it has been used to characterize a selection of PMTs manufactured by Hamamatsu, ranging from 3" to 10" in diameter and featuring regular bialkali photocathodes: R6091 (3"), R6594-ASSY (5"), R5912 (8") and R7081 (10"). Fig.\,\ref{PMTtiming} shows the transit time distribution measured for the 8'' PMT with this setup: Prepulses and early pulses that potentially compromise the arrival times of the first photons, and therefore position and track reconstruction, as well as late pulses are discernible.

\medskip
\noindent\textbf{INFN Milano.} This test stand is based on a picosecond 405\,nm laser (Edinburgh Instruments EPL-405) as light source and an 8-bit 1GHz National Instruments FADC\nomenclature{FADC}{Fast Analog-to-Digital Converter}. Here, Hamamtsu PMTs featuring HQE Superbialkali photocathodes (QE$\approx$35\,\% at peak wavelengths) have been characterized. PMTs with diameters from 5" to 10" were tested.

\medskip
\noindent\textbf{Universit\"{a}t T\"{u}bingen.} The test stand uses fast LED drivers and both FADCs and standard TDC\nomenclature{TDC}{Time-to-Digital Converter}+ADC\nomenclature{ADC}{Analog-to-Digital Converter} read-out. Measurements of TTS and SER were done for 3", 8", 10" and 20" PMTs, including HQE\nomenclature{HQE}{High Quantum Efficiency} tubes of Hamamatsu R7081 (10"), Photonis XP5301 (3") and XP5312 (3").

\medskip
\noindent\textbf{TU M\"unchen} A fourth test stand using a modified EPL-405 ps laser and a fast pulsed LED as light sources and a 10-bit 8\,GHz Acqiris DC282 FADC is currently being set up. 

\medskip\noindent The results of these measurement are currently being analyzed and will lead to a first preselection of PMT series.


\paragraph{Number of PMTs} 

\noindent Currently, the LENA reference design aims at an optical coverage of 30\,\%. Light concentrators will be used in order to enhance the light collection efficiency. A concentrator increasing the effective collection area by a factor of $1.75$ seems realistic, reducing the necessary active photocathode area to 17\,\%. For 8'' PMTs of normal QE, this corresponds to 63\,000 pieces. HQE photocathodes could be employed to further reduce this number. The scaling of the necessary number is demonstrated in Tab.\,\ref{tab::pmtno}.

In general, the use of smaller-sized PMTs offers some advantages: With smaller diameter, the transit time spread, the requirements on the dynamic range and the rate of ionic afterpulses decrease in general. A higher sensor granularity would be beneficial for position reconstruction and tracking. On the other hand, smaller diameters increase the number of sensors and electronic channels drastically, which favors larger PMTs from the economic point of view. Further simulations of the detector behavior are necessary to determine the dependence of the physics potential on the PMT diameter and to derive the optimum cost-performance ratio.

\begin{table}[h]
\begin{tabular}{cccc}
\toprule
Cathode  & Unarmed & Light & HQE-PMTs \\
diameter & PMTs & concentrators & +concentrators\\
\colrule
3'' & 979\,100 & 574\,900 & 410\,600 \\
5'' & 341\,900 & 200\,700 & 143\,400\\
8'' & 114\,600 & 67\,300 & 48\,100\\
10'' & 85\,500 & 50\,200 & 35\,800\\
12'' & 55\,900 & 32\,800 & 23\,500\\
20'' & 22\,400 & 13\,100 & 9\,400\\
\botrule
\end{tabular}
\caption{Total number of PMTs necessary to achieve an OC of 30\,\% in the inner detector (ID) and 1.1\,\% in the outer detector of LENA (as in Borexino), depending on the photocathode diameter and the use of light concentrators increasing the effective detection area by a factor of $1.75$ in the ID. Additionally, the use of HQE photocathodes of 35\,\% peak QE (assuming a collection efficiency of 80\,\% without photoelectron backscattering losses) can further reduce the number of PMTs, in this case the OC in the ID amounts to 21\,\%.}
\label{tab::pmtno}
\end{table}

\subsubsection{Optimization of light detectors}

There are various aspects beyond the selection of an optimum PMT that enter the design of the LENA detector instrumentation. In the following, a short overview of these issues is presented.

\begin{figure}[t]
\centering
\includegraphics[width=0.33\textwidth]{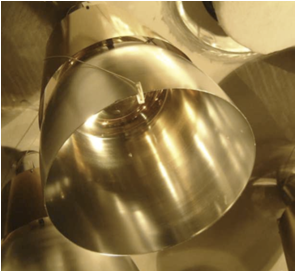}
\caption{Winston cone type light concentrator used in Borexino, increasing the effective photon collection area by a factor of 2.5 (courtesy of Borexino).}
\label{fig::winstoncone}
\end{figure}

\medskip\noindent\textbf{Winston cones.} Compound parabolic concentrators also known as Winston cones are non-imaging light concentrators focussing incident light onto a flat or curved surface \cite{Winston}. Mounted to a PMT, they increase its effective light collection area (see Fig. \ref{fig::winstoncone}). The optimum concentrator shape can be constructed mathematically by rotating the tilted positive branch of a parabola around the $z$-axis. The length of the resulting concentrator determines the area increase $\varepsilon_\mathrm{A}$ but also the field of view of the PMT: Light incident at larger angles than the cutoff angle $\theta_\mathrm{c}$ to the $z$-axis in a first approximation is not collected by the PMT but reflected back into the detector. Winston cones have been widely used in neutrino experiments (Tab.\,\ref{TabLimitsSensor}).

Assuming a fiducial volume of 11\,m radius in LENA (e.g.\,for solar neutrinos, Sect.\,\ref{subsec::solar}), the optimum value of $\theta_\mathrm{c}$ is 50$^\circ$ for 8'' PMTs, corresponding to $\varepsilon_\mathrm{A}=1.71$. First MC simulations have shown that for equal OC and $\varepsilon_\mathrm{A}=2.0$, the average pe yield decreases only by a few percent. However, the spatial dependence of the pe yield increases substantially. Further studies are required.

\medskip\noindent\textbf{HQE photocathodes.} A further way of increasing photocollection efficiency is the use of HQE photocathodes. Conventional PMTs feature a peak QE of not more than 25\,\%. Since a few years, ET Enterprises and Hamamatsu Photonics are commercializing PMTs featuring HQE photocathodes. From Hamamatsu, two new types of HQE cathodes are available, and have already been tested in the Milano and T\"{u}bingen setups: \textit{Super BiAlkali} (SBA) photocathodes feature a peak QE of 35\,\%, while \textit{Ultra BiAlkali} (UBA) cathodes reach a peak QE of up to 43\,\%. SBA cathodes are available for PMTs of up to 10'' diameter, while for UBA cathodes the maximum diameter is 2'' at the moment. However, 12'' SBA PMTs as well as larger UBA PMTs are expected for the next years.

Currently, it is not evident whether HQE PMTs lower the cost per detected pe. The considerably larger price per PMT must be set in relation to the lower amount of cabling and electronic channels needed. Also, performance aspects like the dynamic range per channel or increased afterpulsing probabilities must be taken into account.

\medskip\noindent\textbf{Pressure resistance.} The pressure tolerance required for PMTs is given by the hydrostatic pressure at the bottom of the tank, depending on the scintillator density. 
The minimum requirement is a resistance to 9.8\,bar for LAB and 11.1\,bar for PXE (Sect.\,\ref{subsec::scintillator}). Another 3 to 4\,bar might be necessary to withstand an implosion shock wave as it occurred in Super-Kamiokande. Simulations will be needed to determine precise numbers.

However, the pressure tolerance of available PMTs amounts to only $\sim$7\,bar. The weak spots of the PMT glass bulbs are located at the sharp curvatures at the PMT neck and base. One option is to increase the thickness of the glass casing, which would increase the weight and the amount of radioactivity introduced by the PMTs, and is deemed to be demanding by manufacturers.

\begin{figure}[htb]
\includegraphics[width=0.33\textwidth]{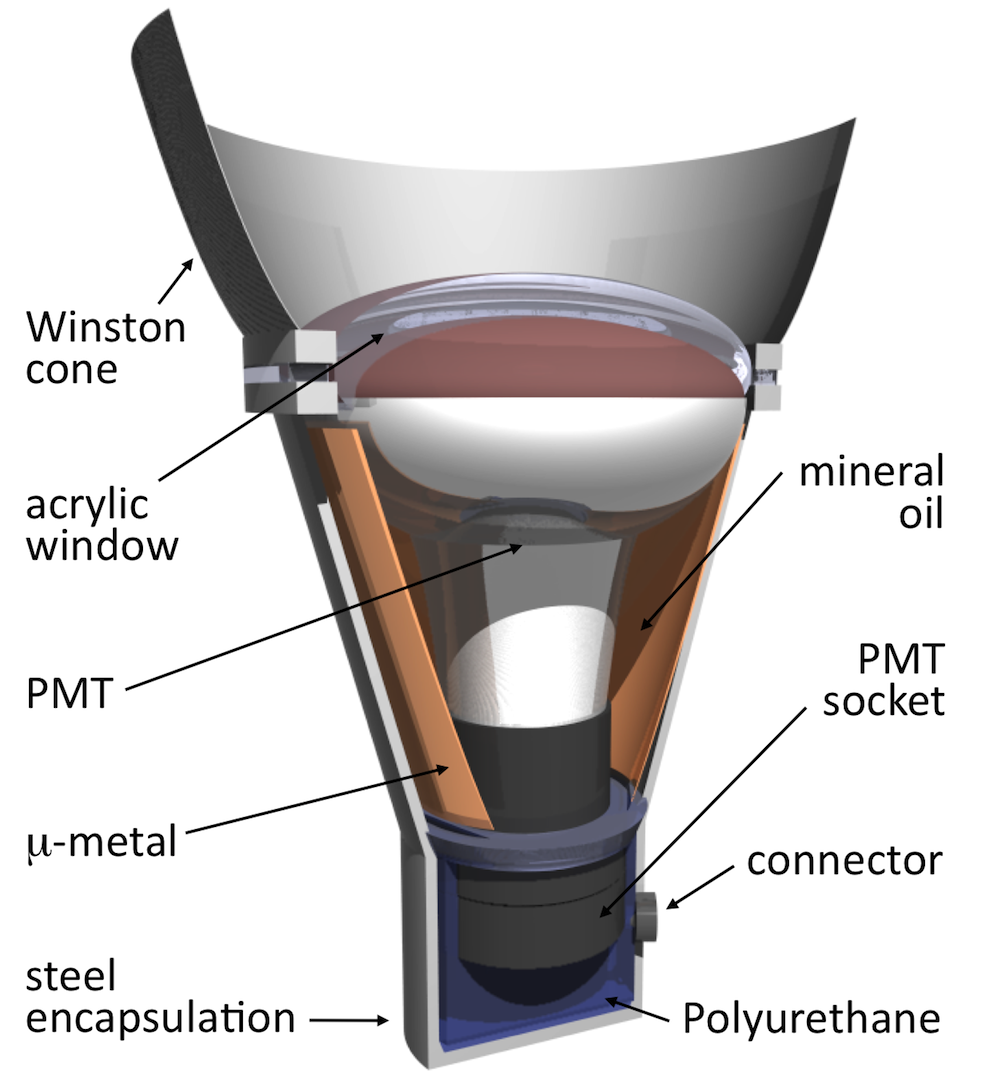}
\caption{Illustration of a LENA PMT encapsulation.}
\label{fig::encapsulation}
\end{figure}

A promising alternative is the use of pressure-withstanding encapsulations housing the PMTs at atmospheric pressure. A widely used approach in HE neutrino physics are spherical glass bulbs. However, the currently preferred solution is an adaptation of the Borexino Outer Detector PMT encapsulations. Their design consists of a conically shaped metal housing enclosing the PMT body combined with a transparent window\footnote{In the Borexino design, this window is a thin PET foil. In LENA, this window will have to absorb the external pressure, so a more resistive material like acrylic might be used.} in front of the photocathode (Fig. \ref{fig::encapsulation}). This design also allows for an easy integration of the light concentrators and the \textmu-metal shielding. Design work based on finite element simulations has recently started and will result in a prototype encapsulation for pressure testing.

\subsubsection{Alternative photosensor types}

In spite of the broad range of photosensors currently available or under development, the number of options for use in LENA is fairly limited. Considering the requirements regarding time resolution, availability within the next decade and long-term reliability, only Silicon Photomultipliers and Hybrid Phototubes are promising alternatives to conventional bialkali PMTs. While their performance is mostly superior, it remains uncertain whether the price per photosensitive area is low enough and if their performance is sufficient in all respects.

\medskip\noindent\textbf{Silicon Photomultipliers (SiPMs)}\nomenclature{SiPM}{SIlicon Photo Multiplier}, also known as Multi-Pixel Photon Counters (MPPCs) are arrays of avalanche photo diodes, operated in limited Geiger mode \cite{HamamatsuMPPC}. Typically there are 100--1000 pixels per {$\mathrm{mm}^2$}, which are electrically connected in parallel, so the total signal is proportional to the number of cells hit by one or more photons.\\
SiPMs have several advantages over conventional PMTs: The QE of SiPMs can reach 70 to {$80\,\%$}, with spectral response very close to the PPO emission spectrum and PDEs over $70\,\%$. spe time resolutions as low as {$60\,\mathrm{ps}$} have been observed \cite{Ronzhin:2010zz}, dominated by the jitter of photons being detected in different cells of the array. With gain similar to conventional PMTs ($10^5$ to $10^6$), their SER is much smaller, which allows exact counting of small numbers of photons at temperatures of 10-15$^{\circ}\mathrm{C}$. Regarding environmental properties, SiPMs are insensitive to magnetic fields, which would allow to magnetize the LENA detector. Furthermore their radioactive contamination can be expected to be extremely low and they are very slim, possibly allowing to reduce the buffer thickness or even omit the buffer, which would permit to increase the target volume substantially or greatly reduce costs. In addition, the pressure tolerance of SiPMs probably is very high. Also, their bias voltage is below {$100\,\mathrm{V}$}, allowing for a much simpler voltage supply system.\\
On the other hand, there are several disadvantages: The size of the active area is very small, currently {$5\times5\,\mathrm{mm^2}$} for the largest commercially available SiPMs. With larger sizes the number of channels decreases and the PDE increases. It is reasonable to assume that {$1\times1\,\mathrm{cm^2}$} SiPMs will become available in the near future and many SiPMs could be combined into local clusters to save channels. However, if one wants to preserve the photon counting ability, the active area per channel is limited to about {$1\,\mathrm{cm}^2$}. Another problem is the dark count, which can approach MHz frequencies, disturbing the reconstruction of events and increasing with active area. Also, temperature stabilization is necessary to prevent drifts in gain. Finally, it is not yet clear, whether the price per area will be competitive to PMTs.

\medskip\noindent\textbf{Hybrid Phototubes (HPTs).}\nomenclature{HPT}{Hybrid Photo Tube} The basic design of HPTs combines a large area hemispherical photocathode for photoelectron conversion, a HV field accelerating these electrons towards a small luminescent screen and a small diameter PMT (1'') reading out the scintillation signal. HPTs have been used successfully in the Lake Baikal Neutrino Telescope, featuring the QUASAR-370 of 15'' diameter produced in the 1990s \cite{Bezrukov:1987,quas4,Lubsandorzhev:2000jd}. However, they are currently out of production.

Compared to conventional PMTs, HPTs would feature a range of advantages: The angular acceptance is $2\pi$, the transit time spread is typically less than 1\,ns, prepulses, ionic AP, fast AP and late pulses occur only on very low levels, the DC is very low and the susceptibility to magnetic fields is greatly decreased. However, it is currently unknown, whether the dynamic range of HPTs is sufficient for use in LENA, as per detected photon about 25\,pe hit the small PMT. One solution would be to use less dynodes as it was done in the TUNKA Air Cherenkov Array experiment for the QUASAR-370G \cite{Lubsandorzhev:2000jd}, which on the other hand could affect the discrimination from noise.

\subsubsection{Conclusions}

\noindent The benchmark value for the photon detection efficiency of LENA is 6\,\%, arising from the product of optical coverage (OC=30\,\%) and PMT photo detection efficiency (PDE=20\,\%). 

Bialkali PMTs currently appear to be the most promising choice for photosensors, basically fulfilling all requirements concerning detection performance and long-term reliability, and low in cost per photoactive area. PMT diameters from 3'' to 12'' are considered. It is foreseen to equip the PMTs with Winston cones, the factor of area increase ranging from 1.6 to 2, and \textmu-metal shielding. If available and cost-effective, HQE PMTs might be used to reduce the necessary OC. Furthermore, a pressure-absorbing encapsulation will be necessary to protect the PMTs at the bottom of the detector from implosion.

Based on the interim results of the performance tests, further tests on the most promising series of Hamamatsu Photonics and ET Enterprises will be conducted with larger numbers. This study will also extend to new releases as the 12'' PMT by Hamamatsu (R11780) and its upcoming HQE version. The middle-term aim is the development of a prototype optical module consisting of PMT, encapsulation, \textmu-metal and light concentrator by the end of 2014.

%% file: readout.tex

%

\noindent In view of the huge number of photosensors needed for the next generation detector the only solution technically available today are PMTs. With a properly designed read-out electronics, PMTs can guarantee the required time resolution, charge resolution and dynamic range. Two different options for readout electronics are described, one relying on FADC readout of all channels, the other featuring customized ASIC boards inside the detector servicing small arrays of PMTs.

\subsubsection{Minimum requirements}

\noindent The broad and rich physics scope of LENA puts rather severe requirements on electronics and data acquisition. The relevant performance parameters are summarized in Tab.\,\ref{tab:PmtCoo}. Particularly, the goal to perform both low energy solar neutrino physics (especially pp neutrinos) and high energy beam neutrino physics force a very large dynamic range on the signal amplitude because the system must work both on single photoelectron mode and with very large signals corresponding to several hundreds of photoelectrons per channel. Moreover, a precise determination of the relative time of each PMT hit with sub-ns resolution is necessary for the spatial reconstruction of the events, for the pulse shape analysis and in order to disentangle the different event topologies that are associated to neutral and charged current interactions of electron, muon and tau neutrinos. Zero dead time is a must to avoid the loss of interesting time correlated events, and a very flexible triggering system is highly desirable, taking into account again the very broad scientific scope and the long expected lifetime of the experiment, which may lead to unanticipated physics.

\begin{table}[htb]
\centering
\begin{tabular}{lcc}
\toprule
Parameter & LE & HE \\
\colrule
Number of channels (8'' PMTs)	& 45\,000	& 45\,000 \\
Time resolution (pulse on-set)& $<$1\,ns	& $<$1\,ns \\
Dynamic range				& 0-30\,pe	& 0-300\,pe \\
Dead time per channel			& $<$100\,ns & FADCs \\
Channel buffer size			& $\sim$100	& FADCs \\
Number of FADC channels		& $\times$	& 10\,000 \\
Sampling rate				& $\times$	& 500\,MS/s \\
Voltage resolution			& $\times$	& 2$\times$8\,bit \\
\botrule
\end{tabular}
\caption{Requirements for the electronic read-out in LENA. The table lists both the minimum requirements for low-energy (LE) neutrino detection and the optimal configuration for high-energy (HE) beam physics. }\label{tab:PmtCoo}
\end{table}

In the following we describe two very different options for the readout electronics: a full FADC option, and a custom ASIC\nomenclature{ASIC}{Application-Specific Integrated Circuit} option. We believe that the first would be preferable, but the second could well meet the basic scientific requirements and be less expensive. For these
reasons we include both.

\subsubsection{Full FADC readout}

\noindent The scientific requirements can be met by the use of suitable FADCs with the right pulse height resolution, sampling speed, and the combination of on-board zero suppression, software trigger and careful synchronization of the boards. 

At the time of writing (many years before the beginning of data taking) there are several products on the market that basically meet these requirements; 8-bits, 2 GS/s FADC boards are already available, and better products are very close to be. This solution is still quite expensive today, but it is very likely that it will not be at the time when LENA will need it. 

The system could work as follows: each PMT should be connected to a simple linear front end that should perform high-voltage decoupling, some amplification and some shaping. The details of this rather standard front end electronics must not be worked out here. 
It is probably cost-effective to imagine this front end built into a custom chip mounted close to the PMTs. The output signal of this front end should be sent to the FADC. The FADC board will sample it at the correct speed (1 GS/s seems appropriate but 2 GS/s is feasible) with 10-12 bit precision (14 bit is already available and probably more than adequate), and store into a local memory. An on-board fast FPGA\nomenclature{FPGA}{Field-Programmable Gate Array} will perform the zero suppression, keeping the data only around valid PMT pulses, and storing also the time stamp of the pulse. Data from each pulse will then be stored into another internal memory ready for read-out.

Each FADC board shares a common distributed clock with all other boards, so that sampling is synchronous throughout the whole detector. The acquisition software will read continually the time stamps (time stamps only at first), perform the triggering functions and decide whether the 'event' should be read out or not. In case the read-out is needed, a custom designed protocol between the FPGA onboard, the FADC and the readout software will retrieve all samples, and the event will be stored on disk.

This architecture should guarantee the maximum possible information (the fully digitized shape of each PMT pulse for the whole detector in a programmable time window around the event), and zero dead time. Also, the use of the time stamps for triggering should keep 
the complete data flow to a very reasonable level. This fact is easy to prove: the PMT activity is largely dominated by dark noise; physics events of any kind contribute a much lower amount of data. A typical large PMT has a room temperature dark noise of the order of 1 KHz. 
Assuming 2 kHz to have some contingency, the expected total activity of LENA is of the order of 10$^8$ Hz (conservative). This means 400\,Mb/s of time stamps to handle, a very reasonable number. The event triggering rate will be dominated by $^{14}$C and will reach 
10$^4$\,Hz with a few hundreds of PMT hits (again conservative); the expected data flow will be of the order of 100\,Mb/s$-$1\,Gb/s, again an easy number to handle. 

Sufficient onboard memory is required to buffer the sampled signals and timestamps. By employing commercially available memory aimed at the consumer market, available FADC modules already offer an onboard memory of 512\,MB or more per channel at reasonable rates. If assuming a dynamic range of 12\,bit and a sampling rate of 2\,GHz with a time window of 300\,ns saved in the buffer, less than 1\,kB of memory is used per trigger. With an average trigger rate of 2\,kHz, 250 seconds of live data can be buffered before data loss occurs. Data can be read continuously from the memory while new events are stored at the same time, thus leading to a dead-time free data acquisition. To prevent data loss in the rare cases where the trigger rate can be up to ten times the dark noise for the duration of several seconds (e.g.\,a galactic Supernova sufficiently near), normally only a fraction of the total buffer size should be used before readout is initiated.  A further possibility is to set a programmable threshold for the trigger rate. If it is exceeded for several seconds, the onboard logic will stop to store the full pulse shape to prevent a buffer overflow; instead, it will save only the timestamp and the energy deposition derived by integrating over the sampled pulses. Based on this, complete event reconstruction is still possible, while the memory usage of each trigger is reduced to $\sim$12\,byte per channel. In this mode, more than 400 seconds of live data could be buffered at a trigger rate of 20\,kHz, even if only 100\,MB are reserved for this operation mode. So, a FADC-based DAQ would record the essential information even under the extreme conditions of a galactic SN neutrino burst, and should be seriously considered.

It is not crucial here to decide whether the FADC board should be commercial or custom. Most likely, commercial boards will be cheap enough to be preferable, but this is not a decision for today. It is however important to distribute the FADC boards close to the PMTs, 
in order to minimize cable length. Data from the FADC chassis will be collected via optical fibers which are of course cheap, simple and reliable. The distance between front end and FADC should be minimized to avoid spurious noise. 

This architecture is good both for the main detector and for the muon veto, which of course makes the system simpler.

A final note on dynamic range: Even with a large number of bits, it may be difficult to guarantee a good linearity for the signal range from single to hundreds of photoelectrons. In this case it should be considered as an option to have a front end with doubled output and two different gains. This will increase the number of channels (not all channels must be duplicated) but would guarantee very good performance. A trade-off between cost and performance is mandatory on this point, and a decision must be taken after a full simulation.

\begin{figure}[ht]
\begin{center}
\includegraphics [width=0.5\textwidth] {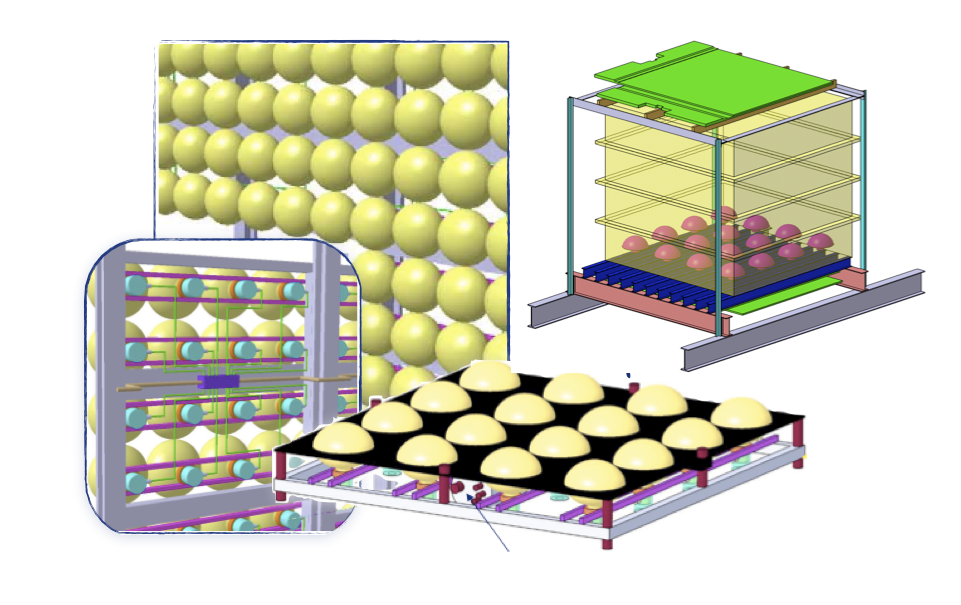}
\caption{On the left the demonstrator of  the PMm$^2$ R\&D program that is going to be tested with its electronics system in the MEMPHYNO prototype (right).}
\label{fig:ReD}
\end{center}
\end{figure}

\subsubsection{Custom ASIC read-out}

\noindent The coverage of a large area with PMTs at a ``low'' cost can be met by a readout electronics integrated circuit (called ASIC) for groups of PMTs. The development of such electronics is the aim of a dedicated French R\&D program, called PMm$^2$ \cite{PMm2:web}. This R\&D program was initially stimulated by the MEMPHYS Water Cherenkov project \cite{deBellefon:2006vq, Memphys:web}.

PMm$^2$ intends to realize a new electronics board dedicated to a grouped acquisition of a matrix of 16 PMTs. Each matrix will have a common board (PARISROC) for the distribution of high voltage and for the signal readout. The circuit under development allows to integrate for each group of PMTs: a high-speed discriminator on the single photoelectron signal, the digitization of the charge (on a 12-bit ADC) to provide numerical signals, the digitization of time (on a 12-bit TDC) to provide time information, a channel-to-channel gain adjustment and a common high voltage. DAQ system, trigger and mechanical integration of the matrix is currently under development in a joint effort by teams of the Laboratoire de l'Accelerateur Lineare (LAL), the Institut de Physique Nucl\`eaire in Orsay (IPNO), and the APC Paris.

To test the system with real physical signals, the Water Cherenkov prototype detector MEMPHYNO is presently under construction at the APC Laboratory in Paris \cite{Memphyno}. The aim is to install a complete 4$\times$4 array of PMTs and the complete electronics and acquisition readout chain in a cubic water tank of 2\,m edge length. A muon hodoscope based on four scintillator planes will provide the trigger for cosmic muons as well as muon track information. Based on this, MEMPHYNO will evaluate the system trigger threshold, the track reconstruction performance in water and the properties of the PMTs. Fig.\,\ref{fig:ReD} displays both the PMm$^2$ matrix and a conceptional view of MEMPHYNO.

\vspace{2cm}


%% file: baspar.tex

%

\noindent The performance of the LENA detector will depend first of all on the quality of the liquid scintillator used, but also on the granularity of and active area covered by photonsensors as well as the time resolution and dynamic range of the readout electronics. Therefore, all these aspects have to be included in a Monte Carlo simulation of neutrino and background events in the detector. For this purpose, a GEANT4-based framework has been set up for LENA, using the geometry described in Sec.~\ref{sec::hardware} and a set of baseline parameters presented in Tab.~\ref{tab::baselineparameters}. Several steps have to be run through in order to obtain the detector response:

\begin{table}
\begin{tabular}{llcc}
\toprule
\textbf{Parameter} & & \textbf{~LAB~} & \textbf{~PXE~} \\
\colrule
\multicolumn{2}{l}{\textit{Target properties}} & & \\
Target height & [m] & \multicolumn{2}{c}{96.0} \\
Target radius & [m] & \multicolumn{2}{c}{13.0} \\
PMT radial distance & [m]& \multicolumn{2}{c}{14.5} \\
Mass & [kt] & 43.8 & 50.3 \\
Number of electrons & [$10^{34}$] & 1.5 & 2.6 \\
Number of free protons & [$10^{33}$] & 3.1 & 2.4 \\
Number of C nuclei & [$10^{33}$] & 2.0 & 2.1 \\
\colrule
\multicolumn{2}{l}{\textit{Light emission}} & & \\
Light yield &  & \multicolumn{2}{c}{$10^4$/MeV} \\
\multicolumn{4}{l}{Birks coefficient $k_{B}$:~[mm/MeV]} \\
~-~electrons & & \multicolumn{2}{c}{0.15} \\
~-~protons & & \multicolumn{2}{c}{0.12} \\
~-~$\alpha$-particles & & \multicolumn{2}{c}{0.107} \\
\colrule 
\multicolumn{2}{l}{\textit{Fluorescence time profile}} & & \\
Time constant $\tau_1$ & [ns] & \multicolumn{2}{c}{4.6} \\
Time constant $\tau_2$ & [ns] & \multicolumn{2}{c}{18} \\
Time constant $\tau_3$ & [ns] & \multicolumn{2}{c}{156} \\
Weight $N_1$ ($e$-like event) &  & \multicolumn{2}{c}{0.71} \\
Weight $N_2$      &      & \multicolumn{2}{c}{0.22} \\
Weight $N_3$      &      & \multicolumn{2}{c}{0.07} \\
\colrule
\multicolumn{2}{l}{\textit{Light propagation}}Ê& & \\
Attenuation length & [m] & \multicolumn{2}{c}{11}\\
Absorption length & [m] & \multicolumn{2}{c}{20}\\
Abs./Reemission length & [m] & \multicolumn{2}{c}{60}\\
Rayleigh scat.~length & [m] & \multicolumn{2}{c}{40}\\
\colrule
\multicolumn{2}{l}{\textit{Light detection}}Ê& & \\
Optical coverage & & \multicolumn{2}{c}{0.3} \\
Quantum efficiency & & \multicolumn{2}{c}{0.2} \\
Time jitter (tts) & [ns] & \multicolumn{2}{c}{1.0} \\
Mean photoelectron yield & & \multicolumn{2}{c}{240\,pe/MeV} \\
\botrule
\end{tabular}
\caption{Baseline parameters of the detector performance entering the LENA simulation. For target masses and interaction center densities, values are given separately for LAB and PXE.}
\label{tab::baselineparameters}
\end{table}

\medskip
\noindent \textbf{Detector environment.} Central to the simulation is the definition of the materials and dimensions making up the detector. In accordance with the design presented in Sec.~\ref{sec::hardware}, the scintillator volume is fixed to 26\,m diameter and 96\,m height, surrounded by a mantle of inactive buffer medium with a width of 2\,m. While earlier studies presume PXE as solvent, more recent works use LAB as baseline. Corresponding material properties like the number densities of electrons and protons are listed in Tab.~\ref{tab::solventproperties}. PMTs are assumed as photo detectors, their photocathode planes located at 14.5\,m radius. At 15\,m a steel tank of 10\,cm separates buffer and outside water volume, adding another 2\,m of (active) shielding. 
 
\medskip
\noindent \textbf{Event vertex.} While most event vertices at low energies are rather simple, mostly one charged leptons and at times a $\gamma$-quantum or free nucleon in the final state, high-energy neutrino interactions may create additional particles like pions and often break up a Carbon nucleus acting as interaction target. GEANT4 is well suited to follow the path of these particles in the target volume, creating a realistic picture of the energy deposition inside the scintillator.

\medskip
\noindent \textbf{Light emission.} The number of photons produced per energy deposit, i.e. the light yield of the scintillator, corresponds to roughly $Y_{p}\approx10^4$\,photons per MeV. The exact value may vary by the order of 10\,\% depending on the scintillator solvent and fluor concentrations (cf.~Tab.~\ref{tab::scintillatorresults}). Moreover, there is a strong dependence of the light output on the d$E$/d$x$ of the ionizing particle, resulting in a quenching for heavy particles (protons, $\alpha$) compared to light ones (electrons). This effect is taken into account by the Birks' formalism, using $k_{B}=0.01$\,cm/MeV \cite{}. 

\medskip
\noindent \textbf{Emission profile.} the light is not emitted instantaneously, but following a time profile described by a superposition of exponential decays. The simulation uses a description based on three components, which are close to the ones of the favored LAB+PPO+bisMSB cocktail. Timing properties depend mostly on the fast fluorescence time $\tau_1$. The relative weight of these components depends on the ionizing particle, protons and $\alpha$'s featuring enhanced secondary components compared to electrons.
  
\medskip
\noindent \textbf{Light propagation.} While the scintillation light is produced in a relatively broad span of wavelengths that is defined by the emission spectrum of the secondary fluor bisMSB, an effective description of the light transport through the liquid can assume that the light is shifted to the wavelength range around 430\,m. As the scintillator transparency increases strongly with wavelength, most of the light is transmitted and detected towards the long-wavelength part of emission spectrum and PMT sensitivity. Photon absorption, reemission as well as Rayleigh scattering are reproduced by the code. The corresponding parameters are chosen conservatively at the lower end of the expectation for LAB, which is also a good approximation for the case of PXE .

\medskip
\noindent \textbf{Light detection.} The baseline design for LENA assumes a photocoverage of 30\,\% and a photosensitivity of 20\,\%. In many cases, it is sufficient to reduces the light yield to 6\,\% and count all photons reaching the detector walls as detected. However, the baseline scenario assumes 13,000 unarmed 20''-PMTs evenly distributed over the inner surface. This choice can be regarded as conservative, as a greater number of smaller PMTs resulting in the same photocoverage will only increase the granularity and resolution for photon arrival times. Other configurations like, e.g., 8''-PMTs equipped with light concentrators) can be implemented, taking also the limited photon acceptance angle into account. Arrival times of individual photons are smeared by a Gaussian of 1\,ns width (1$\sigma$), corresponding to the PMT specifications presented in Sec.~\ref{subsec::pmts}.

\medskip
\noindent \textbf{Electronic readout.} Up to know, the effects of electronics are described only coarsely. However, a baseline design comparable to the one of Borexino is often applied, featuring a time resolution better than the PMT time jitter for the arrival time of the first photon per channel and integrating the number of all subsequent photons  in the same channel to provide an integral charge. For the moment, a similar scheme is used for most of the simulations aiming at particle track reconstruction at sub-GeV energies. However, studies using the equivalent of a FADC recording each individual channel are also on-going (Sec.~\ref{subsec::readout}).

\begin{figure}
\centering{
\includegraphics[width=0.48\textwidth]{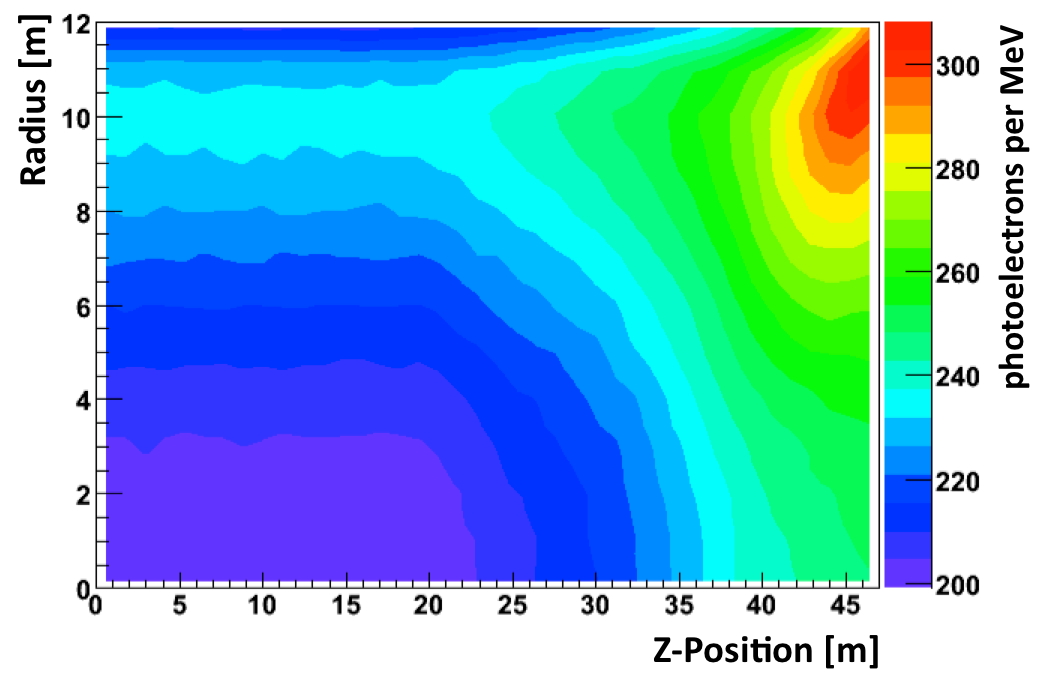}
\caption{Map of the photoelectron yield $Y_{pe}(\bf r)$ that features a considerable dependence on the radial and $z$-positions of the event vertex.}
\label{fig::lightyieldmap}}
\end{figure}

\medskip
\noindent Based on this chain, the photoelectron (pe) yield per deposited energy, $Y_{pe}(\bf r)$ can be obtained as a function of the vertex position inside the target volume. Unlike current spherical detectors were the light yield is approximately uniform over the whole detection volume, a significant dependence is expected for LENA mainly because of the larger detector dimensions and the cylindrical shape. Fig.~\ref{fig::lightyieldmap} shows $Y_{pe}$ in the height-radius plane for one fourth of the detector. For the baseline values shown in Tab.~\ref{tab::baselineparameters}, the mean value for uniformly distributed events is $\langle Y_{pe}\rangle = 238$\,pe/MeV. 

This conservative value will be significantly exceeded if the liquids prove to be more transparent. For an attenuation length of 15\,m, well within the range feasible for LAB, the mean yield increases to $\langle Y_{pe}\rangle = 340$\,pe/MeV. Yields as large as 450\,pe/MeV might be obtained in optimistic scenarios, almost equalling the pe yield achieved in Borexino.

%% file: vertex.tex
\noindent The reconstruction of the primary vertex position as well as the reconstruction of the event time and the visible energy\footnote{The visible energy of a particle in the scintillator is defined as the energy an electron has to deposit in the scintillator to create the same number of photons.} of low energy events is a prerequisite for all further analysis.

Accounting for the spacial and time resolution of LENA, events with visible energies below 10\! MeV can be considered as point-like in space-time. Hence, they are described by a set of five coordinates
\begin{equation*}
 \boldsymbol{X} = (\boldsymbol r, t_0, T)
\end{equation*}
where $\boldsymbol r$ is the position of the event, $t_0$ is the time when the event occurred and $T$ is the visible energy. These parameters are determined by a negative logarithmic likelihood fit to the hit times of the first photons as well as the number of photons detected on each PMT.

The main fit is seeded by a series of previous analysis steps. First, as an estimate for $\boldsymbol r$, the charge barycenter $\boldsymbol r_b$ is determined by an approximate fit to the integrated number of photoelectrons collected on each PMT. Using $\boldsymbol r_b$ as input, the distribution of TOF-corrected first-photon arrival times is used to gain a seed for $t_0$. Finally, an approximate value for $T$ is obtained from the total charge seen in the detector, taking into account the spatial inhomogeneity of the photoelectron yield.

The probability density function (PDF) used in the main fit considers the time resolution and the finite dimensions of the PMTs, the expected dark count rate, the fluorescence times of the scintillator as well as absorption and scattering in the scintillator. Currently, the PDF assumes LENA to be a cylinder of infinite height. Hence it does not describe events close to the lid/bottom of the detector correctly.

The fit based reconstruction has been tested with several datasets of point like events generated with the LENA Monte-Carlo simulation \cite{Moellenberg:2009}. It produces stable fit results without significant systematic shifts for $|z| < 25$\! m and $T\in [200$\! keV, \mbox{$10$\! MeV$]$}. To speed up the simulation as well as the reconstruction, $20''$ PMTs (without Winston cones) have been used throughout the study. PMT resolution is included by a Gaussian transit time spread of 1\! ns (1$\sigma$), dark noise is neglected.

\begin{figure}[tbp]
\centering
\includegraphics[width = 0.48\textwidth]{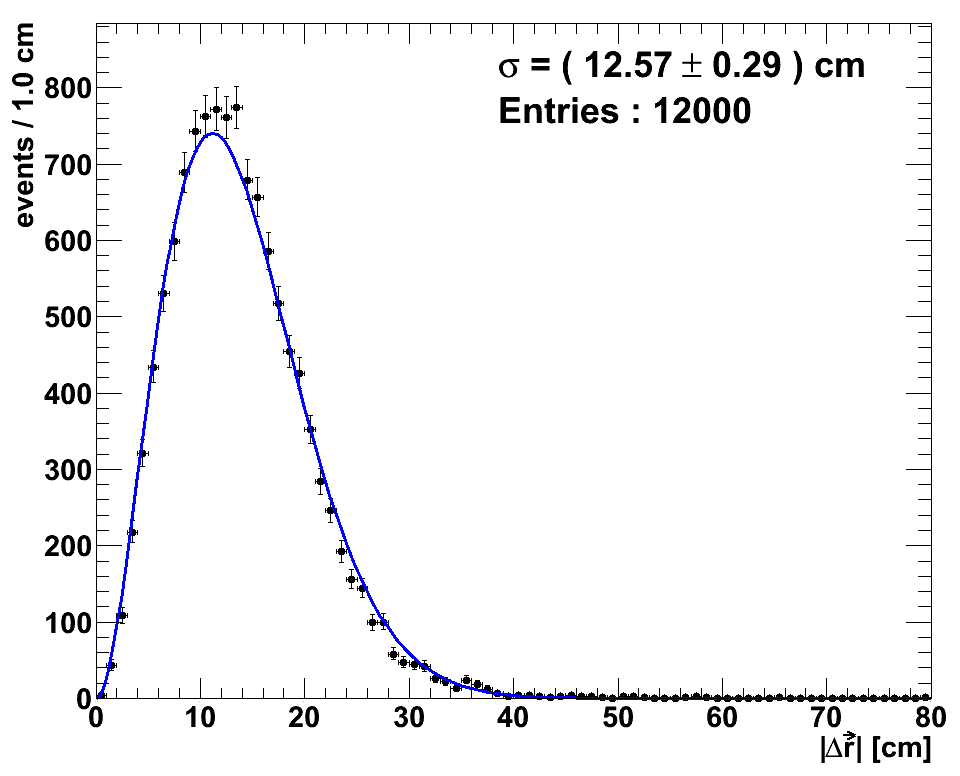}\\
\includegraphics[width = 0.48\textwidth]{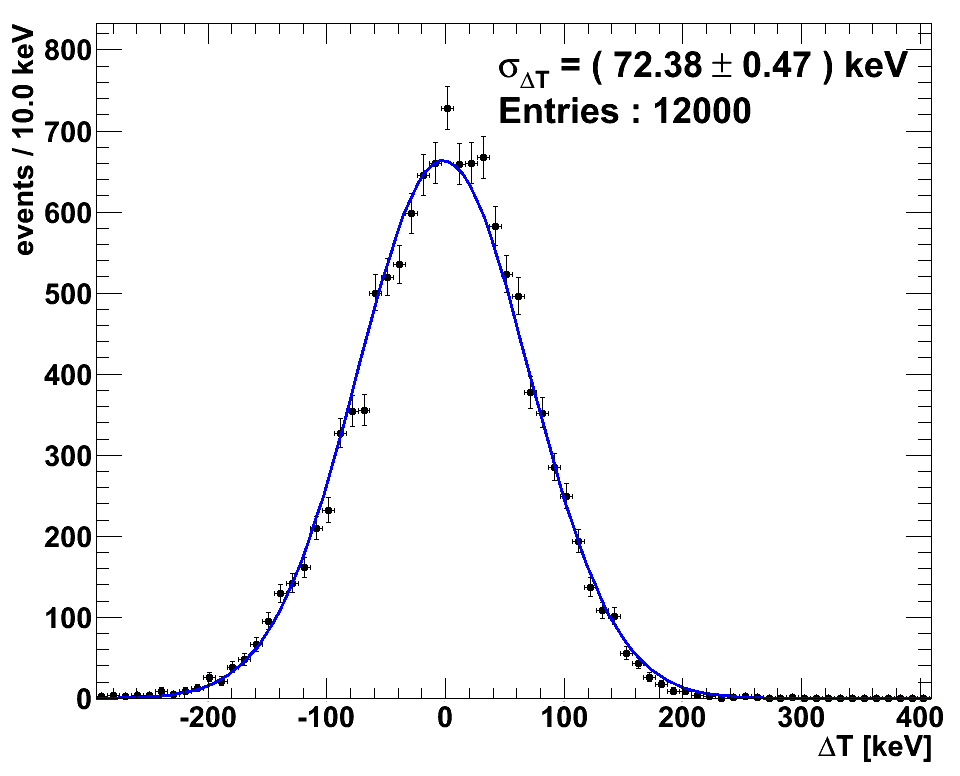}
\caption{Overall performance of the vertex reconstruction at 1\! MeV: The distance of the reconstructed to the true event position is shown in the upper histogram. The lower panel shows the deviation of the reconstructed energy. Both distributions are fitted with the appropriate functions, the resulting resolutions are indicated in the corresponding panels. The simulated events were randomly distributed in a slice of the active volume with $|z| < 25$\! m, the effects of dark noise are neglected.}
\label{fig::subsec::vertex::pointFitResults}
\end{figure}

Figure \ref{fig::subsec::vertex::pointFitResults} shows the results for 12\! 000 electron events of 1\! MeV, distributed randomly in the detection volume (\mbox{$|z| < 25$\! m}). As LENA has a cylindrical shape, the resolution in $z$-direction, $\sigma_z = (9.99\pm0.05)$\! cm, differs from the resolution in $x$- and $y$-direction, \mbox{$\sigma_{x,y} = (8.25\pm0.05)$\! cm}. Furthermore, as no Winston cones are used in the simulation, the photoelectron yield increases with rising radii. Hence, the resolution at the border of the active volume is better than the resolution in the center of the detector. The difference is of order 20\% at \mbox{$T = 1$\! MeV}.

Averaged over the whole detector, the number of detected photoelectrons is $N_{pe} = 220$ at 1\,MeV. The resulting energy resolution follows the expected behavior of $(\Delta E/E) = (N_{pe})^{-1/2}$. Also the spatial resolution improves with energy, but the energy dependence is more complicated.

At low energies, the dark noise of the PMTs has a degrading influence on the vertex reconstruction. For an integral dark rate of 45\,kHz (roughly 4\,kHz per PMT), the resolution is decreased on the percent level. While this effect plays almost no role above 1\,MeV, it becomes much more prominent at lower energies, reaching approximately 30\,\% for electron events of 200\! keV deposited in the center of the detection volume.

In the final experiment, the ambiguities of position and energy reconstruction will be resolved by a calibration campaign based on radioactive sources. Similar campaigns have been conducted in both KamLAND and Borexino detectors (e.g.\,\cite{Busenitz:2009ac}). The  point reconstruction in liquid scintillators has proven to be a very powerful tool in analyses. The future development of the presented algorithm will be focused on including events close to the lid and the bottom of the detector.



%% file: tracking.tex

%

\noindent The reconstruction of particle momenta is a prerequisite for the analysis of atmospheric and accelerator neutrinos. It was realized only recently that liquid scintillator detectors $-$ opposed to general opinion $-$ feature this capability, provided the particle track length exceeds several tens of centimeters. This section reflects the state-of-the-art of Monte Carlo simulations that investigate energy and angular resolution both for single and multiple-particle events.

\subsubsection{Introduction}

The analyses of beam and atmospheric neutrinos require a neutrino detector to be capable of reconstructing both energy and momentum of the incoming neutrino. Depending on the exact task, it may also be necessary to identify the flavor (or antiflavor). At higher energies, charged current interactions will excite resonances and start to scatter inelastically, creating not only a lepton but also pions and heavier hadrons in the end state. Moreover, background signals due to beam contaminations or flavor-insensitive neutral current interactions must be identified and rejected. Detectors that fulfill these requirements are usually highly segmented or feature excellent tracking capabilities (like liquid argon time projection chambers).

However, at energies not exceeding a few GeV, event vertices and backgrounds are less complex, and low-energy neutrino experiments become viable candidates for a far detector. This is most impressively demonstrated by the Super-Kamiokande experiment, that found neutrino oscillations both by analyzing atmospheric neutrinos and the K2K\nomenclature{K2K}{KEK-to-Kamiokande neutrino beam} neutrino beam, and is currently serving as far detector in the T2K\nomenclature{T2K}{Tokai-to-Kamiokande neutrino beam} beam experiment searching for $\theta_{13}$ \cite{Ashie:2005ik,Ahn:2006zza,Le:2009nr}. 

At first glance, it seems unlikely that unsegmented liquid-scintillator detectors might be used in the same way: An imminent feature of water Cherenkov detector is the directional information coded in the orientation of the Cherenkov cone. Using this information is possible even for particles close to the Cherenkov threshold. Opposed to that, scintillation light emitted by low-energy events is distributed isotropically, bearing no directional information at all for the quasi-pointlike events.

However, high-energy particles deposit their energy over macroscopic distances. In liquid scintillator, they will create a track of ionization extending for tens of cm or even meters, leading to a distortion of the spheric light front emerging from the track. As illustrated in Fig.\,\ref{fig::FermatCone}, the superposition of spherical light waves emitted along the particle track creates a light front which resembles the Cherenkov light cone, adding a spherical backward running front to the v-shaped forward front.

The possibility to exploit the inherent directionality has been neglected for a long time, as the deformation of the light front is too small for low-energy neutrinos. Only in the rejection of cosmic background, namely in the muon track reconstruction algorithms of KamLAND and Borexino, the arrival time patterns projected by the light front on the PMTs are exploited \cite{Abe:2009nd,Bellini:2011yd}. Based on this, the orientation of the track can be reconstructed with an astounding accuracy: The Borexino track reconstruction achieves an angular resolution of $3^\circ$ for muons crossing the scintillator volume \cite{Bellini:2011yd}.

Only recently, the basic possibility to reconstruct the track direction by exploiting the peculiar shape of this light front has been brought to attention \cite{Learned:2009rv,2009arXiv0909.4974P}: Since then, tracking algorithms based on the Monte Carlo simulations of GeV neutrino events have confirmed the basic notion that energy and momentum of the end-state particles and finally of the incident neutrino can be resolved. Moreover, the studies demonstrated that the accuracy of the reconstruction could in principle exceed the performance of water Cherenkov detectors due to the much larger light yield.

In the following, two different approaches to the reconstruction of GeV neutrino events in LENA are presented: The first one is a GEANT4\nomenclature{GEANT4}{GEometry ANd Tracking MC platform}-based evaluation of tracking single electrons and muons at sub-GeV energies, the second investigates the possibility to resolve more complex interaction vertices in the 1$-$5\,GeV range, based on a specifically written prototype code \cite{2009arXiv0909.4974P}. 

\subsubsection{Tracking in the sub-GeV range}


\begin{figure}
\centering
\includegraphics[width=0.3\textwidth]{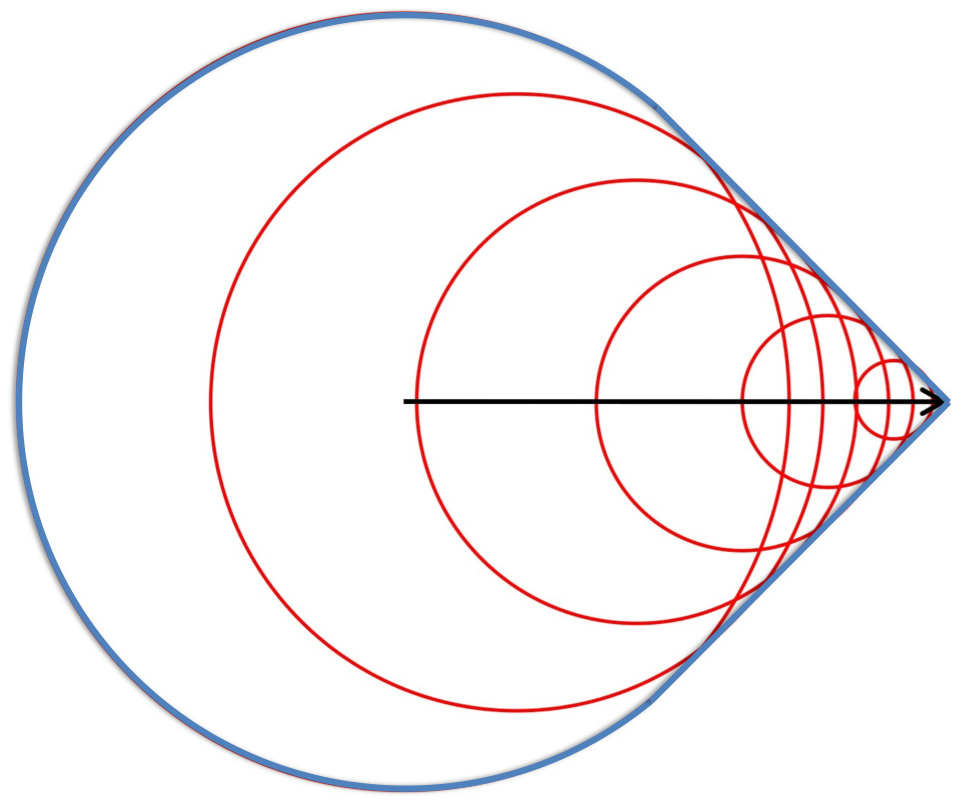}
\caption{Construction of the first photon surface (\textit{blue}) by superposition of spherical waves (\textit{red}) created by a particle transversing the scintillator (\textit{black})} 
\label{fig::FermatCone}
\end{figure}


The reconstruction of a particle track must rely on the projection of the Fermat surface depicted in Fig.\,\ref{fig::FermatCone} on the surface composed by the PMTs mounted to the detector walls. As a start point, the patterns of first photon arrival times and integrated charge per PMT can be exploited.

\begin{figure*}
\begin{minipage}{0.48\textwidth}
\includegraphics[width=0.85\textwidth]{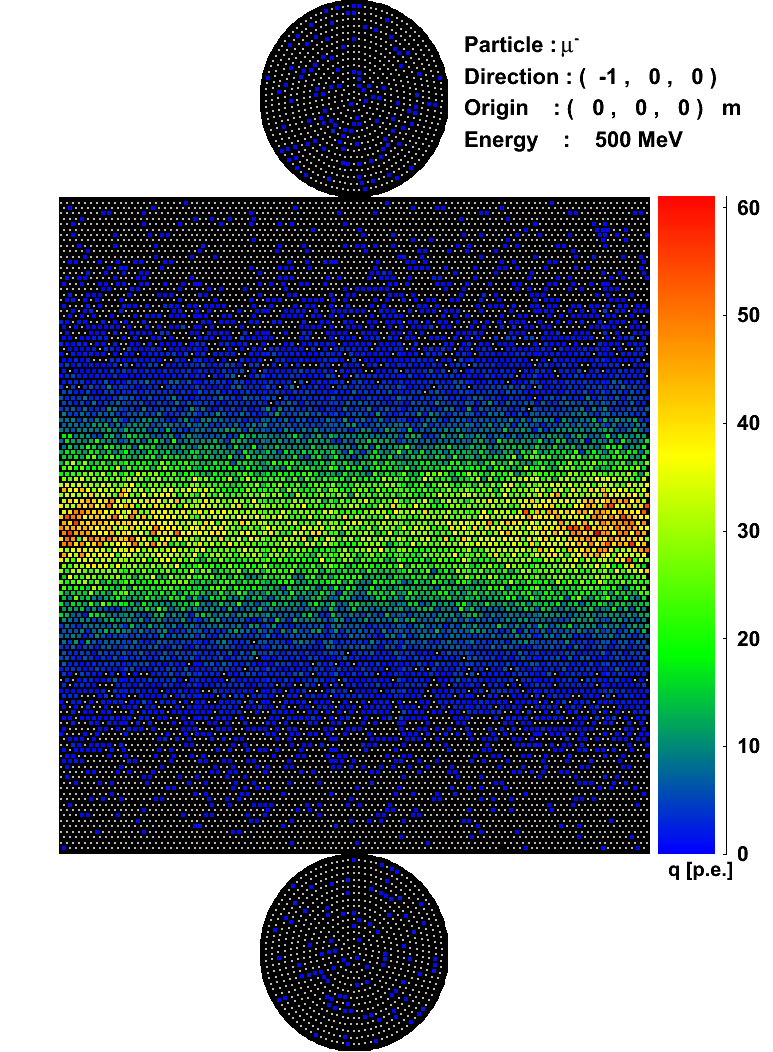}\\
\end{minipage}
\hfill
\begin{minipage}{0.48\textwidth}
\includegraphics[width=0.85\textwidth]{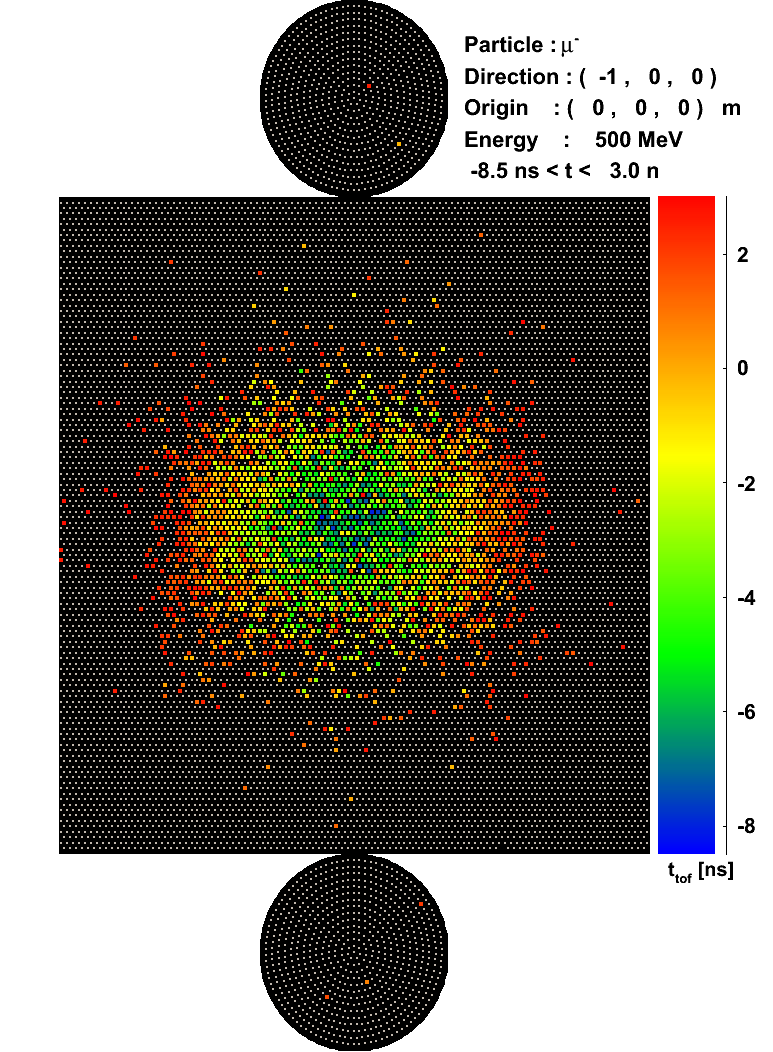}\\
\end{minipage}
\caption{A 500\,MeV muon in LENA. On the left, the color coded information is the charge seen by each PMT, while the hit time of the first photon at each PMT is shown on the right, applying a time of flight correction with respect to the charge barycenter of the track.} 
\label{fig::EventDisplayD}
\end{figure*}

Fig.\,\ref{fig::EventDisplayD} shows an example for a single particle track: A 500 \!MeV muon traveling from the center of the detector towards the wall. The event was created using a Geant4 based simulation of the LENA detector. Depicted in \mbox{figure \ref{fig::EventDisplayD}a)} is the charge distribution which features only a slight asymmetry due to the displacement of the track's center of charge with respect to the symmetry axis of the detector. Nevertheless, the charge signal of the PMTs can be used to obtain the track's barycenter. This allows removing most of the dependence of the photon arrival times on the track position by a time of \mbox{flight (TOF\nomenclature{TOF}{Time Of Flight})} correction with respect to the barycenter.
 The resulting distribution is depicted in \mbox{figure \ref{fig::EventDisplayD}b)}. Only the first 11\,ns (from -8\,ns to +3\,ns of TOF corrected hit time) are shown to enhance clarity. The observed distribution is clearly anisotropic and can be used to get a rough estimate on the track direction.\\

\begin{figure*}
\begin{minipage}{0.32\textwidth}
 \includegraphics[width=\textwidth]{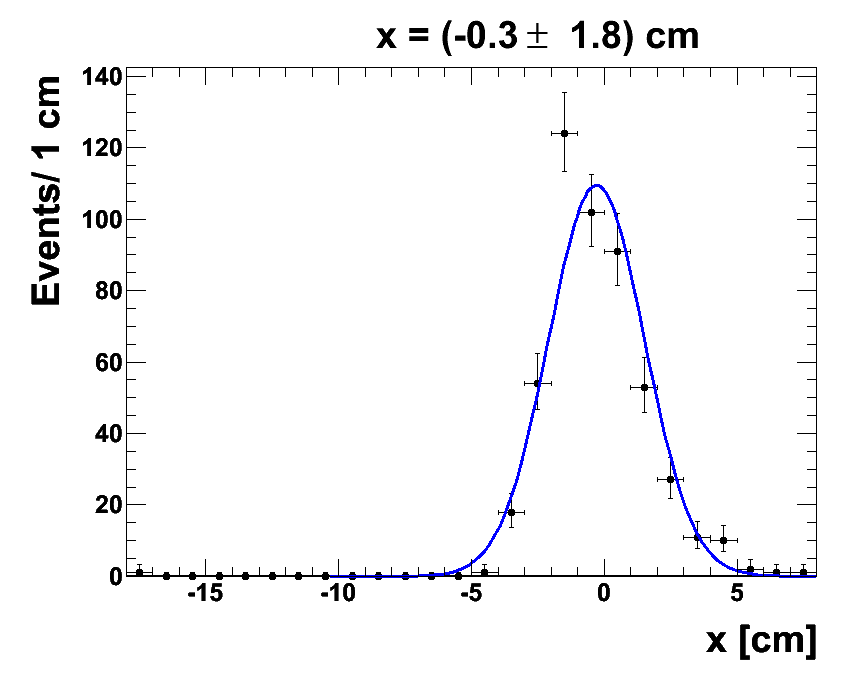}
\end{minipage}
\begin{minipage}{0.32\textwidth}
 \includegraphics[width=\textwidth]{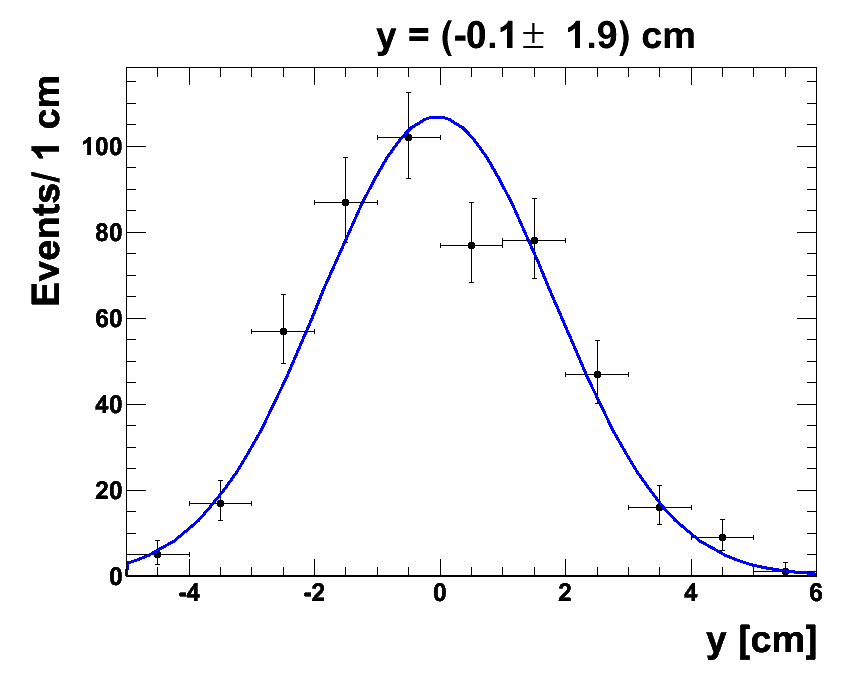}
\end{minipage}
\begin{minipage}{0.32\textwidth}
 \includegraphics[width=\textwidth]{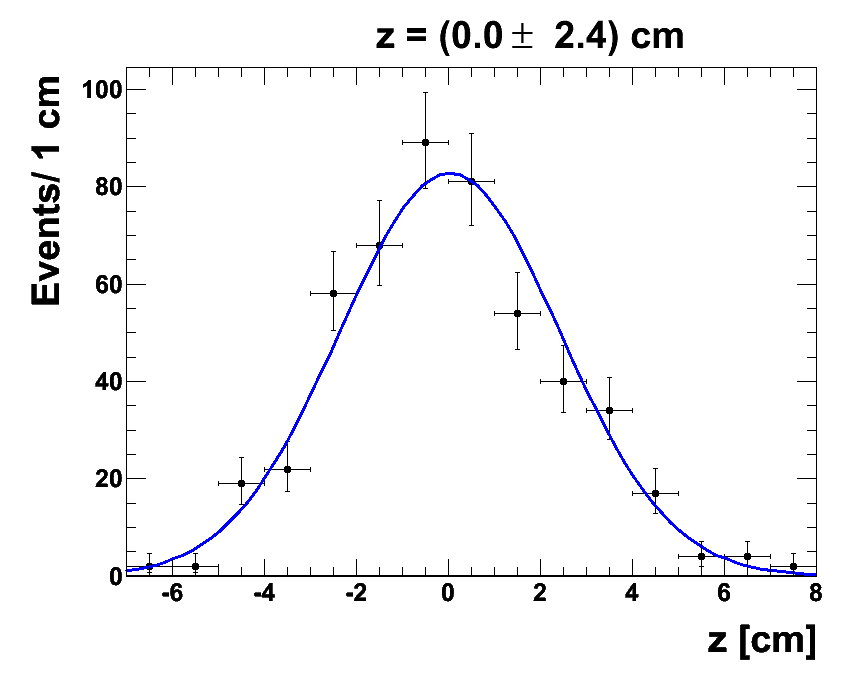}
\end{minipage}
\begin{minipage}{0.32\textwidth}
 \includegraphics[width=\textwidth]{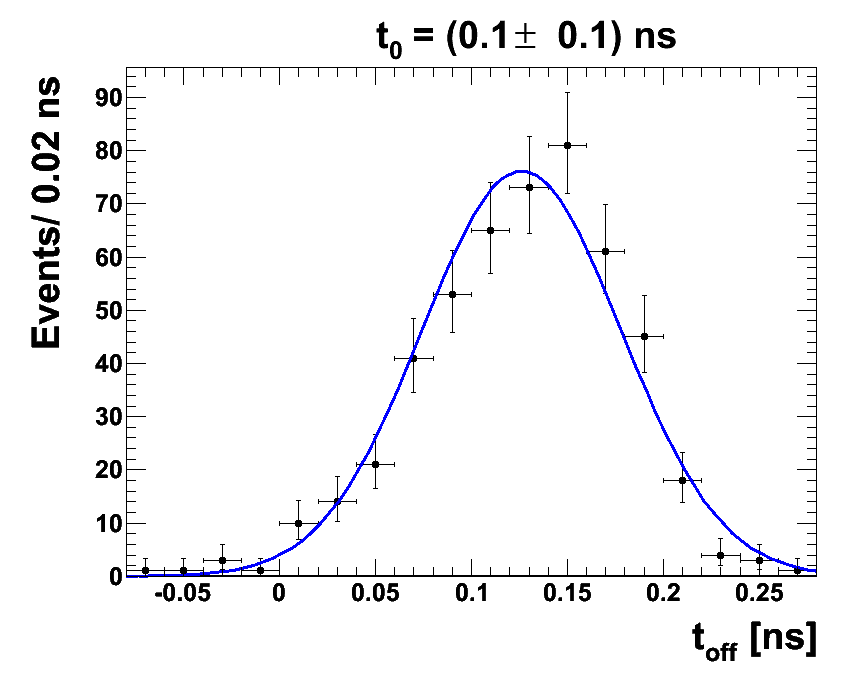}
\end{minipage}
\begin{minipage}{0.32\textwidth}
 \includegraphics[width=\textwidth]{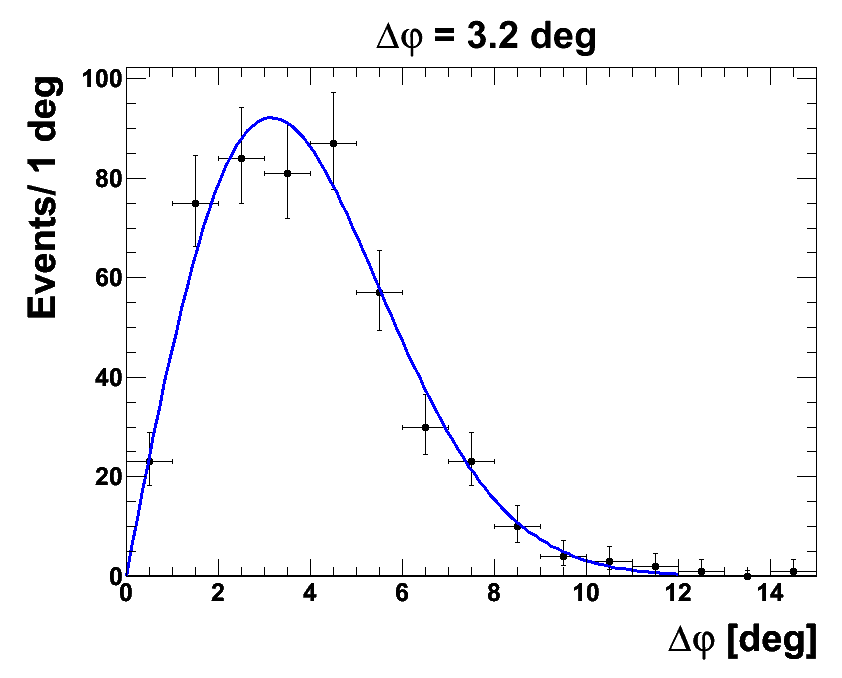}
\end{minipage}
\begin{minipage}{0.32\textwidth}
 \includegraphics[width=\textwidth]{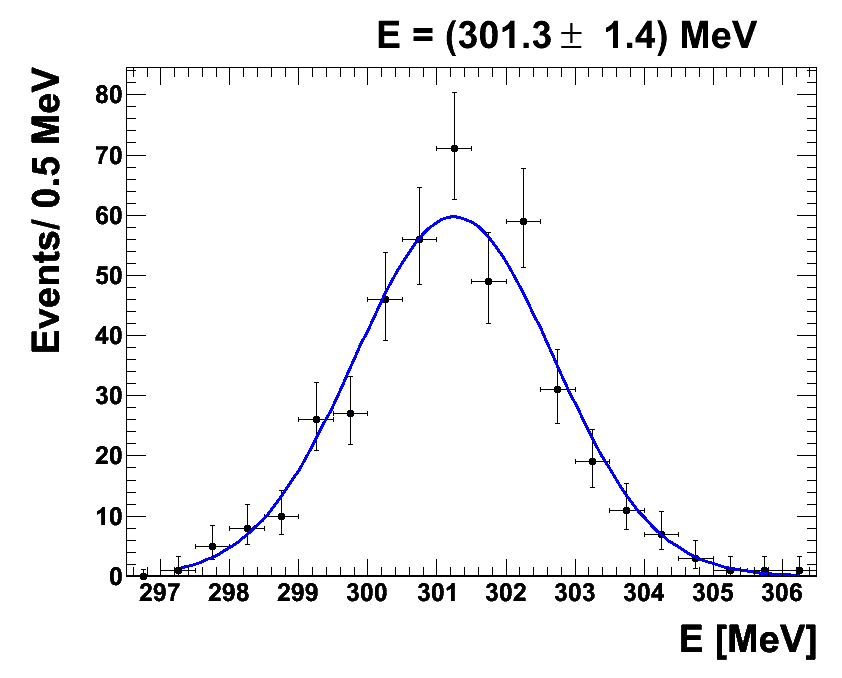}
\end{minipage}
\caption{Results obtained by reconstructing 300\! MeV muons created in the center of the detector and traveling in negative $x$ direction (500 events). The upper row shows the results for the start point of the track, the lower row shows the reconstructed start time (left), the angular deviation of the reconstructed track from the Monte Carlo truth (center) and the kinetic energy of the muon (right).} 
\label{fig::Fitresults}
\end{figure*}

A more precise reconstruction of the track is achieved by determining the track parameters using a negative logarithmic likelihood fit to the integrated charge and the first hit times of each PMT. This is done employing the continuous slowing down approximation i.e. neglecting any kind of statistical fluctuations of the track. The number of parameters required to characterize a track is therefore reduced to seven: The kinetic energy of the particle, the start point of the track, the track direction and the time when the particle was created.\\
\mbox{Fig.\,\ref{fig::Fitresults}} shows the results for single 300\! MeV muons traveling from the center of the detector towards the wall. The resolution obtained for the start point of the track is in the order of a few centimeters and the start time of the event can be determined with sub-nanosecond accuracy. The obtained angular resolution is in the order of a few degrees. The results obtained for electrons are in the same order of magnitude but tend to be slightly worse compared to muons due to the higher statistical fluctuations of electron tracks.\\
Track reconstruction yields useful results for kinetic energies down to 100-200 \!MeV for single muons and down to  $\sim$250\! MeV for single electrons. 
The performance for muons at energies of order 100\! MeV is limited as muons are no longer minimum ionizing which leads to very short track lengths of a few ten centimeters. The low energy limit for electrons on the other hand is due to the increasing statistical deviations of the track from the straight line.\\
Quasielastic CC $\nu_\mu$-events can be reliably distinguished from quasielastic $\nu_{e}$-events, by tagging the decay of the muon produced in the former case. As the spectrum of the Michel electrons exceeds the energies of background processes, most notably the energies released due to neutron capture on hydrogen or carbon, a cut on the visible energy of the muon decay efficiently eliminates background processes mimicking a muon decay. This principle has been verified by simulating 7600 quasielastic $\nu_e$ interactions. As non of these events was wrongly assigned a $\nu_\mu$-event, the probability to wrongly identify a quasielastic $\nu_e$-event as a $\nu_\mu$-event is smaller than $4\cdot 10^{-4}$ at 95\% C.L.. At the same time, the efficiency is roughly $70\%$. This method to reject background is of course insufficient for inelastic interactions where pions are produced. These events, though suppressed at beta-beam energies, require further studies to devise a reliable discimination. 

\subsubsection{Tracking in the 1$-$5 GeV range}

Here we consider a charged current (CC) neutrino event producing a charged lepton. In the lower part of this energy range the scattering is usually quasielastic, but at higher energies single and multi-pion events dominate. Typically a recoil nucleon is emitted but intranuclear collisions and excitations may absorb energy and momentum. 

The reconstruction of the 1--5 GeV event is done with a prototype code that tries to find the test event giving the best fit with the recorded PMT data of the ''true'' event. The code simulates the scintillation light emission from all the secondary particles but no tertiary particles nor nuclear physics.

We found that the single lepton tracks can be reconstructed very well. Also 2 sufficiently long ($>\cal{O}$(50\,cm)) angularly separated tracks can be distinguished easily. Events consisting of 3 tracks can be reconstructed if the tracks are clear, long ($>\cal{O}$(1\,m)) and well separated, though tracks of a few 10\,cm in length remain unseen. Almost parallel tracks are always hardest to distinguish.

Events with 4--5 tracks are very challenging, and can be reliably reconstructed only in special cases, like well-separated tracks longer than a meter, with additional signals from particle decays. The lepton track itself can be distinguished in almost every case. 

For the studied energy range from 1 to 5 GeV, the identification of the lepton flavor shows no ambiguities. No misidentification occurred throughout the testing for straight leptons with any number of secondary particles. The muon and electron signals are very different and there is no way to confuse clear lepton tracks at GeV energies unless there are some rare tertiary processes. Also the statistical fluctuations in photon emission or detection cannot cause such errors with any reasonable probability. 

The position of the interaction vertex can be defined within a few centimeters for most event categories. The length of the longest track (muon) can be measured at $\cal{O}$(10 cm) accuracy or even better. The length scale of the electron shower is found at lower accuracy.

The angular resolution for a long muon track is better than one degree. For electron showers it is a few degrees. For the additional tracks (proton, pions, gammas) the angular resolution is weaker, typically some tens of degrees at one meter track lengths, while for tracks shorter than $\cal O$(50 cm) the directions remain undefined. 

Successive studies point to $\cal O$(1\,\%) deviations between the energies of the reconstructed and the true events. However, the error varies remarkably with different event topologies and the error distribution seems to be far from Gaussian. 

For most applications we may assume 5\,\% energy resolution throughout the regime from 1 to 5 GeV. This value includes already the nuclear physics uncertainties of 1--2\,\%. For events where the neutrino direction is known --- i.e. known neutrino beams --- additional kinematical information is available to improve the energy resolution. 

The recognition of neutral current background ($\nu + X \to \nu + X^* + \pi$) has not been fully demonstrated. It is evident that a substantial fraction of neutral current events could be confused with $\nu_e$ elastic scattering events ($\pi^ 0 \to \gamma\gamma$ as $\nu_e$, $\pi^- $ as $\nu_\mu$). 

The high-energy performance depends primarily on the time resolution of the detector. This suggest to use a fast scintillator with short fluorescence times, and phototubes of low time jitter. Multi-particle tracking will also profit from pulse shape information for individual channels, suggesting the use of FADCs for recording the signals of individual PMTs or PMT arrays (Sec.\,\ref{subsec::readout}. This is particularly important for event recognition and background rejection. 

\subsubsection{Conclusions}

Based on the algorithms presented above, energy and momentum reconstruction for GeV neutrinos in LENA seem well feasible. However, several other aspects must be investigated to obtain a definite result on the final sensitivity of LENA for a long-baseline experiment. The imminent next step is for sure the investigation of neutral current and charged-current pion backgrounds. Nevertheless, the idea of determining the direction of a particle track at degree accuracy in liquid scintillator would have seemed outrageous even a few years ago, let alone the reconstruction of multiparticle vertixes. Once again, the versatility of the liquid scintillation technique has been demonstrated.

%% file: sn.tex


\noindent Measuring neutrinos from the next galactic supernova (SN)
is at the frontier of low-energy neutrino physics and astrophysics.
LENA provides a high-statistics neutrino signal---roughly twice that
of Super-Kamiokande---that can confirm, refute or extend the
standard paradigm of stellar core collapse and determine detailed
neutrino ``light curves" and spectra. Additionally, LENA's superior
energy resolution and various flavor-sensitive detection channels
are particularly advantageous for identifying flavor oscillation
effects that are sensitive to the unknown mixing angle $\theta_{13}$
and the neutrino mass hierarchy.

\subsubsection{Basic picture}

Core-collapse SNe are the spectacular outcome of the violent deaths
of massive stars, including the spectral types II, Ib and Ic
\cite{Burrows:2000mk, Janka:2006fh}. The early universe aside, it is
only here that neutrinos do not stream freely in spite of their weak
interactions and actually dominate the dynamics and energetics. The
basic picture of core collapse is supported by the neutrino
observation from SN~1987A \cite{Hirata:1987hu, Hirata:1988ad,
Bionta:1987qt, Bratton:1988ww, Alekseev:1987ej, Alekseev:1988gp}.
This historical measurement and the early solar neutrino
observations remain the only astrophysical sources detected in
neutrinos. A high-statistics neutrino observation of stellar core
collapse is at the frontier of low-energy neutrino astronomy,
providing an unprecedented wealth of astrophysical and
particle-physics information~\cite{Schramm:1987ra, Raffelt:1990yz,
Raffelt:1999tx, Dasgupta:2008zz, Raffelt:2010zz}.

A core collapse anywhere in the Milky Way and its satellites (such
as the Magellanic Clouds) provides a detailed neutrino light curve
and spectrum. The distance distribution is rather broad with an
average of around 10~kpc \cite{Mirizzi:2006xx}. At this distance, a
SN produces around $10^4$ events in LENA from the dominant inverse
beta decay reaction $\bar\nu_e+p\to n+e^+$. Many existing and
near-future detectors will pick up tens to hundreds of events
\cite{Scholberg:2007nu, Scholberg:2010zz}, whereas statistics
comparable to LENA is provided only by Super-Kamiokande. Moreover,
the high-energy neutrino telescope IceCube at the South Pole will
register roughly $10^6$ uncorrelated Cherenkov photons in excess of
background, providing superior sensitivity to fast signal variations
that are suggested by recent multi-dimensional
simulations~\cite{2009A&A...496..475M, Lund:2010kh,
2011ApJ...728....8B}.

Galactic SNe occur a few times per century as implied by SN
statistics of external galaxies~\cite{vandenBergh:1994,
Cappellaro:1999qy, Cappellaro:2000ez}, the historical
record~\cite{Strom:1994, Tammann:1994ev}, and the galactic abundance
of the unstable isotope $^{26}$Al measured with the INTEGRAL
gamma-ray observatory~\cite{Diehl:2006cf}. The low-energy neutrino
sky has been systematically watched since 30 June 1980 when the
Baksan Scintillator Telescope took up operation. Only SN~1987A was
detected over thirty years, beginning to provide non-trivial
constraints on hypothetical ``invisible'' core-collapse
phenomena~\cite{Alekseev:2002ji}. Still, the neutrinos from about a
thousand galactic SNe are on their way and observing one of them is
a once-in-a-lifetime opportunity.

Readiness for a galactic SN burst is an essential detector
capability. Reaching Andromeda (M31) and Triangulum (M33) at
750~kpc, the next large galaxies in the local group, requires
megaton-class detectors for tens of events. Multi-megaton detectors
would detect a few neutrinos from SNe out to a few
Mpc~\cite{Kistler:2008us}. One could systematically build up an
average SN neutrino spectrum, but such a project is for the more
distant future.  On a shorter term, another realistic opportunity to
detect SN neutrinos is the diffuse SN neutrino background (DSNB)
from all past SNe (Sec.~\ref{subsec::dsnb}).

LENA has about twice the signal statistics of the Super-Kamiokande
water Cherenkov detector. More importantly, it has superior energy
resolution, a lower threshold, and distinguishes inverse beta decay
from other channels by recognizing the final-state neutrons.
(Dissolved gadolinium in water Cherenkov
detectors~\cite{Beacom:2003nk}, currently studied in the EGADS
project at Super-Kamiokande~\cite{Watanabe:2008ru}, will also
provide neutron tagging.) LENA's excellent energy resolution is a
huge advantage for recognizing Earth effects in SN neutrino flavor
oscillations (Sec.~\ref{sec:flavoroscillations}). Moreover, LENA is
complementary to water Cherenkov detectors by including $^{12}$C as
a target nucleus and by sensitivity to elastic scattering on free
protons~\cite{Beacom:2002hs}.

An increased neutron rate in LENA can signify neutrino emission by
thermal processes in the progenitor during its last weeks of pre-SN
evolution~\cite{Odrzywolek:2003vn}. While this effect requires the
star to be close, the red supergiant Betelgeuse at about 200~pc
\cite{Harper:2008} is a possible candidate. The neutrino burst after
collapse would trigger about $10^7$ events in LENA. The data
acquisition system must be able to handle such a case without being
blinded by neutrinos.

\subsubsection{Supernova astrophysics}

Supernovae and the related, though much rarer, long cosmic gamma-ray
bursts are the strongest astrophysical sources of low-energy
neutrinos. These core-collapse events are the final stages of the
evolution of massive stars and as such play a central role in
stellar and nuclear astrophysics~\cite{2002RvMP...74.1015W}. Besides
being the birth sites of neutron stars and stellar-mass black holes,
they are probably the origin of about half of the chemical elements
heavier than iron.

While the basic concept of stellar core collapse and neutron-star
formation was confirmed by the historical measurement of neutrinos
from SN~1987A, our understanding of the detailed processes driving
the core's evolution and ultimately causing the SN blast remains
incomplete and has little empirical underpinning~\cite{Janka:2006fh}.
Observations of the bright electromagnetic spectacle that accompanies
stellar death provide only indirect information about the initiating
mechanism: the center of the explosion is obscured by several solar
masses of intransparent, gaseous ejecta.

The current theory of stellar explosions strongly relies on
numerical modeling that requires empirical support. While the
produced heavy elements somewhat probe the conditions around the
origin of the explosion, only neutrinos and gravitational waves can
escape directly from the densest regions and thus are unique
messengers from the very center. Their time-dependent signal
features carry detailed and complementary information of the
evolving thermodynamical state and of the dynamical motions in the
compact remnant assembling at the heart of the dying star.

Neutrino measurements from a galactic SN together with the detection
of a gravitational-wave burst will trigger a breakthrough in our
understanding of some of the most important questions in stellar
astrophysics: What are the conditions in collapsing cores of massive
stars? Is there a significant amount of rotation? Do magnetic fields
play an important role? How can massive stars succeed to reverse
their catastrophic infall to a powerful explosion? What are the
properties of hot nuclear matter? What are the mass, radius, and
binding energy of the newly-formed neutron star? Does the compact
remnant undergo a phase transition to a more compressed quark-matter
state or even collapse to a black hole? Are SN explosions and
new-born neutron stars the long-sought formation sites of the
heaviest neutron-rich elements that are made by the rapid-neutron
capture process (r-process)?

\begin{figure*}
\hbox to
\textwidth{\includegraphics[height=0.32\textheight]{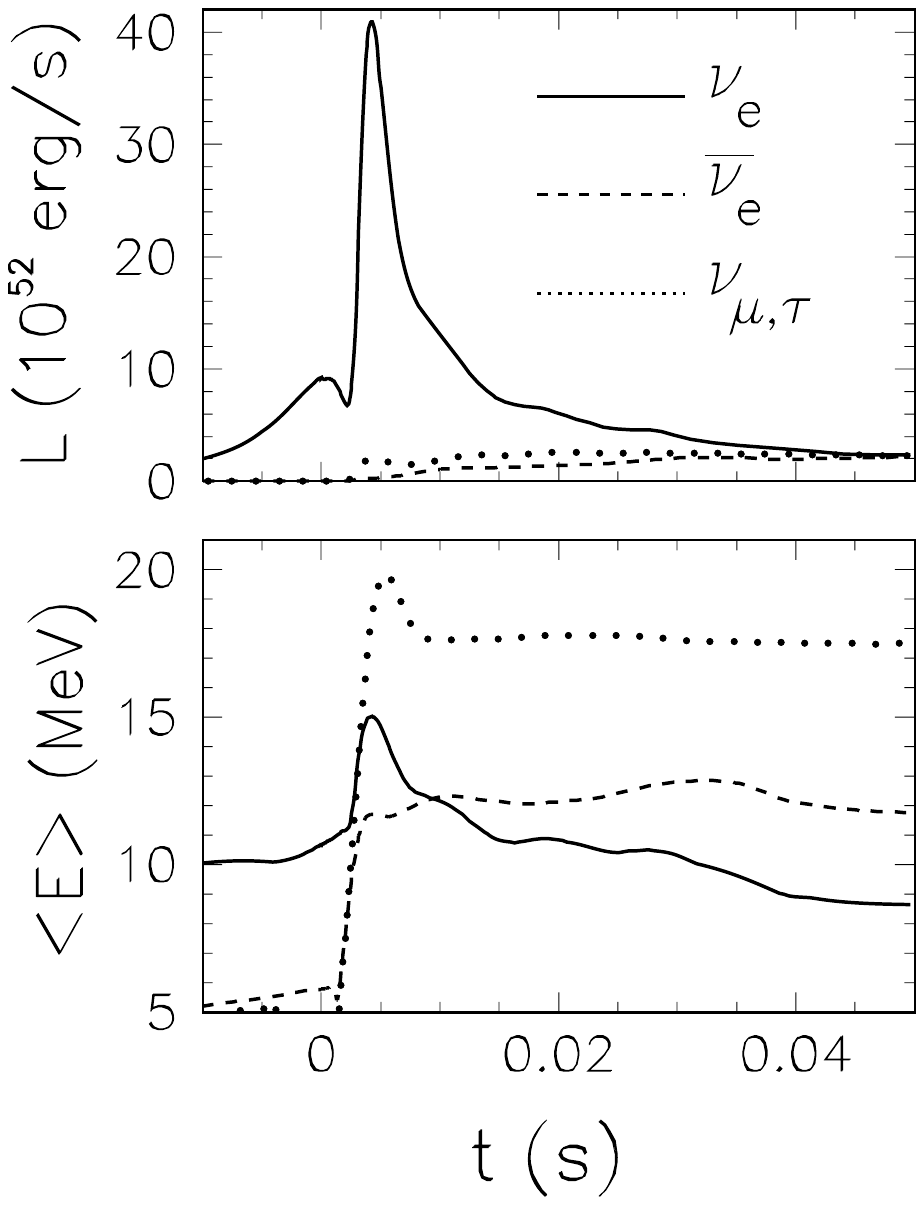}\kern0.2cm
\hfill
\includegraphics[height=0.32\textheight]{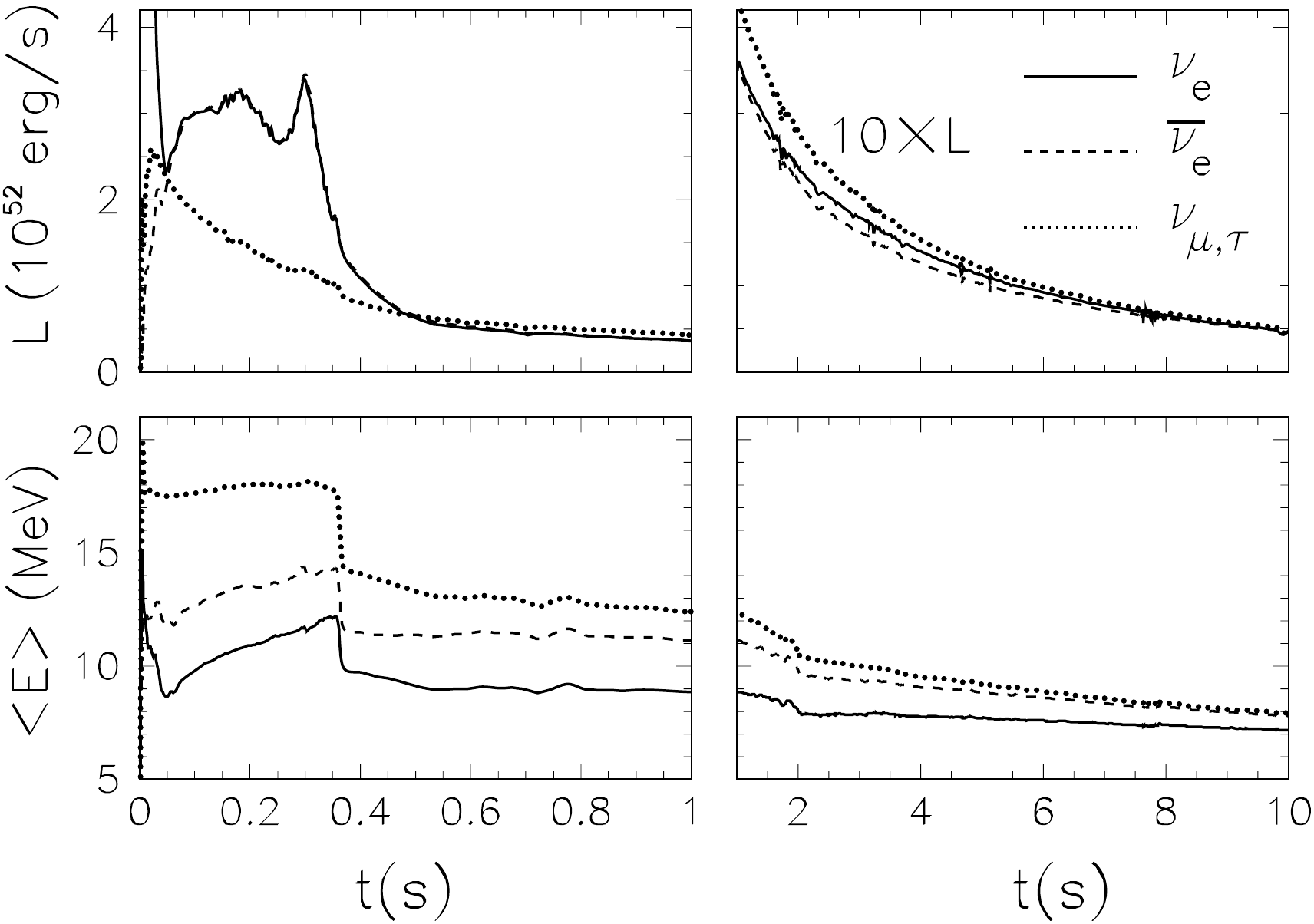}}
\caption{Neutrino signal of a core-collapse SN for a $10.8\,M_\odot$
progenitor according to a numerical simulation of the Basel
group~\cite{Fischer:2009af}. All quantities are in the laboratory
frame of a distant observer. In this spherically symmetric
simulation the explosion was triggered artificially by implementing
enhanced neutrino energy deposition. {\it Left}: Prompt neutrino
burst. {\it Middle}: Accretion phase. {\it Right}: Cooling
phase.}\label{fig:SNneutrinosignal}
\end{figure*}

\subsubsection{Expected neutrino signal}
\label{sec:neutrinosignal}

In spite of many open questions, today's numerical SN models may
well provide a reasonable first guess of the signal characteristics.
Spherically symmetric simulations have recently provided robust
explosions for small progenitor masses of 8--10~$M_\odot$, the class
of electron-capture SNe (or O-Mg-Ne-core SNe) \cite{Fischer:2008rh,
Huedepohl:2009wh, 2006A&A...450..345K, 2008A&A...485..199J,
2006ApJ...644.1063D, 2007ApJ...669..585D}. For more massive stars,
leading to the conventional iron-core SNe, strong deviations from
spherical symmetry caused by large-scale convection and the standing
accretion shock instability (SASI) are probably important, but
full-fledged 3D simulations with sufficiently sophisticated neutrino
transport are only beginning to come into reach.

The expected neutrino signal consists of three main phases
(Fig.~\ref{fig:SNneutrinosignal}), testing different aspects of SN
theory and neutrino flavor oscillations.
\begin{enumerate}
\item Few tens of ms after bounce: Shock break-out and
   deleptonization of the outer core layers, emission of the
   ``prompt $\nu_e$ burst.'' Emission of other flavors only
   begins and that of $\bar\nu_e$ is at first suppressed.
   Largely independent of progenitor mass and equation of
   state~\cite{Kachelriess:2004ds}.
\item Accretion phase, few tens to several hundred ms, depending
   on progenitor mass and other parameters. Shock stalls at
   100--200~km, neutrino emission is powered by infalling
   material. Fluxes of $\nu_e$ and $\bar\nu_e$ much larger (a
   factor of two is not unrealistic) than those of the other
   flavors. Pronounced hierarchy $\langle E_{\nu_e}\rangle
   <\langle E_{\bar\nu_e}\rangle <\langle E_{\nu_x}\rangle$
   with $\nu_x$ representing any of $\nu_{\mu,\tau}$ and
   $\bar\nu_{\mu,\tau}$. Large-scale convection and SASI mode
   build up, leading to strong time variation of the neutrino
   signal.
\item Cooling phase, up to 10--20~s. Neutrino flux powered by
    cooling of the deep core on a diffusion time scale.
    Approximate luminosity equipartition between all species and
    only a mild $\langle E_{\bar\nu_e}\rangle$/$\langle
    E_{\nu_x}\rangle$ hierarchy. Larger $\nu_e$ number flux due
    to de-leptonization.
\end{enumerate}
Of course, completely different signatures can arise if new
phenomena occur. Examples are a late time QCD phase transition or
black hole formation.

A broad range of possible spectral properties of the neutrino signal
have been considered in the literature, but until recently the only
numerical model of multi-flavor SN neutrino emission from bounce to
long-term cooling was provided by the Livermore
group~\cite{Totani:1997vj}. It is only during the past year that
their pioneering work has been superseded by modern long-term
simulations. For the first time hydrodynamic simulations coupled
with modern neutrino Boltzmann solvers in 1D have been carried all
the way to proto-neutron star cooling. The Basel group has evolved
progenitors with different masses up to 10~s after bounce
\cite{Fischer:2009af}. The Garching group has published a similar
simulation for an electron-capture SN~\cite{Huedepohl:2009wh}.

The emerging picture suggests smaller average energies than often
assumed and much less pronounced spectral hierarchies, particularly
during the cooling phase. Time-integrated values in the range
$\langle E_{\bar\nu_e}\rangle=12$--14~MeV, and somewhat larger for
$\nu_{\mu,\tau}$, look reasonable and are in agreement with the
SN~1987A observations and with analytic~\cite{Raffelt:2001kv} and
Monte-Carlo studies of neutrino transport~\cite{Keil:2002in}. We use
such relatively modest energies to gauge our expectations for LENA.
Of course, it is the very purpose of SN neutrino observations to
measure the neutrino flux characteristics independently of
theoretical predictions and it remains quite possible that typical
SNe produce much larger or very different signals. Moreover, one
expects large variations between different SNe, depending for
example on different rates and amounts of accretion.

\subsubsection{Detection channels in LENA}
\label{sec:detectonchannels}

The purpose of a high-statistics SN neutrino observation is to
measure time-dependent features. However, a first impression of the
detector capabilities is gained from integrated detection rates.
Detailed spectral studies will be important and so we assume a range
of different source characteristics. To this end we treat the SN
schematically as a black-body source for all neutrino species. We
assume a total emitted energy of $E_{\rm tot}=3\times 10^{53}$~erg,
equipartitioned among all neutrino species, and Maxwell-Boltzmann
spectra with $\langle E_\nu\rangle =12$, 14 and 16~MeV. Of course,
in a realistic SN one expects flavor-dependent differences that will
be used, for example, to search for flavor oscillations.

LENA's golden detection channel is inverse beta decay ${\bar\nu}_e +
p \to n+ e^+ $. The produced neutron thermalizes and wanders in the
detector until it is captured by a proton, $n + p \to d
+\gamma$\,(2.2\,MeV) after an average time of $\sim$250\,\textmu s. The large
homogeneous detection volume ensures efficient neutron capture and
$\gamma$ detection. Therefore, these events are tagged by the
delayed coincidence between the prompt positron and the $\gamma$-ray
from neutron capture.

\begin{table}[b]
\begin{tabular}{llccc}
\toprule
\multicolumn{1}{l}{Reaction} & Type &\multicolumn{3}{l}{Events for $\langle E_\nu\rangle$ values\vbox to12pt{}}\\
&&12~MeV&14~MeV&16~MeV\\
\colrule
$ \bar\nu_e\,p \to n\,e^+$  & CC & 1.1$\times$10$^4$ & 1.3$ \times$ 10$^4$ & 1.5$\times$ 10$^4$ \\
$\nu\,p \to p\,\nu$ &  NC & 1.3$\times$10$^3$ & 2.6$\times$10$^3 $ & 4.4$\times$10$^3$ \\
$\nu\,e \to e\,\nu $  & NC & 6.2$\times$10$^2$ & 6.2$\times$10$^2$ & 6.2$\times$10$^2$ \\
$\nu\,{^{12}{\rm C}} \to {^{12}{\rm C}^*}\,\nu$\\
~~${^{12}{\rm C}}^* \to {^{12}{\rm C}}\,\gamma$ & NC & 6.0$\times$10$^2$ & 1.0$\times$10$^3$ & 1.5$\times$10$^3$\\
$\bar\nu_e\,{^{12}{\rm C}} \to {^{12}{\rm B}}\,e^+$\\
~~${^{12}{\rm B}} \to {^{12}{\rm C}}\,e^-\,\bar\nu_e$ & CC & 1.8$\times$10$^2$ & 2.9$\times$10$^2$ & 4.2$\times$10$^2$ \\
$\nu_e\,{^{12}{\rm C}}\to{^{12}{\rm N}}\,e^-$\\
~~${^{12}{\rm N}} \to {^{12}{\rm C}}\,e^+\,\nu_e$&CC & 1.9$\times$10$^2$ & 3.4$\times$10$^2$& 5.2$\times$10$^2$\\
\botrule
\end{tabular}
\caption{Expected event rate in LENA for a SN at a distance of
10~kpc, where $\nu$ stands for a neutrino or antineutrino of any
flavor. The NC rates are summed over all flavor channels. Our three
representative values for $\langle E_\nu\rangle$ are taken to be
equal for all flavors. The Birks constant was taken to be
0.010~cm/MeV, and a threshold of 0.2~MeV was assumed to calculate
total number of events for the $\nu\,p \to p\,\nu$
channel.\label{tab:evenrates}}
\end{table}

Three charged-current (CC) reactions measure $\nu_e$ and $\bar\nu_e$
fluxes and spectra while three neutral-current (NC) processes,
sensitive to all flavors, give information on the total flux. Typical
event rates for a generic SN at a distance of 10~kpc are reported in
Table~\ref{tab:evenrates} for our representative cases of $\langle
E_\nu\rangle$. A LAB-based scintillator and a fiducial mass of 44~kt
is assumed, providing about $3.3\times10^{33}$ protons. The $^{12}$C
reactions have a high kinematical threshold ($E>15$~MeV). The
resulting steep energy dependence in principle provides information
on the neutrino spectra.

\begin{figure}
\includegraphics[width=0.9\columnwidth]{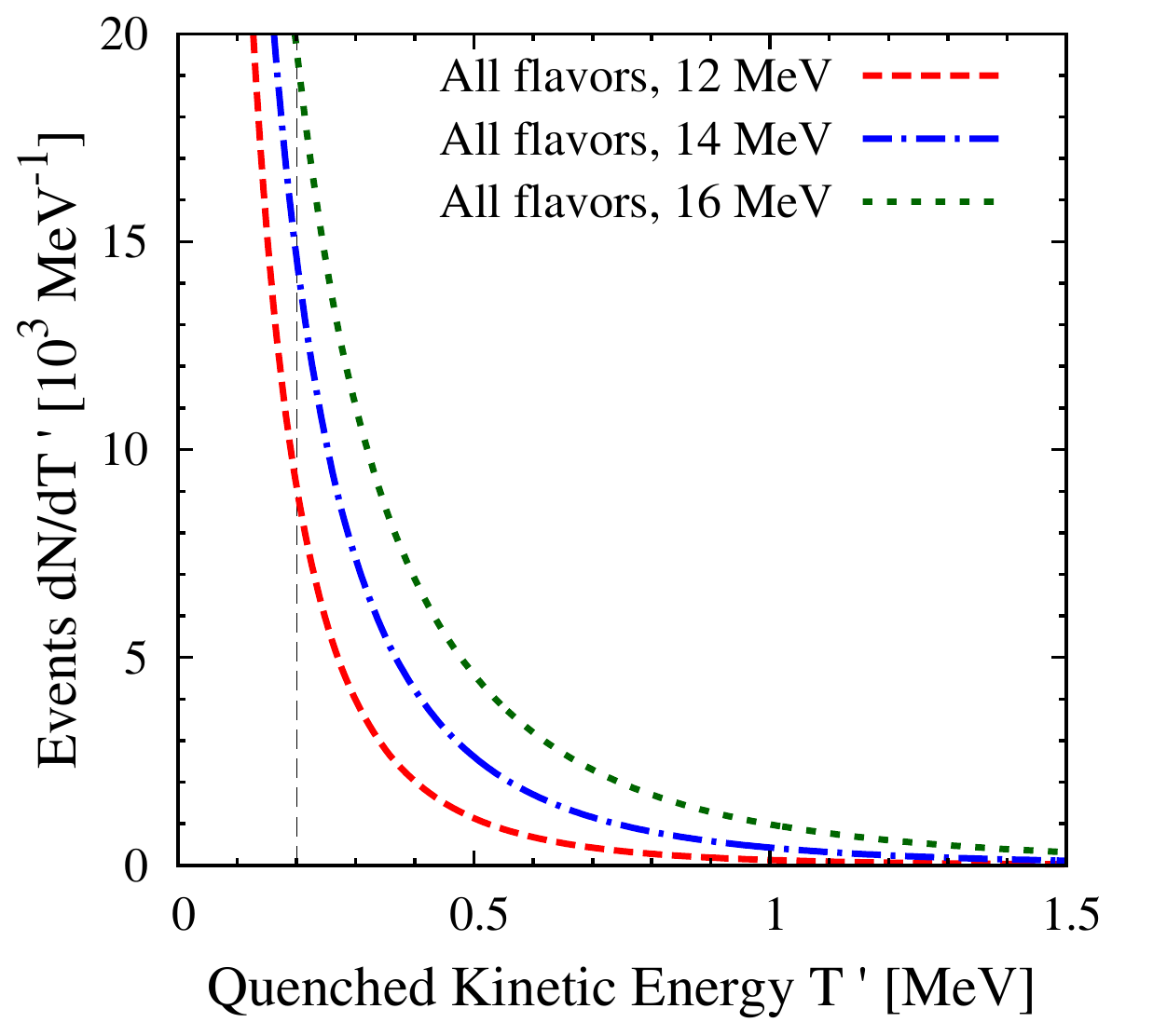}
\caption{Neutrino signal in the neutrino-proton elastic scattering channel
according to Ref.~\cite{Dasgupta:2011wg}
for the typical SN flux parameters mentioned in the text.
The vertical line shows an assumed threshold of 0.2 MeV.}\label{fig:nu-p-events}
\end{figure}

It is particularly difficult to detect the $\nu_{\mu}$,
$\nu_{\tau}$, $\bar{\nu}_{\mu}$ and $\bar{\nu}_{\tau}$ flavors and
measure their energies, because, unlike $\nu_e$ and $\bar{\nu}_e$,
they have only NC interactions. On the other hand, observing these
flavors is essential to disentangling flavor mixing and correctly
estimating the total energy emitted in neutrinos. Directly observing
two spectral components due to flavor mixing in the CC data, one for
$\bar{\nu}_e$ and another for the non electronic flavors, will only
be possible if the average energies are significantly different.
That possibility is not strongly favored by current theory.
Promisingly, one of the main strengths of LENA is its low threshold
that should allow us to observe neutrino-proton elastic scattering
events, which have spectral information and a substantial yield from
these neutrinos~\cite{Beacom:2002hs, Dasgupta:2011wg}. Other NC
channels typically lack spectral information and have lower yields.
The expected spectrum of elastic neutrino-proton scattering events
for typical SN parameters is shown in Fig.{~\ref{fig:nu-p-events}}.
The spectrum has been smeared with the energy resolution given in
the previous section, and the threshold is taken to be 0.2~MeV to
stay above the $^{14}$C background. With a few thousand observed
events, one could reconstruct the non electron flavor neutrino
spectra with almost the same precision as that of
$\bar{\nu}_e$~\cite{Dasgupta:2011wg}. The prompt $\nu_e$ burst alone
will produce around 50 events, depending on the mixing scenario, by
electron scattering and 90, independent of oscillations, by proton
elastic scattering (using the Basel model of
Fig.~\ref{fig:SNneutrinosignal}).

\subsubsection{Astrophysical lessons}

A gravitational-wave signal provides information on non-radial
deformation and non-spherical mass
motions~\cite{2009CQGra..26f3001O}, whereas a high-statistics
neutrino signal allows us to follow directly the different stages of
core collapse without additional assumptions
(Fig.~\ref{fig:SNneutrinosignal}). The prompt $\nu_e$ burst is a
robust and uniform landmark structure of all theoretical
predictions. Because of LENA's capability of distinguishing NC and
CC events, it offers a unique possibility of identifying this
feature. For example, one could estimate the SN distance in the
plausible case that the optical display is hidden behind the dense
gas and dust clouds of a star-forming
region~\cite{Kachelriess:2004ds}. Moreover, one could use the prompt
$\nu_e$ burst in LENA for coincidence measurements with the
gravitational wave burst that may arise at core bounce. Using the
prompt $\nu_e$ burst could provide an even sharper coincidence than
can be achieved with the onset of the $\bar\nu_e$ signal in
Super-Kamiokande~\cite{Pagliaroli:2009qy} and
IceCube~\cite{Halzen:2009sm}. Moreover, the prompt $\nu_e$ burst
could help to find the SN direction by neutrino
triangulation~\cite{Beacom:1998fj}, although the recoil electron
signal in a water Cherenkov detector provides superior pointing
capabilities~\cite{Beacom:1998fj, Tomas:2003xn}.

The magnitude of the $\nu_e$ and $\bar\nu_e$ accretion luminosities
after core bounce (Fig.~\ref{fig:SNneutrinosignal}, middle) depends
on the mass infall rate and thus on the progenitor-dependent
structure of the stellar core, with more massive cores producing
higher luminosities~\cite{2003NuPhA.719..144L, 2006A&A...457..281B}.
Luminosity variations during this phase~\cite{2009A&A...496..475M,
Lund:2010kh, 2011ApJ...728....8B}, accompanied by sizable
gravitational-wave emission at several hundred~Hz
\cite{2009A&A...496..475M, 2009ApJ...707.1173M} would confirm the
presence of violent hydrodynamic instabilities stirring the
accretion flow around the assembling neutron star. Such activity and
a several hundred millisecond delay of the onset of the explosion
are expected within the framework of the delayed neutrino-driven
mechanism. A pronounced drop of the $\nu_e$ and $\bar\nu_e$
luminosities, followed by a close similarity to those of
heavy-lepton neutrinos, would finally signal the end of the
accretion phase and the launch of the outgoing SN blast wave. The
cooling signature of a nascent neutron star is characterized by a
monotonic and gradual decline of the neutrino emission. It would
be prolonged if additional energy was released by phase transitions
in the nuclear matter. Exotic scenarios might feature a
secondary $\nu_e$ burst~\cite{Dasgupta:2009yj} or an abrupt
end of neutrino emission if the collapse to a black hole occured~\cite{Beacom:2000qy}.

LENA could provide even more information: Due to its superior energy
resolution it could help to disentangle source-imposed spectral
features from those caused by neutrino-flavor conversions. Moreover,
detecting significant numbers not only of $\bar\nu_e$ but also of
$\nu_e$ and heavy-lepton neutrinos (Table~\ref{tab:evenrates}) would
yield at least time-averaged spectral information for different
emission channels. Conceivably one could extract information on the
neutron-to-proton ratio in the neutrino-processed SN outflows,
presently also a sensitive result of numerical modeling of a
multitude of complex processes. The relative abundance of neutrons
and protons determines the conditions for nucleosynthesis and are
set by competing $\nu_e$ and $\bar\nu_e$ captures, which in turn
depend delicately on the relative fluxes and spectral distributions
of these neutrinos. A LENA measurement of a SN burst may offer the
only direct empirical test of the possibility for r-processing in
the SN core, except for an extremely challenging in-situ measurement
of r-process nuclei in fresh SN ejecta.

\subsubsection{Particle physics and neutrino properties}
\label{sec:flavoroscillations}

On the particle-physics side, the high-statistics observation of a
SN neutrino burst can provide crucial particle-physics lessons.
Numerous results derived from the sparse SN~1987A data can be
refined. One can also probe more exotic scenarios. Spin-flavor
conversions caused by the combined action of magnetic fields and
matter effects can transform some of the prompt $\nu_e$ burst to
$\bar\nu_e$, leading to a huge inverse-beta
signal~\cite{Akhmedov:2003fu}. Such an observation would provide
smoking-gun evidence for neutrino transition magnetic moments.
Non-radiative decays would also produce a $\nu_e\to \bar\nu_e$
conversion during the prompt burst~\cite{Ando:2004qe}.

Perhaps of greatest interest are flavor oscillations. Neutrinos
propagating through the SN mantle and envelope encounter a large
range of matter densities, allowing for Mikheyev-Smirnov-Wolfenstein
(MSW) conversions driven by the atmospheric neutrino mass difference
and the small mixing angle $\theta_{13}$. Therefore, in principle a
SN neutrino signal is sensitive to the two as yet unknown neutrino
mixing parameters: $\theta_{13}$ and the ordering of the neutrino
masses that could be in the normal (NH) or inverted hierarchy~(IH).

Our understanding of SN neutrino oscillations has recently undergone
a change of paradigm by the insight that the neutrino-neutrino
refractive effect is crucial. These collective (or self-induced)
flavor conversions occur within a few hundred km above the neutrino
sphere; see Ref.~\cite{Duan:2010bg} for a review of the recent
torrent of literature on this topic. The most important
observational consequence is a swap of the $\nu_e$ and $\bar\nu_e$
spectrum with that of $\nu_x$ and $\bar\nu_x$ in certain energy
intervals~\cite{Duan:2006an}. The sharp spectral features at the
edges of these swap intervals are known as ``spectral splits.''
Their development depends on the neutrino mass hierarchy as well as
on the ordering of the flavor fluxes at the source. Therefore, the
split features can depend on time in interesting
ways~\cite{Fogli:2007bk, Dasgupta:2009mg, Friedland:2010sc,
Dasgupta:2010cd, Choubey:2010up}.

The main problem to detect oscillation features is that one can not
rely on detailed theoretical predictions of the flavor-dependent
fluxes and spectra. Therefore, model-independent signatures are
crucial. One case in point is the energy-dependent modulation of the
neutrino survival probability caused by Earth matter effects that
occur if SN neutrinos arrive at the detector ``from
below''~\cite{Lunardini:2001pb}. The appearance of Earth effects
depends on the flux and mixing scenario~\cite{Choubey:2010up}.
Therefore, its detection could give hints about the primary SN
neutrino fluxes, as well as on the neutrino mass hierarchy and the
mixing angle $\theta_{13}$.

The excellent energy resolution of LENA is a particular bonus for
discovering small energy-dependent flux modulations caused by Earth
effects~\cite{Dighe:2003jg}, but of course depends on seeing the SN
shadowed by the Earth. In a far-northern location such as the
Pyh{\"a}salmi mine in Finland the shadowing probability for a
galactic SN is about 58\%, against an average of 50\% for a random
location~\cite{Mirizzi:2006xx}.  A particularly interesting scenario
consists of a large volume scintillator detector in the north to
measure the geo-neutrino flux in a continental location and another
one in Hawaii to measure it from the oceanic crust. The probability
that only one of them is shadowed exceeds 50\% whereas the
probability that at least one is shadowed is about 80\%. Therefore,
Pyh{\"a}salmi and Hawaii are complementary both for observing
geo-neutrinos and Earth matter effects in SN neutrinos.

Additional signatures of flavor conversions can be imprinted by
matter effects of the shock fronts in the SN
envelope~\cite{Schirato:2002tg}. The number of events, average
energy, or the width of the spectrum may display dips or peaks for
short time intervals~\cite{Fogli:2003dw, Tomas:2004gr}. Such
signatures yield valuable information about shock-wave propagation,
the neutrino mass hierarchy and $\theta_{13}$. However, realistic
chances to detect shock features remain unclear. The
flavor-dependent spectral differences in the anti-neutrino channel
are probably small during the cooling phase. Moreover, strong
turbulence in the post-shock regions could affect these
signatures~\cite{Kneller:2010sc}.

\subsubsection{Summary}

A worldwide network of neutrino and gravitational-wave detectors,
constituting the SuperNova Early Warning System (SNEWS)
\cite{Antonioli:2004zb}, will provide early warning and detailed
multi-messenger measurements of the next nearby SN. A
high-statistics neutrino observation, even from a single SN, will go
a long way to answering many fundamental questions about the role of
neutrinos for the astrophysics of core collapse and may shed new
light on the properties of neutrinos and other particles. LENA will
play an exceptional role due to its low energy threshold, excellent
energy resolution, and multi-channel signatures that will allow one
to disentangle flavor-dependent properties of the neutrino signal
and to identify subtle modulations imprinted by Earth effects. LENA
may be the only facility that is able to spot the prompt $\nu_e$
burst and thus the earliest and largely model-independent signature
of stellar death, even yielding an estimate of the SN distance.

%% file: dsnb.tex

%

\noindent The diffuse SN neutrino background (DSNB) from all core
collapse events in the universe is a guaranteed neutrino flux from
cosmological distances.  DSNB $\bar{\nu}_e$ can be detected at
energies above 10 MeV, where the reactor neutrino background
vanishes and atmospheric neutrino backgrounds are small and likely
controllable.  LENA provides about twice the counting rate of
Super-Kamiokande, and together they could collect 5--10 events per
year. Measurement of the average $\bar{\nu}_e$ emission spectrum
will help test models of SNe, variation in emission, and neutrino
properties.

\subsubsection{Basic picture} \label{sec:DSNB-basics}

A great and varied scientific return is expected from the
observation of a nearby SN (Sec.~\ref{subsec::sn}), but such events
are rare in the Milky Way. The guaranteed DSNB flux provides a way
to detect SN neutrinos without a fortuitous
burst~\cite{BisnovatyiKogan:1984, Krauss:1983zn, Dar:1984aj,
Woosley:1986aa, Totani:1995rg, Malaney:1996ar, Hartmann:1997qe,
Kaplinghat:1999xi, Fukugita:2002qw, Ando:2002zj, Ando:2002ky,
Strigari:2005hu, Lunardini:2005jf, Yuksel:2007mn,
Chakraborty:2008zp, Galais:2009wi, Beacom:2010kk, Lunardini:2010ab, Vissani:2011kx}. DSNB signals
depend on three ingredients. First, the cosmic core collapse rate,
about 10 per second in the causal horizon; this is determined by
astronomical measurements that are already precise and quickly
improving. Second, the average SN neutrino emission, which is
expected to be comparable for all core collapses, including those
that fail and produce black holes (for which it may be even larger,
as discussed below); this is the quantity of
fundamental interest. Third, the detector capabilities, including
the energy dependence of the cross section and detector backgrounds;
Super-Kamiokande and LENA should be able to detect DSNB
$\bar{\nu}_e$.

Detecting the DSNB is important even if a Milky Way burst is
observed. DSNB $\bar{\nu}_e$ will provide a unique measurement of
the average neutrino emission spectrum to test SN simulations.
Comparison to data from SN~1987A and an eventual Milky Way SN will
test the variation between core collapses.  While the statistics of
DSNB events will be low, like those of SN~1987A, this data will
more effectively measure the exponentially falling tail of the
spectrum at high energies.  The DSNB is also a new probe of stellar
birth and death: its energy density is comparable to that of photons
produced by stars, but the DSNB is unobscured and has no known
competition from astrophysical sources. Finally, the DSNB data will
test flavor mixing and more exotic particle properties.

The importance of running LENA to detect the DSNB should not be
underestimated.  If Super-Kamiokande does not add gadolinium, or
does but encounters technical problems, LENA could be the only
experiment to detect the DSNB.  If both experiments are successful,
their data, based on detection by inverse beta decay above 10 MeV,
would be similar.  Having two independent experiments would be very
valuable, as this will be a challenging measurement.  In addition, LENA
provides about twice the counting rate of Super-Kamiokande.  Collecting
statistics at a combined three times higher rate than Super-Kamiokande
alone could have a decisive impact on the physics that could be extracted.

\subsubsection{DSNB signals} \label{sec:DSNB-signals}

The DSNB event rate spectrum follows from a line of sight integral
for the radiation intensity from a distribution of distant sources.
After integrating over all angles due to the isotropy of the DSNB
and the transparency of Earth, it is, in units ${\rm s}^{-1}~{\rm
MeV}^{-1}$,
\begin{eqnarray}
\frac{dN_{\rm vis}}{dE_{\rm vis}}(E_{\rm vis})&=&
\int_0^\infty
\left[R_{\rm SN}(z) \phantom{\frac{a}{b}} \!\!\!\! \right]
\left[(1+z) \phantom{\frac{a}{b}} \!\!\!\! \varphi[E_\nu (1 + z)] \right]
\nonumber\\
&&\quad{}\times
\left[ N_T \, \sigma(E_\nu) \phantom{\frac{a}{b}} \!\!\!\! \right]
\left[\left| \frac{c \, dt}{dz} \right| dz\right]\,.
\end{eqnarray}
On the right hand side, the ingredients are ordered as described
above. The first is the comoving cosmic core-collapse rate, in units
Mpc$^{-3}$ yr$^{-1}$; it evolves with redshift.  The second is the
average time-integrated emission per SN, in units MeV$^{-1}$;
redshift reduces emitted energies and compresses spectra.  The third
is the number of targets times the detection cross section; this
does not need to be under the integral.  The last term is the
differential distance, where $| \mathrm{d}t / \mathrm{d}z |^{-1} =
H_0 (1+z) [\Omega_\Lambda + \Omega_m(1+z)^3]^{1/2}$; the
cosmological parameters are taken as $H_0 = 70$ km s$^{-1}$
Mpc$^{-1}$, $\Omega_\Lambda = 0.7$, and $\Omega_m = 0.3$. (The
cosmology and the SN rate derived from star formation rate data are
really one combined factor proportional to the ratio of the average
luminosity per galaxy in SN neutrinos relative to stellar photons.)
The left hand side is the DSNB spectrum in visible energy $E_{\rm
vis}$; the relation to neutrino energy $E_\nu$ depends on cross
section and detector specifics.  We next consider details of the
three main ingredients.

\medskip
\noindent{\bf Cosmic SN rate.}
The cosmic core collapse rate is precisely
known~\cite{Hopkins:2006bw, Horiuchi:2008jz, Horiuchi:2011zz}.
The redshift range
relevant for the DSNB depends on energy, with lower energies probing
higher redshifts.  For detected energies above 10 MeV, most DSNB
neutrinos are emitted at redshifts $z < 1$, where the astronomical
data are most precise.  The best determinations of the core collapse
rate come from predictions based on measured star formation rates
and related observables such as the extragalactic background light
~\cite{Horiuchi:2008jz}. As massive stars are short-lived, the
redshift evolution of the core collapse and star formation rates
must be the same.  The relative normalization depends on just the
minimum mass for core collapse, about $8
M_\odot$~\cite{Smartt:2008zd}; the predicted rate depends only
weakly on the assumed stellar initial mass function because star
formation data primarily sample massive stars.  Comparable neutrino
fluxes are expected for ordinary SNe and those that are faint,
obscured, or even
failed~\cite{Sumiyoshi:2007aa,Sumiyoshi:2008zw,Fischer:2008rh,
Nakazato:2008vj}, so the DSNB does not depend much on the outcomes,
though it may be larger than assumed here.
Measured SN and predicted core collapse rates are in reasonable
agreement, and the data will quickly improve~\cite{Lien:2009db,
Lien:2010yb, Horiuchi:2011zz}.

The local core collapse rate is $R_{\rm SN}(z = 0) = (1.25 \pm 0.25)
\times 10^{-4}$ Mpc$^{-3}$ yr$^{-1}$~\cite{Horiuchi:2008jz}. The
evolution of the comoving rate, roughly the rate per galaxy, has a
strong and clear rise of one order of magnitude between $z = 0$ and
$z = 1$ and then a slow and eventually steepening decline at higher
redshift.  Taking into account the variation of the uncertainties
with redshift, the uncertainty on the DSNB due to that on the core
collapse rate is presently $\pm 40\%$~\cite{Horiuchi:2008jz}. This
will decrease quickly with new data, so that the focus of DSNB
measurements will be on the neutrino emission
parameters~\cite{Yuksel:2005ae}.

\medskip
\noindent {\bf SN neutrino emission.}
While we have some information about neutrino emission from SN~1987A
and SN simulations, detecting the DSNB is necessary to measure the
average emission per SN.  It is typically assumed that the total
energy in neutrinos is $3 \times 10^{53}$~erg, that each flavor
carries $1/6$ of this, and that the spectra are quasi-thermal with
temperatures of several MeV.  But the total energy, its partition
among flavors, and spectral distributions and average energies may
be different or show more variation than expected. Uncertainties
include those due to the collapse
mechanism~\cite{1966ApJ...143..626C, 1976Ap&SS..41..287B,
1979PhLB...83..158M, 1985ApJ...295...14B, Blondin:2002sm,
2003ApJ...584..954A, Burrows:2005dv}, the effects of progenitor
mass, rotation, and magnetic fields~\cite{Yamada:1993tr,
Fryer:1999he, Thompson:2004if}, the neutron star equation of
state~\cite{Pons:1998mm, Horowitz:2004yf, Burrows:2004vq,
Langanke:2007ua}, and effects due to neutrino properties
(Sec.~\ref{sec:neutrinosignal}).

The neutrino emission is parameterized here with a Maxwell-Boltzmann
thermal spectrum, $\varphi(E_\nu) = E_{\rm tot} \, [E_\nu^2 / (2
T_\nu^3)] \, \exp(-E_\nu/T_\nu)$, where the total energy and
temperature (average emitted energy $\langle E_\nu \rangle  = 3
T_\nu$) are for $\bar{\nu}_e$ after neutrino flavor oscillations,
which occur in the SN and not en route.  Following
Sec.~\ref{sec:neutrinosignal}, the nominal expectation for
$\bar{\nu}_e$ might be $E_{\rm tot} = 0.5 \times 10^{53}$ erg and
$T_\nu = 4$ MeV, with large uncertainties.  Further, we do not know
if SN~1987A was a typical core collapse or if present SN simulations
are correct.  Measurements of the DSNB are needed to help decide.

The DSNB signal may be larger than assumed here, due to unusual core
collapse outcomes that could be disproportionately important due to
their larger-than-average neutrino emission. The most interesting
possibility is prompt black hole formation, as this is expected to
have a nonzero rate even in standard scenarios \cite{Heger:2002by,
O'Connor:2010tk}, and present constraints allow even larger rates
\cite{Beacom:2000qy, Kochanek:2008mp, Smartt:2008zd, Lien:2010yb,
Horiuchi:2011zz}. Even though the neutrino emission can be cut off,
it is expected to be enhanced before that, such that the
time-integrated total and average neutrino energies can be larger
than usual \cite{Sumiyoshi:2007aa, Sumiyoshi:2008zw, Fischer:2008rh,
Nakazato:2008vj}.  Other possibilities include emission from the
hot, magnetized corona of a proto-neutron star or accretion
disk~\cite{RamirezRuiz:2005cf} or from fallback~\cite{Fryer:2007cf}.
The DSNB will thus be especially valuable for probing outcomes that
may not occur for a Milky Way core collapse and the corresponding
extreme physical conditions in such collapses~\cite{Lien:2010yb,
Keehn:2010pn}.

\medskip
\noindent{\bf Detector capabilities.}
The detection channel in LENA and Super-Kamiokande is inverse beta
decay, $\bar{\nu}_e + p \rightarrow e^+ + n$, while other DSNB
neutrino interactions have smaller detectable
rates~\cite{Lunardini:2008xd}. The positron kinetic energy is close
to that of the neutrino, $T_e \simeq E_\nu - 1.8$ MeV, with a nearly
isotropic distribution.  The low-energy neutron will thermalize and
then register its presence by radiative capture. The time and space
coincidence between positron and neutron suppresses detector
backgrounds.

LENA has $2.9 \times 10^{33}$ free protons in 44~kt of scintillator
(Super-Kamiokande has $1.5 \times 10^{33}$ in 22.5~kt of water). The
cross section is $\sigma(E_\nu) \simeq 9.42 \times 10^{-44} {\rm\
cm}^2 \, (E_\nu/{\rm MeV} - 1.3)^2$ at lowest order; we use the
corrections to the cross section and kinematics from
Refs.~\cite{Vogel:1999zy, Strumia:2003zx}. In LENA, the visible
positron energy is its kinetic energy plus the annihilation energy
with an electron, $E_{\rm vis} = T_e + 2 m_e c^2$ (Super-Kamiokande
defines visible energies via the positron total energy, $E_{\rm vis}
= T_e + m_e c^2$). The effects of energy resolution on the DSNB
spectrum are negligible.

\begin{figure}[!t]
\includegraphics[width=0.8\columnwidth]{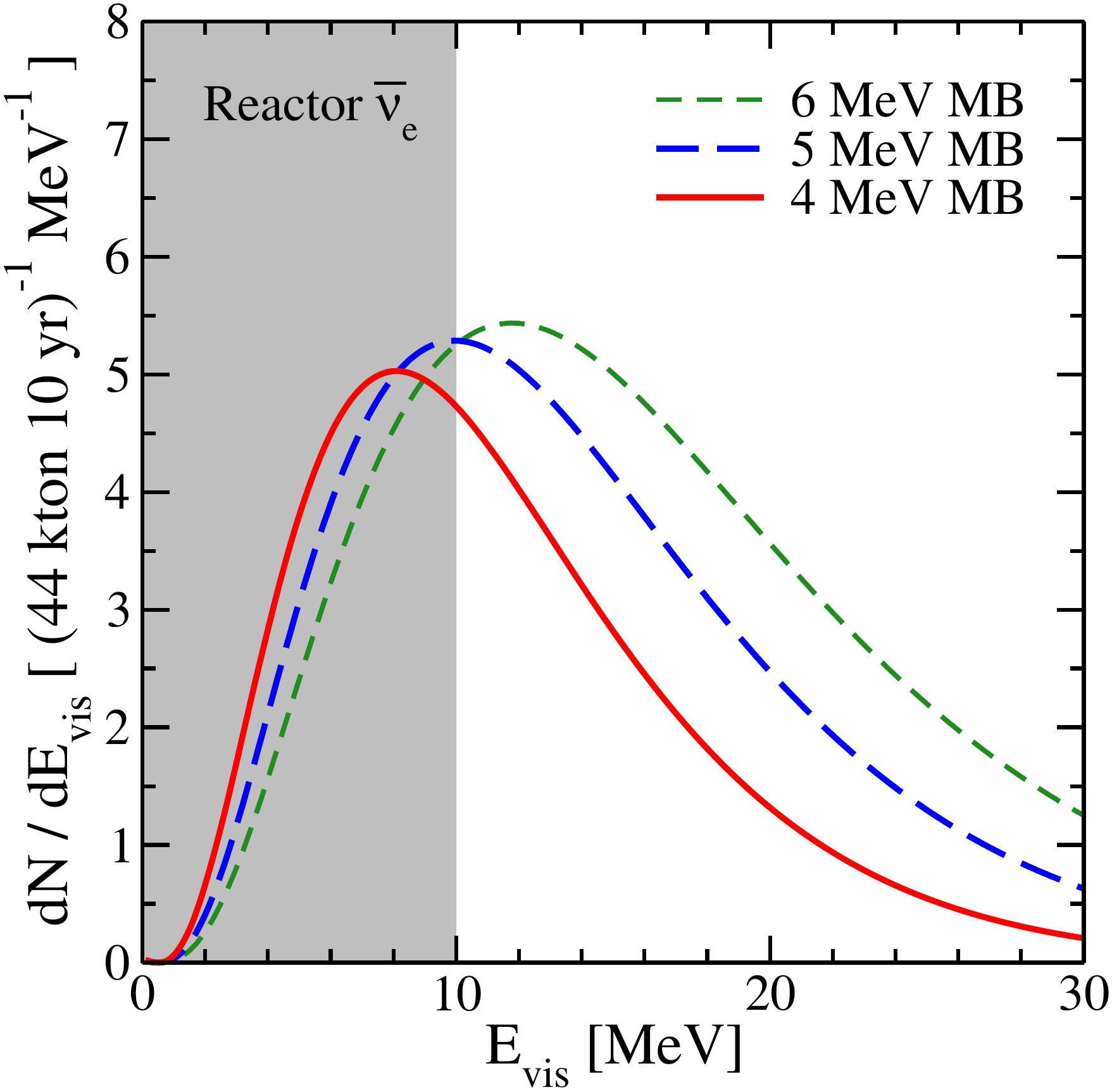}
\caption{DSNB signal spectra in LENA, with labeled lines indicating
assumed $\bar{\nu}_e$ emission spectra, which are of the
Maxwell-Boltzmann form with the indicated temperatures and which
have a total energy of $0.5 \times 10^{52}$ erg.  Below 10 MeV, the
reactor neutrino background (spectrum not shown) is overwhelming.
At higher energies, atmospheric neutrino backgrounds should be small
and controllable.} \label{fig:DSNB}
\end{figure}

Figure~\ref{fig:DSNB} shows the expected DSNB signal spectrum in
LENA, following the above details and assuming perfect detection
efficiency.  A range of SN $\bar{\nu}_e$ emission spectra,
parameterized by changes in $T_\nu$, are shown.  There are also
uncertainties due to $E_{\rm tot}$ and the assumed spectrum shape.
(Astrophysical uncertainties---those due to the SN rate alone---are
neglected because they are small and quickly decreasing.)  The
minimal allowed case is close to $E_{\rm tot} = 0.5 \times 10^{53}$
erg, $T = 4$ MeV, as this DSNB prediction is comparable to that
obtained from a direct non-parametric reconstruction of the high
energy SN~1987A data~\cite{Yuksel:2007mn} (see
also~Ref.~\cite{Fukugita:2002qw, Lunardini:2005jf} for previous
analyses of SN~1987A data applied to the DSNB).  Maximal allowed
cases (not shown), based on just the 2003 Super-Kamiokande limit,
are close to $E_{\rm tot} \sim 1 \times 10^{53}$ erg, $T_\nu = 6$
MeV, or $E_{\rm tot} \sim 2 \times 10^{53}$ erg, $T_\nu = 4$
MeV~\cite{Horiuchi:2008jz}.

The $\bar{\nu}_e$ emission parameters can thus be directly measured
from the DSNB spectrum if the atmospheric neutrino backgrounds can
be controlled.  The spectra in Fig.~\ref{fig:DSNB} contain 70, 55,
and 35 events in the range 10--30 MeV for ten years of LENA running,
corresponding to a statistical uncertainty of 12--17\%.  Data from
Super-Kamiokande will also help, especially if it begins running
with gadolinium soon.  The spectrum shape will help break the
degeneracy between $E_{\rm tot}$ and $T_\nu$.  In effect, $E_{\rm
tot}$ will be probed best by the lower energy data, where the
$T_\nu$ dependence is weakest, and $T_\nu$ will be probed best by
the higher energy data, where the falloff is exponential.  Precise
data will also test new physics
scenarios~\cite{Ando:2003ie,Fogli:2004gy}.

Theory and data from a future Milky Way SN will be needed to relate
these $\bar{\nu}_e$ emission parameters with flavor mixing effects
included to other SN parameters such as the total energy emitted by
all flavors and the spectra before neutrino flavor mixing.

\subsubsection{LENA detector backgrounds}
\label{sec:DSNB-backgrounds}

As both positron and neutron emerging from an inverse beta decay are
observable in liquid scintillator, single-event backgrounds can be
suppressed very effectively. The only remaining backgrounds are due
to other sources of $\bar\nu_e$, namely reactor and atmospheric
$\bar\nu_e$ that are irreducible. They limit the DSNB detection
window to 10--30~MeV, the exact range depending on the detector site
\cite{wur09phd}. Other backgrounds are signals mimicking the fast
coincidence signal: Cosmogenic $\beta$n-emitting isotopes, fast
neutrons from the surrounding rocks, and NC interactions of
higher-energy atmospheric neutrinos. However, these backgrounds can
be identified, either by their production, location or pulse shape.
These reducible backgrounds, strategies for their rejection and the
accompanying loss in DSNB detection efficiency are outlined below.

The delayed coincidence signal of inverse beta decay enables LENA to
easily reject what is the predominant backgrounds for DSNB detection
in water Cherenkov detectors, i.e.\ low energy muons produced in CC
reactions of atmospheric neutrinos, solar neutrinos and spallation
products of cosmic muons.

\medskip
\noindent\textbf{Cosmogenic \boldmath{$\beta n$}-emitters.} are
unstable isotopes produced by cosmic muons crossing the detector.
They mimic the $\bar\nu_e$ coincidence by the prompt emission of the
electron followed by the emission of a neutron from an excited state
of the daughter nucleus. Fortunately, only $^9$Li
($\mathrm{T_{1/2}=178\,ms}$) has a large enough Q-value to add to
the background. If no cuts are applied, the $^9$Li rate is of the
same order of magnitude as the DSNB signal. Due to its short
half-life, $^9$Li is easily associated to its parent muon: It is
sufficient to veto a cylinder with 2\,m radius around each muon
track for 1\,s ($\sim$5\,$T_{1/2}$), to reduce the residual $^9$Li
rate to about 2\%, while the introduced dead time corresponds only
to $\sim 0.1\%$ of the total measurement time \cite{wur09phd}.

\medskip
\noindent\textbf{Fast neutrons.} are produced in the surrounding rock
by cosmic muons that pass the detector undetected. There is a chance
for these neutrons to propagate into the target volume. The
coincidence is mimicked by a prompt signal due to elastic scattering
of protons, while the delayed signal is caused by the capture of the
stopped neutron. The fast neutron background rate was analyzed by
Monte-Carlo simulations. As most neutrons will stop at the verge of
the scintillator, a fiducial volume cut greatly reduces the rate.
For 10~m fiducial radius, $\sim$0.2 fast neutron events per year are
expected \cite{Moellenberg:2009}. However, proton recoils of
neutrons feature a different typical pulse shape than positron
signals. This allows for an alternative approach to distinguish
neutron events more effectively, reducing the fast neutron
background to 0.12 events per year in the nominal  fiducial volume
of 12~m radius.

\medskip
\noindent\textbf{NC reactions of atmospheric neutrinos.} might prove
to be the most dangerous background: Besides the intrinsic
background of atmospheric $\bar\nu_e$'s, atmospheric neutrinos at
higher energies knock out neutrons from {$^{12}$C} in the
scintillator. Neutron scattering off protons or particles emitted in
the de-excitation of the remaining nucleus cause a prompt signal,
while the neutron is later captured, mimicking the signal
coincidence. MC simulations point towards a background rate about
10--20 times higher than the expected DSNB signal
\cite{Efremko:2009}. Several strategies have been devised to cope
with this background: A possible way is to search for the delayed
$\beta^+$ decay of the residual $^{11}$C that remains after the
neutron knock-out. If the $^{11}$C nucleus is created in its ground
state, this is a very effective strategy, reducing the background by
a factor of 2. However, if an excited $^{11}$C state is created, it
will mostly de-excite via proton, neutron and alpha emission. In
this case, the only way of discrimination is a pulse shape analysis
of the prompt signal. This rejection profits not only from the intrinsic differences in light emission of protons and $\alpha$-particles compared to the $\bar\nu_e$-indudced positrons but e.g. also from the spatially extended energy deposition of high-energetic $\gamma$-quanta. The discrimination power as well as the remaining DSNB detection efficiency is currently evaluated in MC
studies. Preliminary results indicate that in spite of a painful
loss in efficiency, a signal-to-background ratio greater than unity
can be obtained \cite{Moellenberg:2009}. However, further MC studies
as well as laboratory measurements on proton quenching in liquid
scintillator are needed to quantify this result.

While the latter background is absent in water Cherenkov detectors,
their inability to detect the delayed neutron signal makes them
vulnerable to solar neutrinos, the decay of invisible muons and all
kinds of spallation products. Nevertheless, much improved
Super-Kamiokande sensitivity to DSNB $\bar{\nu}_e$ is expected by
the time LENA comes into operation. Their 2003 limit is already
strong and will improve with further data~\cite{Malek:2002ns}. If
gadolinium is added, Super-Kamiokande will reject detector
backgrounds above 10~MeV and will cleanly collect a few DSNB signal
events per year~\cite{Beacom:2003nk}.

\subsubsection{Summary}

The DSNB is a very promising astrophysical neutrino source, with at
most a factor of a few improvement in flux sensitivity required for
a first detection. This will directly probe the neutrino emission
per core collapse via the measured $\bar{\nu}_e$ spectrum above
about 10\,MeV. This spectrum averages over all core-collapse
outcomes, including some which may be relatively rare but which may
be of disproportionate importance due to larger-than-usual neutrino
emission.  The main advantages of LENA are its large size, native
ability to detect neutrons to tag $\bar{\nu}_e + p \rightarrow e^+ +
n$ events, and low detector backgrounds and consequent low energy
threshold.  LENA may make a first detection of the DSNB and would
significantly increase statistics over Super-Kamiokande alone,
leading to more decisive probes of the average neutrino emission per
core collapse, a key comparison point for SN models and a Milky Way
SN.

%% file: solar.tex
%
%

%

\noindent Solar neutrino research is a mature field that has
accumulated a long series of outstanding achievements. Originally
conceived as a powerful tool to investigate the Sun's deep interior,
solar neutrinos provided the first indication for neutrino flavor
oscillations and thus contributed in crucial ways to discover and
analyze this profound phenomenon. Solar experiments provide 
sensitivity to $\Delta m_{12}^{2}$ and especially $\theta_{12}$, with the
fascinating prospect of a possible positive indication regarding the
value of the subleading $\theta_{13}$ angle, connecting solar and
atmospheric sectors of the Pontecorvo-Maki-Nakagawa-Sakata (PMNS)
neutrino mixing matrix. These achievements are the starting point of
the LENA solar neutrino program because high-statistics measurements
will resolve energy spectra and possible time variations in
unprecedented detail.


\subsubsection{Introduction}
\noindent\textbf{The Standard Solar Model (SSM\nomenclature{SSM}{Standard Solar Model}).} The effort to develop a model able to reproduce fairly accurately the solar physical characteristics, as well as the spectra and fluxes of the several produced neutrino components, was led for more than forty years by the late John Bahcall; this effort culminated in the synthesis of the so called Standard Solar Model (SSM) \cite{Bahcall:2004fg}, which represents a true triumph of the physics of 20$^{\rm th}$ century, leading to extraordinary agreements between predictions and observables.

However, over the last years such previous excellent agreement has been seriously compromised by the downward revision of the solar surface heavy-element content from $Z=0.0229$ \cite{Grevesse:1998bj} to $Z = 0.0165$ \cite{Asplund:2004eu}, leading to a severe discrepancy between the SSM and the helioseismology results. Resolution of this puzzle would imply either to revise the physical inputs of SSM or to modify the core abundances. 

In 2009 a complete revision of the solar photospheric abundances for nearly all elements have been done \cite{Asplund:2009fu}. This revision includes a new three dimensional hydrodynamical solar atmosphere model with an improved radiative transfer and opacities. The obtained results give a solar metallicity $Z = 0.0178$. In \cite{Serenelli:2009yc} the three different sets of solar abundances: GS98 \cite{Grevesse:1998bj}, AGS05 \cite{Asplund:2004eu} and AGS09 \cite{Asplund:2009fu} have been used originating two different, low metallicity or high metallicity, versions of the solar model.

\medskip
\noindent\textbf{Neutrino oscillations and the MSW effect.} Solar neutrino oscillations are governed by $\Delta m_{12}^{2}$ and $\theta_{12}$ of the PMNS mixing matrix. At low energies, $\nu_e$ survival probabilities are described well by vacuum oscillations. However, at energies above $\sim$1\,MeV, matter effects first pointed out by Mikheyev, Smirnov and Wolfenstein (MSW) \cite{Mikheev:1986gs, Wolfenstein:1977ue} enhance the conversion $\nu_e\rightarrow\nu_{\mu,\tau}$, leading to a further suppression of the $\nu_e$ rate observed in terrestrial detectors. By now, this MSW-LMA\nomenclature{LMA}{Large Mixing Angle oscillation scenario} oscillation scenario is well confirmed by solar neutrino experiments for vacuum- and matter-dominated regimes. However, the vacuum-matter transition region from 1 to $\sim$5\,MeV remains to be explored and might hold evidence for non-standard neutrino physics.


\subsubsection{Experimental status}

For almost 40 years, solar neutrino detectors have accumulated a large amount of data. Beyond the confirmation of thermonuclear fusion as the solar energy source, the comparison of the experimental results to the accurate predictions of the SSM on solar neutrino flux and spectrum led to the establishment of the LMA-MSW oscillation scenario in the solar sector.


In the 70's and the 80's the Chlorine radiochemical experiment at Homestake \cite{Cleveland:1998nv} and the Cherenkov detector Kamiokande \cite{Hirata:1989zj} played the fundamental role to  establish on solid rock basis what became known as the Solar Neutrino Problem (SNP\nomenclature{SNP}{Solar Neutrino Problem}), i.e. the persisting discrepancy between the measured and predicted solar neutrino flux.

In the early 90's the scene was dominated by the second generation of radiochemical experiments based on Gallium, GALLEX/GNO\nomenclature{GNO}{Gallium Neutrino Observatory} and SAGE\nomenclature{SAGE}{Sowjet-American Gallium Experiment} \cite{Hampel:1998xg,Altmann:2005ix, Abdurashitov:1999bv}, which not only reinforced the physics case of Homestake and Kamiokande, thus further strengthening the SNP, but also marked the first detection ever of the overwhelming flux of the pp\nomenclature{pp}{Proton-Proton fusion} neutrinos. This milestone result represented the first direct proof of the nuclear burning mechanism as the actual stellar energy generating engine.

Later, the final assault to the SNP was launched by the three real time experiments Super-Kamiokande\nomenclature{Kamiokande}{KAMIOKA Nucleon Decay Experiment}, SNO\nomenclature{SNO}{Sudbury Neutrino Observatory} (whose measurements were decisive to solve the SNP) and Borexino\nomenclature{Borexino}{derived from BORon EXperiment}, which were complemented in their effort by the reactor neutrino experiment KamLAND\nomenclature{KamLAND}{KAMioka Liquid-scintillator Anti-Neutrino Detector}:

\medskip
\noindent\textbf{Super-Kamiokande.} Designed to detect the Cherenkov light emerging from the elastic scattering of the incoming neutrinos off the electrons of the water acting as detection medium, Super-Kamiokande started its operation in 1996 at the Kamioka mine in Japan. It is a gigantic detector, with its 50 ktons of water viewed by more than 10\,000 20-inch phototubes.  

Over 15 years of measurement, the experiment returned consistent data on the $^{8}$B-$\nu$ flux and spectrum \cite{Hosaka:2005um, Cravens:2008zn} above a detection threshold of 5\,MeV. For the latest phase III of data taking, a flux of (2.32$\pm$0.04$_{stat}$$\pm$0.05$_{syst}$)$\times$10$^{6}$\,cm$^{-2}$s$^{-1}$ has been determined from neutrino-electron scattering \cite{Abe:2010hy}. However, the low-energy upturn in the $^{8}$B $\nu_e$ spectrum that is predicted by the MSW-LMA solution has not been observed. The new phase IV of data taking will feature a lower threshold of $\sim$4\,MeV and so will further explore the vacuum-matter transition region.   


 
 

\medskip
\noindent\textbf{SNO.} Located in a mine in Ontario, SNO exploited the same Cherenkov technique of Super-K, with the difference that the detection medium was heavy water. The neutral and charge current neutrino reactions on deuterium provided the experiment with a powerful tool enabling to measure concurrently the total $^{8}$B all flavor neutrino flux (neutral current) and the electron neutrino only flux (charge current). The SNO result verified unambiguously that the SNP was due to the conversion between different neutrino flavors, as implied by the MSW paradigm. 

The experiment provided also an accurate measure of the total $^{8}$B flux. 
SNO progressed through three steps (pure heavy water, salt and $^{3}$He counters) that returned consistent result. The most recent low energy threshold (LETA\nomenclature{LETA}{Low Energy Threshold Analysis of SNO}) joint analysis of the phase I and II data results in a $^{8}$B flux of $(5.140^{+0.160}_{-0.158}(stat)^{+0.132}_{-0.117}(syst))$$\cdot$10$^{6}$\,cm$^{-2}$s$^{-1}$ \cite{Aharmim:2009gd}, in good agreement with both high- and low-metallicity predictions of the SSM. The detector is now empty, ready to be filled with liquid scintillator for the future SNO+ data taking phase.


\medskip
\noindent\textbf{Borexino.} While the two Cherenkov experiments focused their investigations to the high energy portion of the $^{8}$B spectrum, Borexino \cite{Arpesella:2008mt} at Gran Sasso lowered for the first time the research range of a real time solar neutrino experiment down to few hundreds keV, based on the much larger light output of the scintillation technique. This allowed to test the LMA vacuum oscillations below 1\,MeV, in contrast to the matter-dominated regime probed by Super-Kamiokande and SNO.

Achieving the ultra-low radioactive background conditions required for the detection of the $^{7}$Be-$\nu$s (0.862 MeV) poses an enormous technological challenge. However, the necessary techniques for purification of the scintillator and the selection and assembly of low-background materials were developed in an extensive R\&D program, culminating in the clear detection of the $^{7}$Be-$\nu$ recoil electrons. The corresponding evaluation of the $\nu_e$ survival probability was in good agreement with MSW-LMA and SSM predictions, obtaining a value of (5.18$\pm$0.51)$\times$10$^9$\,cm$^{-2}$s$^{-1}$ for the total $^{7}$Be flux.


As further proof of the powerful flexibility of the scintillation technique, Borexino performed a measurement of the $^{8}$B-$\nu$ flux above 3\,MeV, achieving the currently lowest threshold in a spectral measurement. The result corresponds to a flux of (2.4$\pm$0.4$_{stat}$$\pm$0.1$_{syst}$)$\times$10$^{6}$\,cm$^{-2}$s$^{-1}$ \cite{Bellini:2008mr}, in excellent agreement with the Super-Kamiokande measurements.

\medskip
\noindent\textbf{KamLAND.} The current understanding of the experimental solar neutrino results in term of the neutrino oscillation paradigm heavily relies also on the outcomes of the KamLAND \cite{Abe:2008ee} reactor neutrino experiment (Sec.\,\ref{subsec::reactor}). By comparing the theoretically expected antineutrino spectrum from a number of nuclear power plants at different distances to the experimental site, with the measured spectrum, KamLAND was able to detect in the latter the clear imprinting of the oscillation effect, thus independently ruling out the other possible explanations for the solar neutrino deficit. 

\begin{figure} [t!]
\centering
\includegraphics[width=0.44\textwidth]{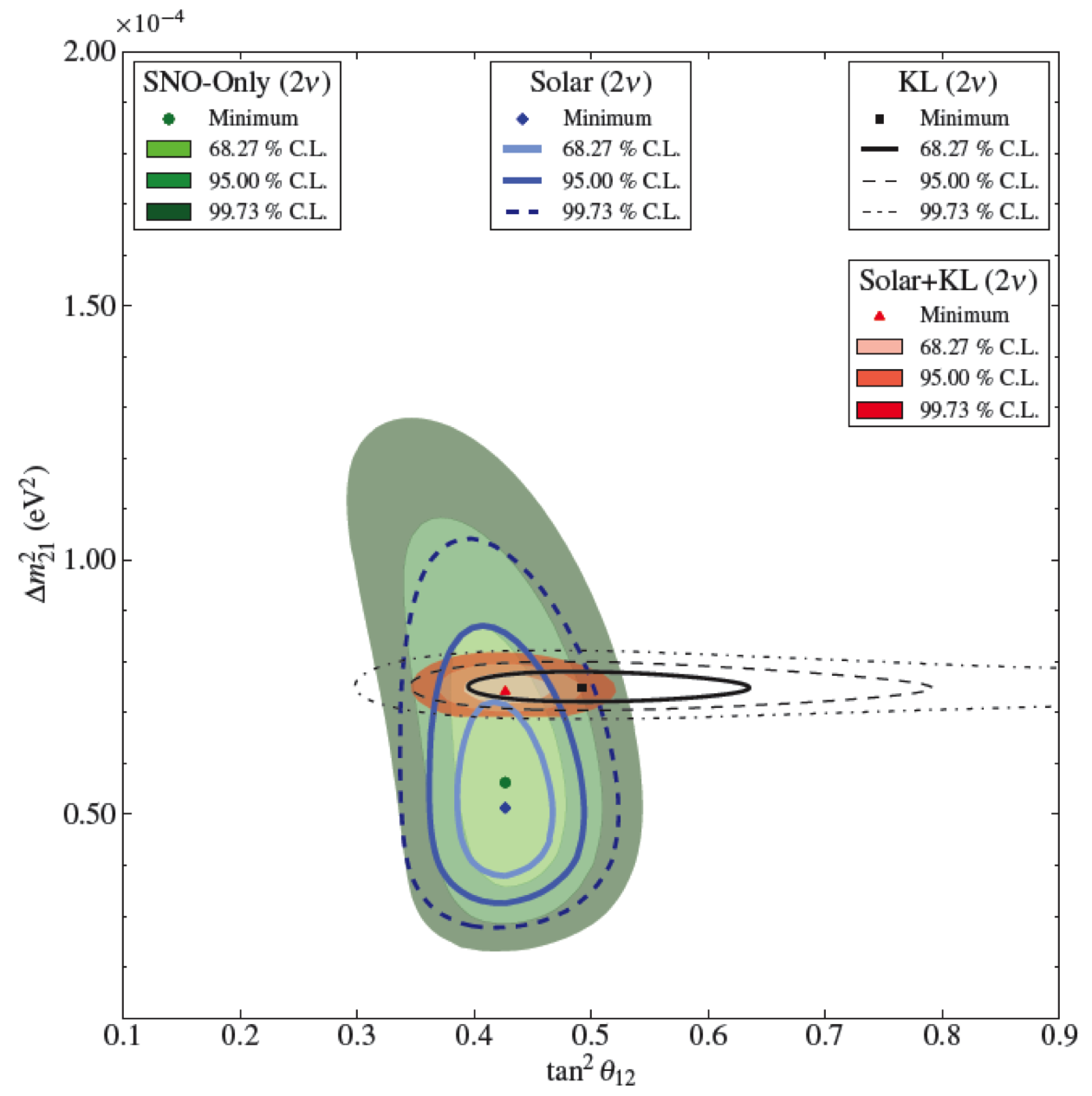}
\caption{Global solar and KamLAND data oscillation analysis \cite{Aharmim:2011vm}.}
\label{fig:osc-analysis}
\end{figure}

The collective analysis of the data from all the solar neutrino experiments performed so far, plus those coming from KamLAND, puts stringent limits on the  $\Delta m^{2}_{12}$  and $\theta_{12}$ oscillations parameters. Fig.\,\ref{fig:osc-analysis}, from reference \cite{Abe:2008ee}, shows the allowed region in the parameters space, stemming from a two flavor oscillation analysis. In a first approximation, the strongest constraint on $\Delta m^{2}_{12}$ comes from KamLAND, while the limit on the mixing angle derives from the solar data.

\subsubsection{LENA observables and capabilities}


Despite the impressive successes accumulated in this field in the past, still additional and important insights can be expected from the detection of solar neutrinos. With a first measurement of pep\nomenclature{pep}{Proton-Electron-Proton fusion} and CNO\nomenclature{CNO}{Carbon-Nitrogen-Oxygen fusion cycle} neutrino fluxes, Borexino and the upcoming SNO+ experiment will probe oscillations in the MSW transition region and solar metallicity, respectively. Even a direct measurement of the fundamental pp-$\nu$ might be within reach of Borexino. 

However, the high-statistics data collected by a gigantic scintillation detector like LENA would allow a precise determination of SSM neutrino rates and MSW-LMA oscillation probabilities. The benchmark experience to be taken as reference is that of Borexino, to date the only liquid scintillator experiment which has successfully accomplished the detection of low-energy solar neutrinos. While the expected performances of a large 50\,kt detector will not equal those of the smaller 0.3 ktons Borexino detector, especially because of the likely inferior photoelectron yield, the extremely high neutrino event rate resulting from the huge target volume will enable not only a detailed study of the features of the neutrino spectrum, but will allow also a thorough investigation of even small time modulations possibly embedded in the recorded flux.

Again building on the Borexino experience, we may anticipate the need of a smaller fiducial volume, compared to other measurements, in order to cope successfully with the external gamma rays background, mainly from the photomultiplier tubes. MC\nomenclature{MC}{Monte Carlo simulation} simulations point to a fiducial mass of $\sim$30\,kt for pep, CNO and low-energy $^{8}$B-$\nu$ detection, while the fiducial mass adopted for $^{7}$Be-$\nu$s and high-energy ($E>5$\,MeV) $^{8}$B-$\nu$s will be 35\,kt or more. 

Table \ref{tab:ratesinLENA} lists the expected rates in 30\,kt for the neutrinos emitted in the pp chain and the CNO cycle, using the most recent solar model predictions. This evaluation refers to a detection threshold set at about 250\,keV, a lower threshold being severely hindered by the intrinsic abundance of $^{14}$C in the scintillator components.

\begin{table} 
\begin{tabular}{lcccr}
\toprule
Source 		& Channel & EW [MeV] & $m_\textrm{fid}$ [kt] & Rate [cpd]  \\
\colrule
pp			& $\nu e\to e\nu$	& $>$0.25		& 30		& 40	 \\		
pep			& 					& 0.8$-$1.4		& 30		& 2.8$\times$10$^2$ \\	
$^{7}$Be		&					& $>$0.25		& 35		& 1.0$\times$10$^4$ \\
$^{8}$B    	& 					& $>$2.8			& 35		& 79 \\
CNO			& 					& 0.8$-$1.4		& 30		& 1.9$\times$10$^2$ \\
\colrule
$^{8}$B 		& $^{13}$C			& $>$2.2 		& 35		& 2.4 \\  	
\botrule	
\end{tabular}
\caption{Expected solar neutrino rates in LENA. The estimates are derived from the existing Borexino analyses \cite{Arpesella:2008mt,Bellini:2008mr} as well as expectation values for the respective energy windows (EW) for observation \cite{wur09phd,Dangelo:2006,Ianni:2005ki}. The quoted fiducial masses ($m_\textrm{fid}$) in LAB are based on a Monte Carlo simulation of the external $\gamma$-ray background in LENA \cite{moel-priv}.}
\label{tab:ratesinLENA}
\end{table}


\medskip
\noindent\textbf{Spectral measurements.} Based on \cite{wur09phd} and a fiducial mass of 30\,kt, about 40 pp neutrino-induced electron backscattering events per day are expected above the threshold: despite the non-negligible rate, it is hard to anticipate the capability to distinguish them from the huge tail of the $^{14}$C, especially taking into account the limited resolution that can be expected in this energy range. On the other hand, $^{7}$Be detection will occur with enormous statistics, the prediction amounting to almost $10^{4}$ $^{7}$Be recoils events per day in 35\,kt. In the assumption that  background levels similar to those of Borexino will be achieved, such a high statistics will permit a measurement of the $^{7}$Be flux with accuracy unprecedented in neutrino physics. In particular, accurate search for temporal variations in the detected rate will be possible (see below).

 
Roughly in the energy range between 1 and 2 MeV, detection of CNO and pep solar neutrinos can occur. A major background to this effort will be formed by cosmogenic $^{11}$C beta decays, induced by the muon-induced knock-out of neutrons from $^{12}$C. The $^{11}$C production rate is mainly a function of the rock overburden shielding the detector. If LENA will be operated at the intended depth of 4000\,mwe (meters water equivalent), the ratio of the CNO/pep-$\nu$ signal to $^{11}$C background rate would be 1:8, a factor 3 better than for example at the depth of Gran Sasso. About 500 CNO-$\nu$s per day will provide valuable information on solar metallicity, especially if the contributions from the individual subfluxes can be disentangled. In Borexino, the $\beta$-decays of $^{210}$Bi originating from $^{210}$Pb dissolved in the scintillator prove to be the most severe background after $^{11}$C subtraction. As $^{210}$Pb is rather long-lived, a special effort will be needed to keep the initial contamination of the scintillator as low as possible. Furthermore, the measurement of the pep neutrino flux could be exploited for a precision test of the $\nu_e$ survival probability in the MSW-LMA transition region. The transition onset can be probed via the low energy portion of the $^{8}$B neutrinos spectrum, through the accurate detection of the expected spectral ``upturn". The detection threshold for $^{8}$B-$\nu$s might be lowered even further than in Borexino, as the background due to the penetrating 2.6\,MeV $\gamma$-rays from external {$^{208}$Tl} decays can be avoided by adjusting the fiducial volume. 
 
Finally, the charged-current reaction on $^{13}$C should be mentioned. The channel is only accessible to $\nu_e$s and virtually background-free due to the delayed coincidence of the $^{13}$N back decay. The reaction threshold is 2.2\,MeV, allowing for a precise measurement of the $\nu_e$ survival probability of $^8$B-$\nu$s in the MSW transition region. About 8$\times$10$^2$ counts per year are expected.

\medskip
\noindent\textbf{Search for time-variations.} The enormous amount of solar neutrino events collected in LENA will offer the possibility to search for temporal modulations in the neutrino flux arriving at Earth, especially regarding the $^{7}$Be signal. Various processes that might cause such processes have been suggested: Apart from the annual modulation induced by the eccentricity of the terrestrial orbit, changes in the survival probability of solar $\nu_e$s might be induced by fluctuations of the solar matter density and magnetic field \cite{Chauhan:2005pn}, or by the transit of terrestrial matter \cite{Bahcall:1997jc} before reaching the detector. Even solar neutrino production rates might vary in the course of the solar cycle of about 11 years \cite{kra90}, or might be subject to short-term variations correlated to the oscillation of the solar core temperature due to helioseismic waves \cite{Collaboration:2009qz}.

Currently, the best limits on periodical $\nu$ flux variations arise from the Super-Kamiokande and SNO experiments, excluding modulation amplitudes of more than 10\,\% in the $^{8}$B-$\nu$ signal \cite{Collaboration:2009qz,Yoo:2003rc}. The investigated range of modulation periods extends from the order of hours to years. However, due to the $\sim$10$^{4}$ $^{7}$Be-$\nu$ events per day available in LENA, the sensitivity to modulations at low amplitudes is expected to be far greater: The MC analyses performed in \cite{Wurm:2010mq} point to a 3$\sigma$ discovery potential for amplitudes as low as 0.5\,\%, covering a period range extending from tens of minutes to a hundred years or more. This will allow to probe the high-frequency regions associated to helioseismic g-modes, but also to test the temporal uniformity of solar fusion processes on long time scales.

%% file: idms.tex

%

\noindent Dark matter particles might be abundantly present in the
  Universe and able to annihilate (decay) efficiently into Standard Model
  particles, in particular neutrinos, in regions where they are highly
  concentrated. We consider these annihilations (decays) in the
  galactic halo and show how LENA could be used to set general limits
  on the dark matter annihilation cross section and on the dark matter
  lifetime.
  
\subsubsection{Introduction}

With the next generation of neutrino experiments we will enter the era
of precision measurements in neutrino physics.  These detectors, and
specifically LENA, thanks to their great capabilities, might also be
used to test some of the properties of the dark matter (DM\nomenclature{DM}{Dark Matter}).

DM is copiously present in the Universe, having been produced in its
very first instants. In the simplest case, the DM particles were in
equilibrium in the Early Universe thermal plasma, decoupling when
their interactions become too slow compared with the expansion of the
Universe.  After decoupling, a thermal distribution remains as a relic
which constitutes the DM we observe today.  For the simplest
assumption of thermal freeze-out, which holds in most of the models of
DM, the annihilation cross section required to reproduce the observed
amount of dark matter is given by $\langle \sigma_{\mathrm{A}} v
\rangle = 3$$\times$$10^{-26} \, {\rm cm}^3/{\rm s}$.  Subsequently,
within the framework of cold DM, structure forms hierarchically with 
DM collapsing first into small haloes and eventually giving rise to
larger ones, as galaxies and clusters of galaxies.  Large
concentrations of DM emerge, for example in the center of galaxies,
such that the DM particles could annihilate efficiently and produce
detectable fluxes of Standard Model (SM\nomenclature{SM}{the Standard Model of particle physics}) particles, such as photons,
neutrinos, positrons and antiprotons.  Such particles could be produced
also if DM is not stable but decays with a lifetime longer than the
age of the Universe in order to be present today.  Among these particles, neutrinos are the least
detectable ones.  Therefore, if we assume that the only SM products
from the DM annihilations (decays) are neutrinos, a limit on their
flux, conservatively and in a model-independent way, sets an upper
(lower) bound on the DM annihilation cross section (lifetime).  This
is the most conservative assumption from the detection point of view,
that is, the worst possible case (see \cite{PalomaresRuiz:2007ry} for a discussion on the implications of other decay modes for DM decays).  Any other channel (into at least
one SM particle) would produce photons and hence would give rise to a
much more stringent limit.  The bounds so obtained are on the total
annihilation cross section (lifetime) of the DM particle and not only
on its partial annihilation cross section (lifetime) due to the
annihilation (decay) channel into neutrinos.

In this section, and following and reviewing the approach of \cite{Beacom:2006tt, Yuksel:2007ac, PalomaresRuiz:2007eu,PalomaresRuiz:2007ry}, we consider this case and evaluate the
potential neutrino flux from DM annihilation (decay) in the whole
Milky Way, which we compare with the relevant backgrounds for
detection.  In such a way, we obtain an estimate of the sensitivity on 
the DM annihilation cross section (lifetime) by LENA.

\subsubsection{Searching neutrinos from MeV DM}

For energies below $\sim$200\,MeV, information on the direction of the
incoming neutrino is very poor if the detection is via interactions
with nucleons, as is the case in LENA.  Thus, we take the flux averaged
over the entire galaxy.  The differential neutrino or antineutrino
flux per flavor from DM annihilation (decay) averaged over the whole
Milky Way is given by 
\begin{equation}
\frac{ d \Phi}{d E_\nu} = {\cal P}_k (E_\nu, m_\chi) \, R_\odot \,
\rho_0^k \, {\cal J}_{avg, k}\, ,
\label{dkflux}
\end{equation}
where $m_\chi$ is the DM mass, $\rho_0 =  0.3$\,GeV\,cm$^{-3}$ is a
normalizing DM density at $R_\odot=8.5$\,kpc (the distance from the Sun
to the galactic center), and ${\cal   J}_{avg, k}$ is the average over
the whole galaxy of the line of sight integration of the DM density
(for decays, $k=1$) or of its square (for annihilations, $k=2$), which
is given by 
\begin{eqnarray}
{\cal J}_{avg, k}  &=& \frac{1}{2 \, R_\odot \, \rho_0^k}  \,
\int_{-1}^1 \, \int_{0}^{l_{\rm max}} \,  \rho (r)^k \, dl \, d(\cos
\psi'),\nonumber\\
r &=& \sqrt{R^2_\odot -  2 l R_\odot \cos \psi' + l^2}, \nonumber\\
 l_{\rm max} &=& \sqrt{(R_{\rm vir}^2 - \sin^2 \psi R^2_\odot)} + R_\odot \cos \psi,
\label{Javg}
\end{eqnarray}
$R_{\rm vir}$ being the halo virial radius.  Commonly
used profiles \cite{Moore:1999gc, Navarro:1995iw, Kravtsov:1997dp}
tend to agree at large scales, although they may differ significantly
in the inner part of galaxies.  The overall normalization of the flux
is affected by the value of ${\cal J}_{avg, k}$, scaling as
$\rho^k$. For instance, for different profiles \cite{Moore:1999gc,
  Navarro:1995iw, Kravtsov:1997dp}, astrophysical uncertainties can 
induce up to a factor of $\sim$100\,($\sim$6) for
annihilations \cite{Yuksel:2007ac, PalomaresRuiz:2007eu}
(decays \cite{PalomaresRuiz:2007ry}).  For concreteness, in what
follows we present results using the Navarro, Frenk and White (NFW)
profile \cite{Navarro:1995iw}, with ${\cal J}_{avg, 1} =
2$ \cite{PalomaresRuiz:2007ry} and ${\cal J}_{avg, 2} =
5$ \cite{Yuksel:2007ac}.

All the dependence on the particle physics model is embedded in ${\cal
  P}_k$ as
\begin{eqnarray}
{\cal P}_1 & = & \frac{1}{3} \, \frac{dN_1}{dE_\nu} \, \frac{1}{m_\chi
  \tau_\chi} \hspace{5mm} {\rm for \, \, decays \, \, and} \nonumber\\
{\cal P}_2 & = & \frac{1}{3} \, \frac{dN_2}{dE_\nu} \, \frac{\langle
  \sigma_{\rm A} v \rangle}{2 \, m_\chi^2} \hspace{5mm} {\rm for \, \,
  annihilations}, \nonumber
\end{eqnarray}
where the neutrino or antineutrino spectrum per flavor is given by
\begin{eqnarray}
\frac{dN_1}{dE_\nu} & = & \delta (E_\nu - \frac{m_\chi}{2})
     \hspace{5mm} {\rm for \, \, decays \, \, and} \hspace{5mm} \nonumber\\
\frac{dN_2}{dE_\nu} & = & \delta (E_\nu - m_\chi) \hspace{5mm} {\rm
  for \, \, annihilations}.\nonumber
\end{eqnarray}
The factor of 1/3 arises from the assumption that the annihilation or
decay branching ratio is the same for the three neutrino flavors.
This is not a very restrictive assumption, for even when only one
flavor is produced, a flux of neutrinos in all flavors is generated by
the averaged neutrino oscillations between the source and the detector.

The neutrinos produced in DM annihilations (decays) travel to the
Earth where they can be revealed in present and future neutrino 
detectors.  Importantly, the signal is monoenergetic, allowing to 
distinguish it from backgrounds continuous in energy.  The number of
signal neutrino events is given by
\begin{equation}
{\cal N} \simeq \sigma_{\rm det} \ \phi \ N_{\rm target} \ t \,
\epsilon,
\label{nevents}
\end{equation}
where the detection cross section $\sigma_{\rm det}$ needs to be
evaluated at $E_\nu = m_\chi$ ($E_\nu = m_\chi/2$) for annihilations
(decays), the total flux of neutrinos or antineutrinos is given by
$\phi$, $N_{\rm target}$ indicates the number of target particles in
the detector, $t$ is the total time-exposure, and $\epsilon$ is the
detector efficiency for this type of signal.  From
Eq.\,(\ref{nevents}), and assuming the annihilation cross section
required to reproduce the observed amount of DM, $\langle
\sigma_{\mathrm{A}} v \rangle = 3 \times 10^{-26} \, {\rm cm}^3/{\rm
  s}$, (or a lifetime $\tau_\chi \sim 10^{24}$\,s) we expect a few events
for an exposure of 1\,Mt$\cdot$yr, requiring large detectors such
as LENA.

\subsubsection{MeV Dark Matter search in LENA}

It has been shown \cite{PalomaresRuiz:2007eu, PalomaresRuiz:2007ry}, particularly for Super-Kamio\-ka\-nde, that already present large neutrino detectors severely constrain the DM properties, namely annihilation cross section and lifetime and, depending on the DM profile assumed, exclude an important part of the parameter space (see Fig.\,1 in \cite{PalomaresRuiz:2007eu}, Fig.\,1 in \cite{PalomaresRuiz:2007ry} and also Fig.\,1 in 
\cite{PalomaresRuiz:2008ef}).

Here we describe the analysis performed in \cite{PalomaresRuiz:2007eu} for the physics reach of the LENA
detector.  The excellent background rejection allows for a significant
improvement on present bounds.  At
the energies of interest, few tens of MeV, the inverse beta-decay
cross section ($\bar{\nu}_e p\rightarrow n e^+$) is by two orders
of magnitude larger that the $\nu-e$ elastic scattering cross section.
The advantage of the LENA detector is the excellent energy resolution
and the fact that the inverse beta-decay reaction can be clearly
tagged by the signal in coincidence of the positron annihilation
followed by a delayed 2.2\,MeV photon, which is emitted when the
neutron is captured by a free proton.  Thus, the only relevant
backgrounds for these events come from reactor, atmospheric and
diffuse supernova $\bar\nu_e$ interacting with free protons in
the detector: The flux of reactor $\bar\nu_e$'s below $\sim 10$\,MeV
represents a background by orders of magnitude higher than the
expected neutrino flux from DM annihilations (decays) \cite{Wurm:2007cy}.  
The Diffuse Supernova Neutrino Background (DSNB), although not yet detected, might potentially represent a background in the interval $\sim$10-30\,MeV. Atmospheric $\bar\nu_e$'s constitute the dominant background in the energy range above $\sim$30\,MeV. The normalization of the flux depends on the location of the detector \cite{Barr:1989ru,Honda:1995hz, Liu:2002sq, Battistoni:2005pd}, more specifically on the geomagnetic latitude, varying roughly within a factor of 2 \cite{Wurm:2007cy}. Moreover, NC reactions on carbon will create a background that is not yet well determined. See Sect.\,\ref{subsec::dsnb} for a closer discussion of these backgrounds.  


\begin{figure}[t]
\centerline{\includegraphics[width=0.47\textwidth]{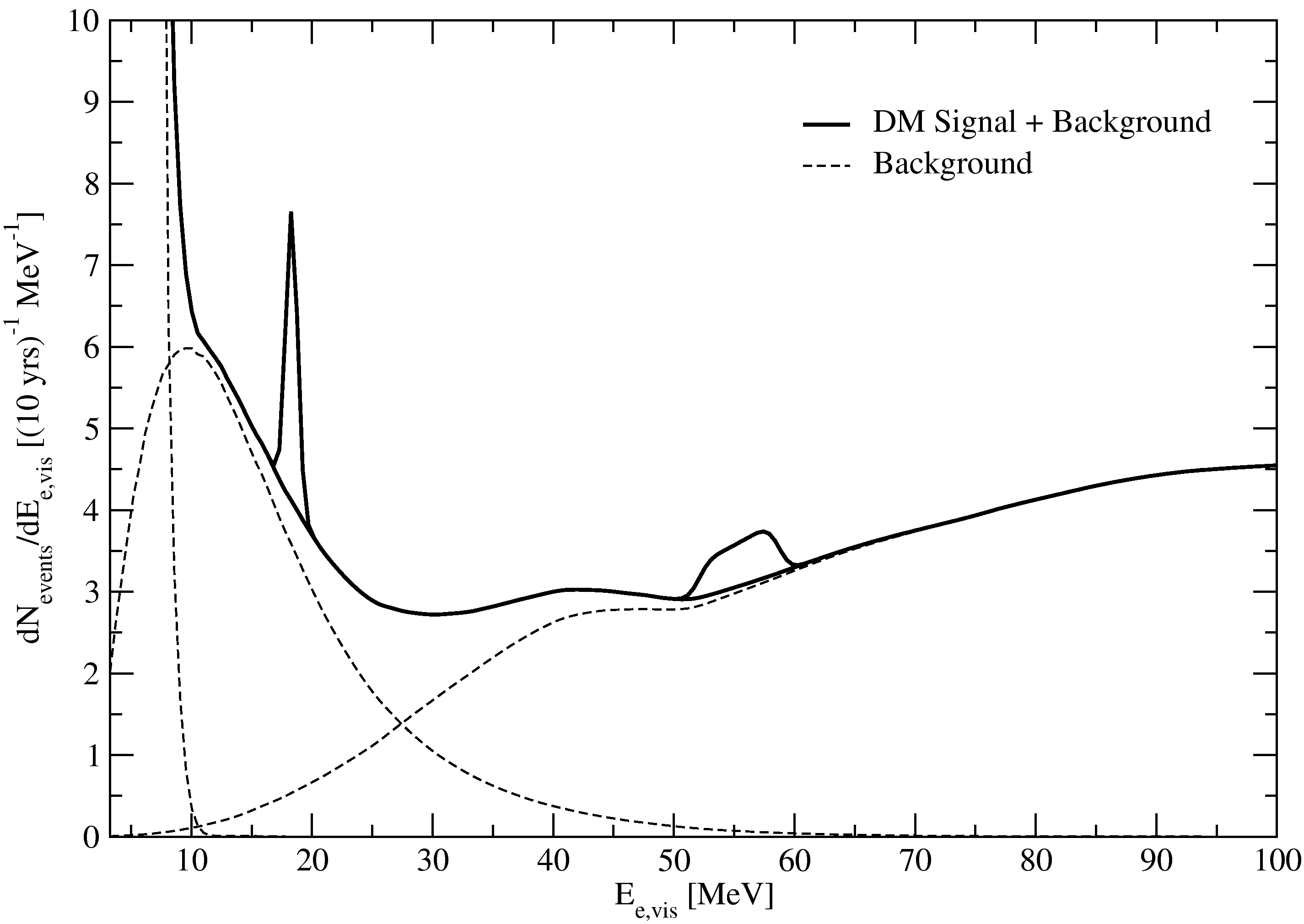}}
\caption{Expected signal in the LENA detector, in
  Pyh\"asalmi, after 10 years of running for two values of
  the DM mass, $m_\chi = 20 \, (40)$\,MeV and $m_\chi = 60 \,
  (120)$\,MeV for annihilations (decays) for $\langle
  \sigma_{\mathrm{A}} v \rangle = 3$$\times$$10^{-26} \, {\rm cm}^3/{\rm
    s}$ ($\tau_\chi = 8.9$$\times$$10^{23}$\,s and $\tau_\chi = 2.7 \times
  10^{24}$\,s, respectively for each mass).  Dashed lines represent the
  individual contributions of each of the three types of
  background events in this type of detector (reactor $\bar{\nu}_e$,
  DSNB and atmospheric neutrinos). The solid lines represent the
  backgrounds plus the expected signal from DM annihilation (decay) in
  the Milky Way.  Taken from \cite{PalomaresRuiz:2007eu}.}
\label{LENA10years}
\end{figure} 

In Fig.~\ref{LENA10years} (taken from \cite{PalomaresRuiz:2007eu}), the expected signal and background spectra are shown for LENA in Pyh\"asalmi after 10 years of data taking. Other locations return similar results. The assumed scintillator mixture is 20\% PXE
(C$_{16}$H$_{18}$) and 80\% Dodecane (C$_{12}$H$_{26}$) for a fiducial
volume of 50$\times 10^3$ m$^3$, which amounts to 3.3$\times$10$^{33}$
free protons. The rates and spectra are calculated using a gaussian energy
resolution function of width \cite{Wurm:2007cy}
\begin{equation}
\sigma_\mathrm{LENA} = 0.10 \, {\rm MeV} \, \sqrt{E/{\rm MeV}}.
\end{equation}
Two values for the DM mass are depicted: $m_\chi = 20 \, (40)$\,MeV and $m_\chi = 60 \, (120)$\,MeV for annihilations (decays) for $\langle \sigma_{\mathrm{A}} v \rangle = 3$$\times$$10^{-26} \, {\rm cm}^3/{\rm s}$ ($\tau_\chi = 8.9$$\times$$
10^{23}$\,s and $\tau_\chi = 2.7$$\times$$10^{24}$\,s, respectively for
each mass).  In the case of low values of the DM mass, even with the
small rate predicted, a rather easy discrimination between signal and
background could be possible.  For higher values of the masses, the
energy of the initial neutrino cannot be precisely reconstructed from
the measured positron energy.  Therefore, the signal at these masses
has a spread over an interval of $\sim$10\,MeV, degrading the
sensitivity of the detector.

\subsubsection{Conclusions}

Determining the DM identity and its properties is one of the
fundamental questions to be answered in the future in astroparticle
physics.  In regions of the Universe where DM is highly concentrated,
such as the center of galaxies, DM particles can annihilate (decay)
efficiently producing observable fluxes of SM particles.  Large
neutrino detectors might be able to observe the neutrino so produced
providing bounds or measurements of the DM mass and annihilation cross
section (lifetime).  It should be noted that neutrinos are the least
detectable particles of the SM and therefore provide the most
conservative bounds on DM annihilations (decays).  Importantly the DM
neutrino signal is mono-energetic, allowing for an enhanced
discrimination between signal and continuous backgrounds.

LENA would be particularly suited to these searches thanks to the
large size, the excellent energy resolution and the good background
discrimination.  For DM masses in the few tens of MeV, it could
observe a signal if these MeV particles exist and the annihilation
cross section is the one required to reproduce the observed amount of
DM (or its lifetime $\tau_\chi \sim 10^{24}$\,s).  In particular, the
LENA detector would have the capability to find a positive signal at
$\sim$2$\sigma$ in a large part of the mass window of interest.  A
null signal in LENA would indicate that, if DM particles with mass 
$\sim$10--100\,MeV exist, then they must live longer
than $\sim$10$^{24}$\,s and they were not produced thermally or the
annihilation cross section at freeze-out was velocity-dependent.  A
positive signal would imply that DM is constituted by particles with
masses in the tens of MeVs, would measure its mass and would determine
the cross section which was relevant at DM freeze-out in the Early
Universe (or its lifetime), for a given halo profile.

%% file: geo.tex

%

\noindent Geoneutrinos (geo-$\nu$s) are $\bar\nu_e$s produced inside the Earth during $\beta$-decays of naturally occurring radioactive elements. They are direct messengers of the abundances and distribution of radioactive elements within our planet, information strongly constraining all geochemical and geophysical models. Geoneutrinos have been successfully detected by the liquid-scintillator experiments KamLAND and Borexino. However, the geological information contained in these measurements is still limited, mostly because of low statistics. A multi-kT detector like LENA, featuring the radiopurity already achieved by Borexino, would address several questions of geological importance. This section presents the event and background rates expected for LENA (both in Pyh\"asalmi and Fr\'ejus), and projects the precision at which the total geo-$\nu$ flux as well as the U/Th ratio could be measured.

\subsubsection{Introduction}

Geo-$\nu$s originate from the $\beta$-decays of radioactive elements in the Earth's crust and mantle, predominantly from $^{40}$K and nuclides in the chains of $^{238}$U and $^{232}$Th. These neutrinos probe direct information about the absolute abundances and distribution of these radioactive elements inside the Earth. Their measurement quantifies the radiogenic contribution to the total heat flux of the Earth, constraining geochemical and geophysical models of the planet. This information provides constraints on the many and complex processes that operate inside the Earth, including the generation of the Earth's magnetic field, mantle convection, and plate tectonics. In addition, determining the absolute abundances of refractory elements (i.e.\,U and Th) in the planet provides insight into its origin and formation.

Estimates of the Earth's surface heat flux emerge from temperature gradient measurements from $\sim$40\,000 drill holes distributed around the globe. Using these data, geophysical models typically conclude that the present surface heat flux is 47$\pm$2\,TW \cite{Davies:2010}. This conventional view has been challenged by an alternative flux estimate of 31$\pm$1\,TW \cite{Hofmeister:2005}. A significant contributor to this heat flux comes from the heat producing elements, K, Th and U, with its flux proportion dependent upon their absolute abundance inside the Earth. The many models that describe the composition of the Earth come from cosmochemical, geochemical and geophysical observations and predict a range of abundances and distributions of these elements \cite{McDonough:1995,Lyubetskaya:2007, Anderson:2007, Javoy:2010}.

The Earth has a silicate shell, the Bulk Silicate Earth (BSE\nomenclature{BSE}{Bulk Silicate Earth model}), surrounding a metallic core, with the core being an iron-nickel mixture (with proportions set from cosmochemical constraints) often considered to contain negligible quantities of Th and U \cite{McDonough:2003, Mann:2010}. The BSE describes the primordial, non-metallic Earth condition that followed planetary accretion and core separation, prior to its differentiation into a mantle, oceanic crust, and continental crust. Elements excluded from the Earth's core are referred to as lithophile and those that accreted onto the Earth in chondritic proportions are
the refractory elements. 

Chondritic meteorites are undifferentiated samples with refractory element abundances in equal proportion and record a high temperature condensation characteristic of the cooling nebular. An important guide to predicting planetary compositions is given by the compositional match between chondritic meteorites and the solar photosphere on a one to one basis, over 5 orders of magnitude for the non-gases, based on an equal atom abundance of silicon.

Thorium and uranium are refractory lithophile elements and contribute equally $\sim$80\,\% of the total radiogenic heat production of the Earth, while the remaining fraction is due to $^{40}$K, a volatile element (assuming Th$/$U$\sim$4 and K$/$U$\sim$10 000) . During mantle melting and because of their chemistry and size, K, Th and U are quantitatively partitioned into the melt and depleted from the mantle. Thus, the continental crust, has over geologic time, been enriched in these elements and has a sizable fraction (about half) of the planet's inventory, producing radiogenic power of 7.3$\pm$2.3\,TW (2$\sigma$) \cite{Rudnick:2003}.

The range of BSE models predicting the Th, U, and K abundances (Tab.\,\ref{tab:BSE}) translates to radiogenic heat contributions of 12--30\,TW, and thus allow other possible heat sources to make up the total surface heat flux. Additional heat might might originate from accretion, gravitational contraction, latent heat from phase transitions, or from a (rather exotic) nuclear reactor in the core/core-mantle boundary (CMB\nomenclature{CMB}{Core-Mantle Boundary}). Systematic errors in both geochemical and geophysical models are not very well known and the validity of several assumptions on which they are based is not proven. Thus, observations of the planetary geo-$\nu$ flux will yield transformational insights into the Earth's energy budget.

\begin{table*}
\begin{tabular} {lllccc}
\toprule
Authors & & & ~~~~~$a(\rm U)\,\mathrm{[ng/g]}$~~~~~ & ~$a(\rm{Th})/a(\rm U)$~ & $H_\mathrm{M}(\rm U+\rm{Th})\,\mathrm{[TW]}$ \\
\colrule
Turcotte and Schubert & (2002) & \cite{Turcotte:2002}  &   31  &  4.0 &  19 \\
Anderson & (2007) & \cite{Anderson:2007}              &   28  &  4.0 &  17 \\
Palme and O'Neill & (2003) & \cite{Palme:2003}      &   22  &  3.8 &  12 \\
Allegre et al. & (1995) & \cite{Allegre:1995}           &   20  &  3.9 &  11 \\
McDonough and Sun & (1995) & \cite{McDonough:1995}      &   20  &  3.9 &  11 \\
Lyubetskaya and Korenaga & (2007) & \cite{Lyubetskaya:2007}   &   17  &  3.7 &  7  \\
Javoy et al. & (2010) & \cite{Javoy:2010}                 &   12  &  3.5 &  3   \\
\botrule
\end{tabular}
\caption{Uranium content $a($U$)$ in the Bulk Silicate Earth, the Th/U
ratio and the radiogenic heat production in the mantle ($H_{\mathrm{M}}$) due
to U and Th according to different authors.}
\label{tab:BSE}
\end{table*}

Typically, based on geophysical calculations, parameterized convection models of the mantle require higher radiogenic heat contributions ($\sim$70\,\% of the total heat flux) in order to describe the Earth's cooling history in terms of a balance of forces between thermal dissipation and mantle viscosity. In contrast, geochemical models using cosmochemical and geochemical observations predict the BSE abundance of U and values for Th/U and K/U, 4 and 1.4$\times$10$^4$, respectively, with an uncertainty of $\sim$10\,\%. Consequently, a geochemist's view of the Earth predicts that its budget of heat producing elements in the BSE are up to a factor of $\sim$3 lower than the models predicted by geophysicists. Thus, the relative contribution of the radioactive power to the total planetary heat flux is poorly known.

The first ideas regarding geo-$\nu$s are from the sixties \cite{Eder:1965} but only recently the first experimental results from large volume scintillator detectors are available. KamLAND and Borexino have observed geo-neutrinos with similar precision~\cite{Araki:2005qa,Abe:2008ee,Bellini:2010hy, Gando:2011} . Borexino has significantly greater radiopurity and lesser reactor flux than KamLAND, allowing Borexino to make the measurement in ~5\% of the exposure of KamLAND. In addition, both experiments have placed limits ($<$3\,TW) on the potential contribution of a putative geo-reactor deep in the Earth's interior.

In spite of these first successful experimental results, it has not been possible yet to neither discriminate among several predictions concerning the radiogenic heat production nor measure Th/U ratio of the observed energy spectra. Even the very low background measurement performed by Borexino has limited power of geological predictions due to slowly accruing statistics. Several future experiments, as for example SNO+ project in Canada, have among their aims geo-$\nu$ measurements. However, a real breakthrough in this field would come with a very large volume detector at 50\,kt scale, like LENA.

In liquid scintillator detectors, the $\bar{\nu}_{e}$ are detected via
the inverse beta decay, with a kinematic threshold of 1.8\,MeV (Sect.\,\ref{subsec::dsnb}). The characteristic time and spatial coincidence of prompt $e^+$ and delayed neutron events offers a clean signature. Since 1\,kt of liquid scintillator contains about $10^{32}$ free protons (the precise value depending on the chemical composition) and the exposure times are of order of a few years, the events rates are conveniently expressed in terms of a Terrestrial Neutrino Unit (TNU\nomenclature{TNU}{Terrestrial Neutrino Unit}), defined as one event per $10^{32}$ target protons per year.

The geo-$\nu$ flux produced from U and Th inside the Earth is some $10^{6}$\,/cm$^{2}$s. Due to neutrino oscillations, the anti--neutrino flux arriving at the detector in the electron flavor will be smaller than that produced: for our calculation we have considered an asymptotic survival probability $\langle P_\mathrm{ee}\rangle = 0.57$ following the best fit obtained in \cite{Strumia:2005tc}. Only a small fraction (about 5\,\%) of $\bar{\nu}_{e}$ from the $^{238}$U and $^{232}$Th series are above the inverse beta-decay reaction threshold, while those from $^{40}$K decays are below this threshold. Geo-$\nu$s originating from different elements can be distinguished - at least in principle - due to their different energy spectra, as only $\nu$s from the uranium chain contribute at energies $E_{e^+} > 2.25$\,MeV. The exact spectrum depends on the shapes and rates of the individual decays within U and Th chains, and on the abundances and spatial distribution of U and Th in the crust and in the mantle. 

The geo-$\nu$ signal spectrum extends to $E_{e^+}\approx2.6$\,MeV. However, $\bar\nu_e$ from nuclear power plants represent a background for geo-$\nu$ detection, $E_{e^+}$ extending up to $\sim$10\,MeV. 
In the following the expected geo-$\nu$ and reactor $\nu$ signals at Pyh\"asalmi and Fr\'ejus sites are discussed. The potential of the LENA project to achieve geologically interesting
results is discussed as well.

\subsubsection{The geoneutrino signal}

Different calculations for geo-$\nu$ production have been presented in the literature \cite{Mantovani:2003yd, Fogli:2005qa, Enomoto:2007, Dye:2010vf}): all models rely on the geophysical $2^{\circ} \times 2^{\circ}$ crustal map of \cite{Bassin:2000} and on the density profile of the mantle as given by the Preliminary Reference Earth Model (PREM\nomenclature{PREM}{Preliminary Reference Earth Model}) \cite{Dziewonski:1981xy}. For the calculation of geoneutrino signal we adopt the values of U and Th abundance recommended in \cite{Plank:1998} for the sedimentary layers and the values reported in \cite{Rudnick:2003} for the upper, middle and lower crust (Tab.\,\ref{tab:Abundances}). The 1$\sigma$ uncertainties for the upper and middle crust are from \cite{Rudnick:2003}. For the lower crust, we adopt an uncertainty indicative of the spread of published values.

The composition and the circulation inside the Earth's mantle is the subject of a strong and so far unresolved debate between geochemists and geophysicists: assuming that a spherical symmetry holds and U and Th abundances do not decrease with depth, the extreme predictions for the signal are obtained by placing U and Th in a thin layer at the bottom and distributing it with uniform abundance over the mantle. For a fixed total U mass in the BSE model, $m(\mathrm{U}) = 0.8$$\times$10$^{17}$\,kg, and a fixed ratio of the elemental abundance Th/$\rm U = 3.9$ \cite{McDonough:2003}, the contribution to the geo-$\nu$ signal from the crust and the mantle is obtained by using the proximity argument presented in \cite{Fiorentini:2007te}: the minimal (maximal) contributed flux is obtained by placing uranium and thorium as far (close) as possible to the detector.

\begin{figure*}
\includegraphics[width=0.46\textwidth,height=0.25\textheight]{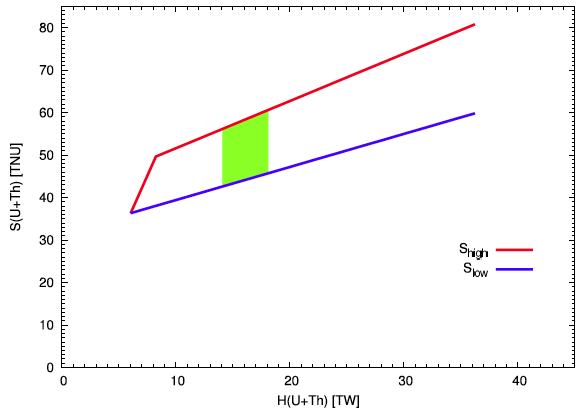}
\hfill
\includegraphics[width=0.46\textwidth,height=0.25\textheight]{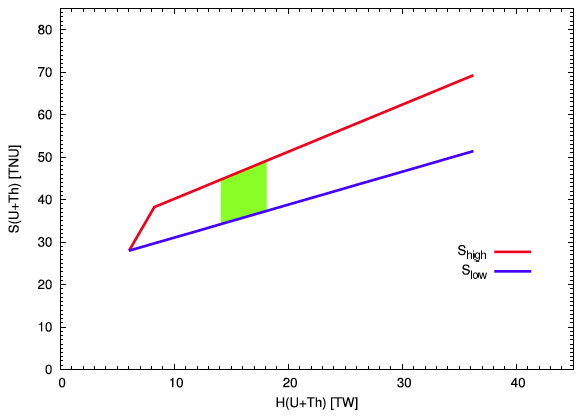}
\caption{The expected geo-$\nu$ signal at Pyh\"asalmi (left) and
Fr\'ejus (right) as function of radiogenic heat due to U and Th in the
Earth $H(\rm U+ {\rm Th})$. The area between the red and blue lines denotes the
region allowed by geochemical and geophysical constraints. The green
region is allowed by the BSE model according to \cite{McDonough:2003}.}
\label{Fig:rette}
\end{figure*}

Our prediction for geoneutrino signal is obtained by the mean of these extremes, assigning an error so as to encompass both of them. For the central value of BSE model, the expected signal from U and Th at Pyh\"asalmi is 51.3$\pm$7.1\,TNU. At Fr\'ejus, it is 41.4$\pm$5.6\,TNU: the accuracy of about 14\,\% corresponds to "3$\sigma$". In a target mass of 44\,kt, corresponding to 2.9$\times$10$^{33}$ free protons, we expect a geo-$\nu$ signal of the order of $10^{3}$ events/year. In Fig.\,\ref{Fig:rette}, the expected signal $S({\rm U}+{\rm Th})$ from U and Th geo-$\nu$s at Pyh\"asalmi and Fr\'ejus is shown as a function
of the radiogenic heat production rate $H({\rm U}+{\rm Th})$.
 
For a given total uranium mass in the Earth, $m($U$)$, corresponding to a fixed radiogenic heat production $H({\rm U}+{\rm Th})$, the minimal and maximal signals are provided by the terrestrial models consistent with available geochemical and geophysical observational data and by proximity argument \cite{Fiorentini:2007te}. The uncertainty band is wide because the signal is dominated by contribution from the crust: a refinement of the reference model taking into account the regional contribution is appropriate. Considering that some 50\,\% of the signal from the crust originates from a region within 200\,km from both detectors, a better geological and geochemical description of the regions surrounding the detectors is needed for a more precise estimate of the geoneutrino signal. A comprehensive study of the crust near the detector is essential for resolving global models.

\begin{table} 
\begin{tabular} {llcc}
\toprule
Reservoir & Units & $a(\rm U)$ & $a({\rm Th})$ \\
\colrule
Sediments     &	\textmu g$/$g  &	1.68 $\pm$ 0.18 &	6.91 $\pm$ 0.8 \\		
Upper Crust   &	\textmu g$/$g  &	2.7  $\pm$ 0.6   &	10.5 $\pm$ 1.0 \\		
Middle Crust  &	\textmu g$/$g  &	1.3  $\pm$ 0.4   &	6.5  $\pm$ 0.5  \\		
Lower Crust   &	\textmu g$/$g  &	0.6  $\pm$ 0.4   &	3.7  $\pm$ 2.4  \\
Oceanic Crust &	\textmu g$/$g  &	0.1  $\pm$ 0.03  &	0.22 $\pm$ 0.07 \\	
\botrule
\end{tabular}
\caption{ U and Th mass abundances in the Earth's reservoirs \cite{Plank:1998,Rudnick:2003}.}
\label{tab:Abundances}
\end{table}

\subsubsection{Reactor neutrino background}

\begin{figure*}[t!]
\includegraphics[width=0.46\textwidth,height=0.26\textheight]{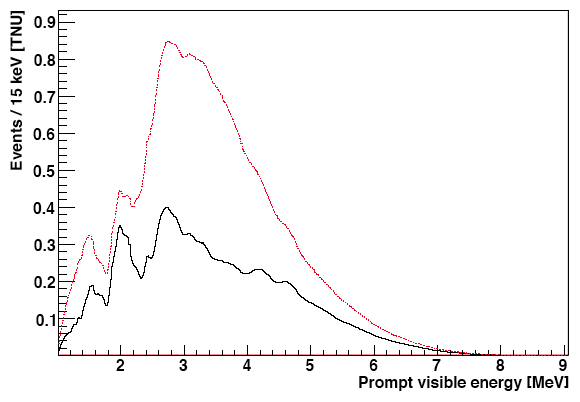}
\hfill
\includegraphics[width=0.46\textwidth,height=0.26\textheight]{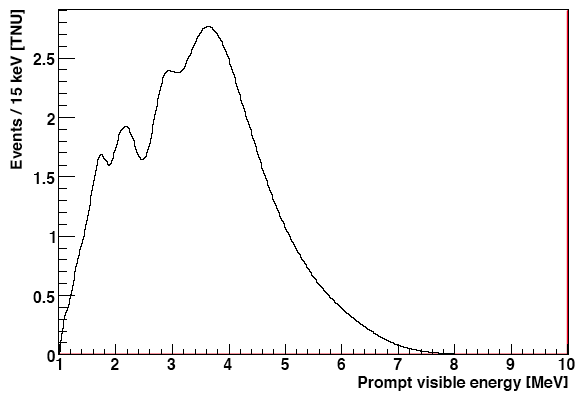}
\caption{The expected reactor $\bar\nu_e$ signal at Pyh\"asalmi (left) and Fr\'ejus (right). The red spectrum (left) is when future reactors Finnish reactors are taken into account. Such spectra would be measured by an ideal detector with $10^{32}$ free target protons in one year.}
\label{Fig:reactors}
\end{figure*}

The expected reactor $\bar\nu_e$ flux was calculated based on the same assumptions as described in \cite{Bellini:2010hy}. The neutrino oscillations parameters ($\Delta m^2_{12}$ = 7.65~$\cdot 10^{-5}$ eV$^2$; $\sin^2\theta_{12}$ = 0.304) from Ref.~\cite{Schwetz:2008er} were used. The monthly load factor of year 2009 was considered for all 493 world nuclear reactors \cite{IAEA}. The expected reactor $\bar\nu_e$ signal at Fr\'ejus and Pyh\"asalmi and shape of the oscillated spectrum are given in Tab.\,\ref{tab:Signal} and Fig.\,\ref{Fig:reactors}, respectively.

For Pyh\"asalmi site also the case of possible future reactors was considered, assuming a typical 80\,\% load factor. In particular, the Olkiluoto-3 power plant is under construction and should be operating in 2013 with 4.3\,GW thermal power. At the same site, 360\,km from Pyh\"asalmi, the construction of Olkiluoto-4 reactor with power up to 1.8\,GW was approved. In addition, the construction of additional reactor with power up to 4.9\,GW is under approval (Pyh\"ajoki site 130\,km from Pyh\"asalmi is among possible sites).

\begin{table} 
\begin{tabular} {lcc}
\toprule
Location &  Signal 1-10\,MeV & Signal 1-2.6\,MeV \\
         &   [TNU]   & [TNU] \\
\colrule
Pyh\"asalmi & 70.9$\pm$3.8 &  20.8$\pm$1.1  \\  
Pyh\"asalmi$^{*}$ & 145.9$\pm$7.7 & 37.3$\pm$1.9 \\ 
Fr\'ejus &  554$\pm$29.4 &  145$\pm $7.7 \\
\botrule
\end{tabular}
\caption{ Expected reactor $\bar\nu_e$ signal. ($^{*}$future
Finnish reactors are taken into account). The corresponding energy spectra are shown in  Fig.\,\ref{Fig:reactors}.\medskip\medskip}
\label{tab:Signal}
\end{table}

\subsubsection{Determining the geoneutrino flux}

LENA, thanks to its large volume, would be a real breakthrough in the field of geo-$\nu$ detection. Within the first year, geologically significant results could be obtained. Independently from the final location, within the first year LENA would measure the total geo-$\nu$ flux at the level of few percent, by far more precise than the current experiments as Borexino or KamLAND could reach.

We simulated the expected energy spectrum at Pyh\"asalmi (with and without future Finnish reactors) and Fr\'ejus, based on 2.9$\times$10$^{33}$ target protons and the expected reactor $\bar\nu_e$ and geo-$\nu$ fluxes described above. A light yield of 400 photoelectrons/MeV at the upper limit of the achievable range was assumed. The expected spectra were convoluted with the consequent energy dependent resolution.

LENA aims to reach, and possible exceed the radio-purity of Borexino detector. Therefore, the no-background approximation is reasonable, since Borexino has shown that the final $\bar\nu_e$ spectrum is almost background free (the total background is less than 2\,\% of the total
$\bar\nu_e$ spectrum \cite{Bellini:2010hy}).

\begin{figure*}[t!]
\includegraphics[width=0.46\textwidth,height=0.25\textheight]{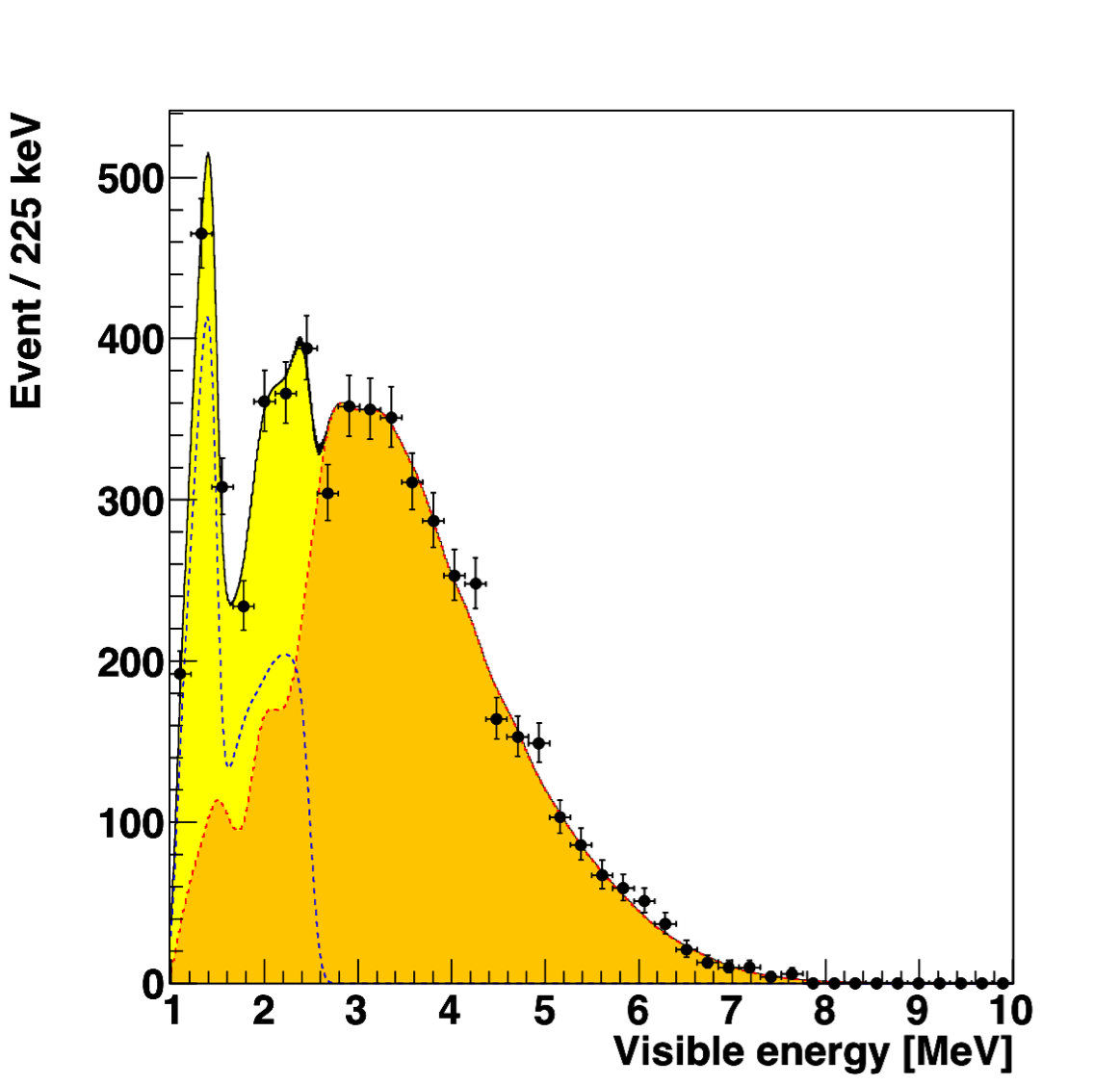}
\hfill
\includegraphics[width=0.46\textwidth,height=0.25\textheight]{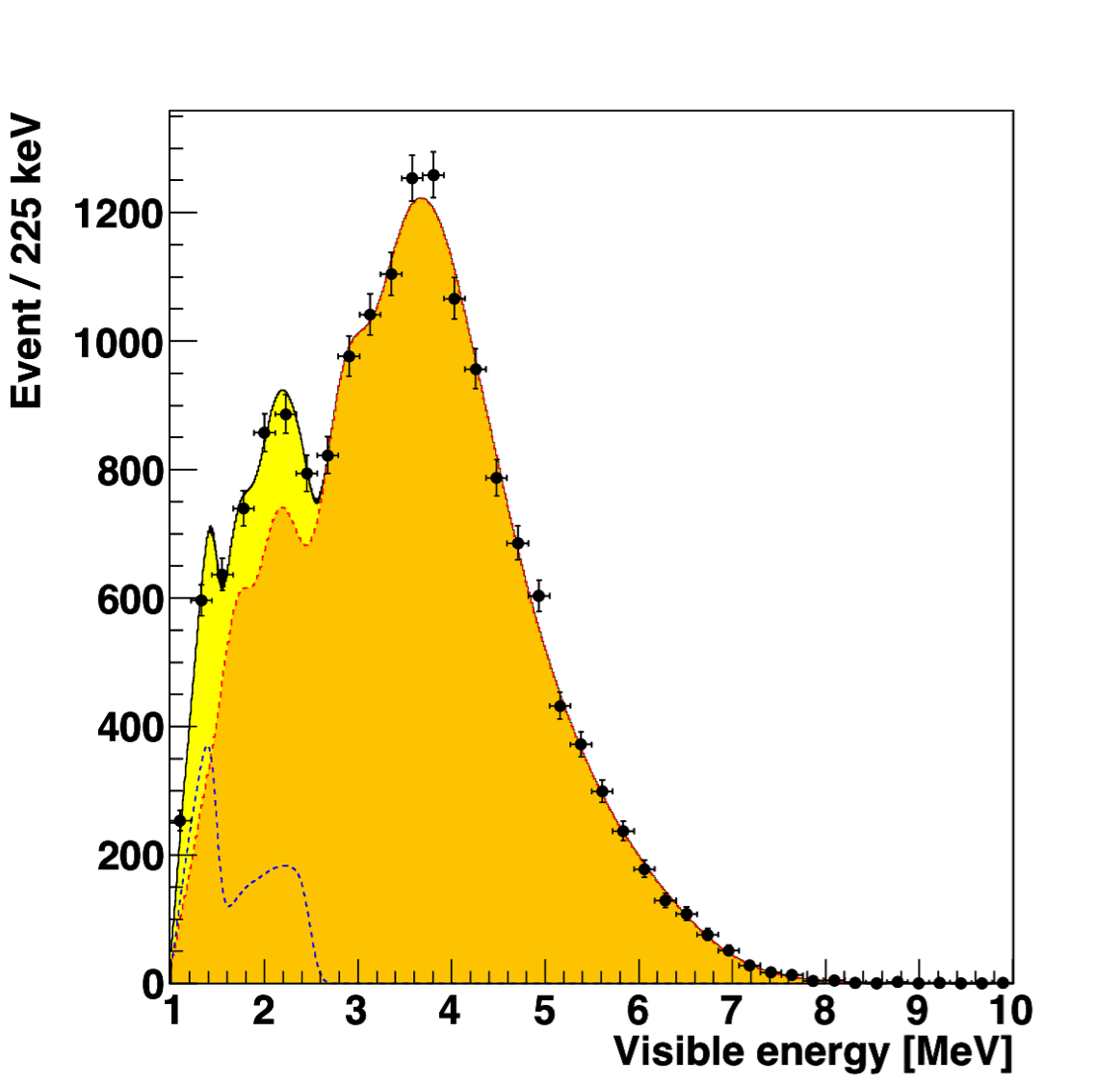}
\caption{ The expected oscillated visible energy spectrum at
Pyh\"asalmi (left, future reactors in Finland are considered) and
Fr\'ejus (right) for 1 year statistics and 2.9$\times$10$^{33}$ target protons, considering a light yield of 400 pe/MeV. 
The contribution of the reactor $\bar\nu_e$s is shown by the filled orange area while that of geo-$\nu$s  by the dashed blue line. The yellow
area isolates the contribution of the geo-$\nu$s in the total signal.}
\label{Fig:spectrum}
\end{figure*}

The shape of the expected spectra is shown in Fig.~\ref{Fig:spectrum}. The chondritic U/Th ratio was assumed. The final precision of the geo-$\nu$ flux measurement which could be reached after 1, 3, and 10 years is given in Tab.\,\ref{tab:FluxPrecision}. Systematic uncertainties on the expected reactor background flux are not considered here and might limit the sensitivity if they surpass the statistical error of the measurement.

\begin{table} 
\begin{tabular} {cccc}
\toprule
Live time & Pyh\"asalmi & Pyh\"asalmi$^{*}$ & Fr\'ejus \\
\colrule
1\,yr  &  3\% & 4\% & 6\% \\
3\,yrs &   2\% & 2\% & 3\% \\
10\,yrs & 1\% & 1\% & 2\% \\
\botrule
\end{tabular}
\caption{ Expected precision in the measurement of the total geo-$\nu$ flux
($^{*}$future Finnish reactors are taken into account). Details in text.}
\label{tab:FluxPrecision}
\end{table}

\subsubsection{Potential to measure the U/Th ratio}

The precision at which the U and Th fluxes, as well as their ratio, could be measured at Pyh\"asalmi and Fr\'ejus sites is summarized in Tab.\,\ref{tab:UThPrecision}.  In these calculations, the assumptions taken into account are the same as in the previous section. However, instead of the single geo--neutrino component (with the U and Th ratio fixed to the chondritic value)  the two individual contributions of  U and Th chains are left free in the fit.

\begin{figure*}[t!]
\includegraphics[width=0.46\textwidth,height=0.275\textheight,angle=0]{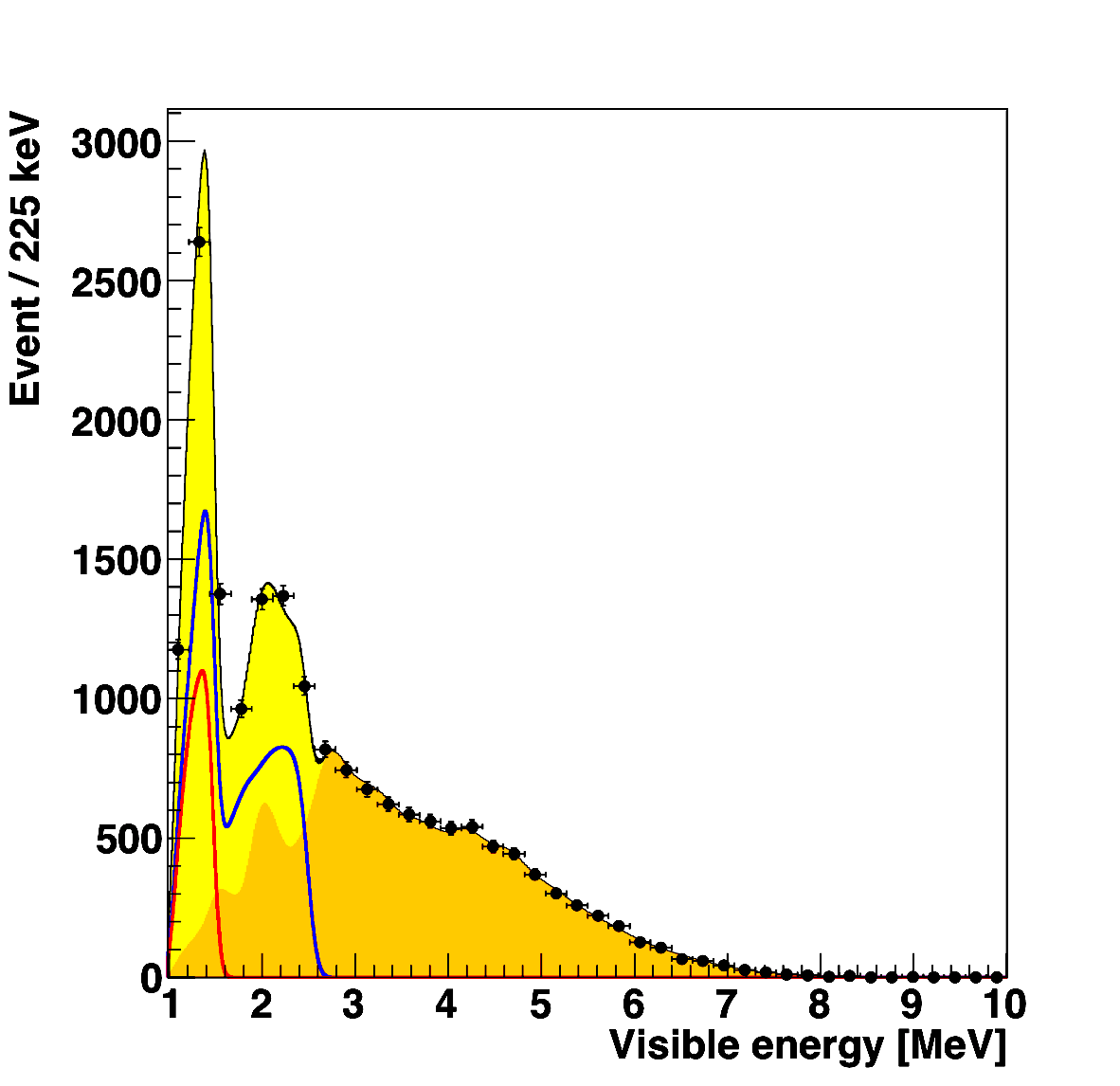}
\hfill
\includegraphics[width=0.46\textwidth,height=0.275\textheight,angle=0]{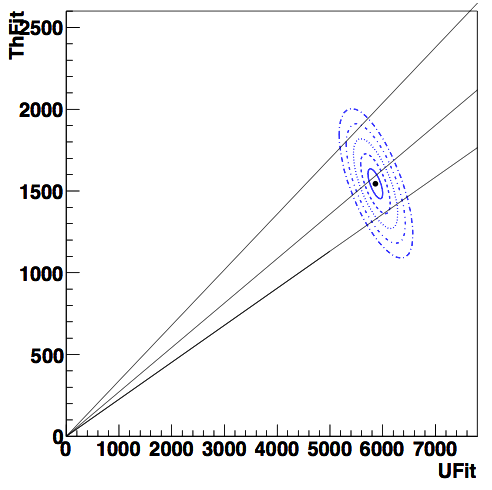}
\caption{ Left: Expected 5-year visible energy spectrum  for 2.9$\times$10$^{33}$ target protons and light yield of 400 pe/MeV at
          Pyh\"asalmi site (future Finnish reactors not considered). The contribution of the reactor $\bar\nu_e$s is shown by the filled orange area while the yellow area isolates the contribution of the geo-$\nu$s in the total signal. The data were generated with the chondritic U$/$Th ratio while in the fit the contribution from U (blue line) and Th (red line) were fit individually.
Right: Corresponding contour plots (1,2,3,4, and 5 $\sigma$ C.L. outwards from the best fit black point, respectively) for the absolute U
          and Th number of events resulting from the fit. Each of the three solid lines corresponds to a fixed U$/$Th ratio. The central
          line corresponds to chondritic U$/$Th ratio,  while the two external  lines correspond to this ratio changed by $\pm$20\,\%.}
\label{Fig:UThfree}
\end{figure*}

~\\~\\
The Pyh\"asalmi site is strongly preferred for this measurement. An example of a possible 5-yr energy spectrum of reactor and U and Th geo-$\nu$s is shown in Fig.\,\ref{Fig:UThfree}, together with the U vs Th contour plot resulting from the fit.

\begin{table} 
\begin{tabular} {lrrrr}
\hline
\toprule
Location & Live time       & U flux   & Th flux   & U/Th \\
          &   [yrs]       &  [\%]   & [\%]      & [\%] \\
\colrule
Pyh\"asalmi &  1            &  6    & 12      &  17    \\
 &  3            &  3    & 8       &  10   \\
 &  10           &   2   & 4        &  5  \\
Pyh\"asalmi$^{*}$ & 1       &   7    & 14 &   21    \\
 & 3       &   4    & 8  &   11    \\
 & 10      &   2    & 4  &    6    \\
Fr\'ejus &            1       &  14    & 25 &   35    \\
 &            3       &  9     & 12 &   20    \\
 &            10      &  4     &  7  &  11     \\
\botrule
\end{tabular} 
\caption{ Expected precision in the measurement of the U and Th geo-$\nu$ flux and in the U$/$Th ratio. ($^{*}$ future Finnish reactors are taken into account). Details in text.}
\label{tab:UThPrecision}
\end{table}

\subsubsection{Directionality}

In the inverse beta decay reaction the neutron is scattered roughly in forward direction. Thus, it is possible to obtain directional information about the $\bar\nu_e$ event by measuring the displacement between the neutron and the $e^+$ event \cite{Hochmuth:2005nh}. As the average displacement is with 1.9$\pm$0.4\,cm \cite{Apollonio:1999jg} rather small compared to the neutron and $e^+$ position reconstruction uncertainty, the $\bar\nu_e$ direction can only be reconstructed with a large uncertainty. But, by analyzing a large number of geo-$\nu$ events, it is still possible to extract information about the $\bar\nu_e$ angular distribution~\cite{Apollonio:1999jg}. From the angular distribution of the geo-$\nu$ events, the differential radial distribution of terrestrial radio-nuclides could be determined \cite{Fields:2004tf}. This would be important to differentiate between the geo-$\nu$s coming from the mantle and those from the crust.

\subsubsection{Backgrounds}

LENA aims to achieve a similar or even better radiopurity level than realized in Borexino. Here, we briefly describe the background sources relevant for $\bar\nu_e$ detection.

\medskip
\noindent\textbf{$^9$Li-$^8$He:} $^9$Li ($T_{1/2}=178$\,ms) and $^8$He ($T_{1/2}=119$\,ms) are $\beta^{-}$-neutron emitters, that are produced by cosmic muons crossing the detector.  Borexino measured 15.4 events/(100\,t$\cdot$yr) \cite{Bellini:2010hy}, which scales to about 1500 events per year in LENA, considering a reduced muon flux by a factor of 5. This background can be reduced to about 1 event per year in LENA, if a 2\,s cut after every detected muon is applied. As $^9$Li and $^8$He are produced close to the muon track, it is possible to reduce the dead time from $\sim$6\,\% to $\sim$0.1\,\%, if only a cylinder with 2\,m radius around the muon track is vetoed.

\medskip
\noindent\textbf{Fast Neutrons:} Cosmic muons that pass the detector can produce fast neutrons. These neutrons have a large range and can reach the Inner Vessel of LENA without triggering the muon veto. In the scintillator they can mimic $\bar\nu_e$ events, as they give a prompt signal due to scattering off protons and a delayed signal caused by the neutron capture on a free proton. The fast neutron background rate was analyzed with a MC simulation \cite{Moellenberg:2009}. In the geo-$\nu$ energy region, less than 10 events per year are expected. Compared to the expected signal this background is negligible.

\medskip
\noindent\textbf{$^{13}$C($\alpha$,n)$^{16}$O:} Neutrons can also be produced in the scintillator by $^{210}$Po $\alpha$ decays and subsequent $^{13}$C($\alpha$,n)$^{16}$O reactions. If the radio-purity level of Borexino is reached in LENA, about 10$\pm$1 events per year are expected in LENA \cite{Bellini:2010hy}.

%% file: reactor.tex

%


Experiments with reactor anti-neutrinos have a long and successful tradition in neutrino physics. KamLAND was the first reactor neutrino experiment to observe a deficit in the flux, confirming the Large Mixing Angle MSW solution to the solar neutrino problem. Consequently, the experiment performed the most precise measurement of the oscillation parameter $\Delta m_{21}^2$. A series of experiments at distances of several meters to $\sim$1\,km to the reactor core(s) has led to the current upper bound on the mixing angle $\theta_{13}$, dominated by the result of the CHOOZ experiment \cite{Apollonio:2002gd}. A new generation of experiments with a multi-detector setup composed of detectors near and far ($\sim$1\,km) to the cores aim to measure or constrain further the value of $\theta_{13}$. The use of liquid scintillator detectors for reactor neutrino detection is a perfectly established technique and regardless which location is finally chosen for the realization of LENA, there will be a measurable reactor neutrino signal.

While anti-neutrinos from nuclear reactors form a considerable background for the detection of geoneutrinos and the DSNB in LENA, their signal also offers the opportunity to perform a high-statistics study of neutrino oscillation effects, especially to improve the knowledge on the parameters that drive solar neutrino oscillations. Each reactor provides a high intensity, isotropic source of anti-neutrinos with a well-known initial spectrum, resulting from $\beta^-$ decays of fission products ($^{235}$U, $^{238}$U, $^{239}$Pu, $^{241}$Pu being the four main isotopes). The overall emitted anti-neutrino spectrum is computed\footnote{New calculations of reactor anti-neutrino spectra from these measurements, including information from nuclear databases, have been performed in \cite{Mueller:2011nm}. While the shapes of the spectra and their uncertainties are comparable to that of the previous analysis, the absolute flux normalization is shifted by about +3\% on average.} from measurements of beta spectra at ILL for $^{235}$U, $^{239}$Pu, and $^{241}$Pu and theoretical calculations for $^{238}$U (see e.g. \cite{Huber:2004xh} for a polynomial parametrization). During reactor operation, the abundance of $^{235}$U decreases, while that of $^{239}$Pu and $^{241}$Pu increases. If the fuel evolution of all reactors is known, this burn-up effect can be taken into account. For the sensitivity studies presented here, a typical averaged isotopic composition is used. The contribution of stored spent fuel elements to the detected signal is considered negligible, as they contain mainly long lived emitters with a low Q value.

In the LENA detector, anti-neutrinos are detected through the inverse $\beta$-decay process on free protons $\bar{\nu}_e+p\rightarrow e^+ + n$ (energy threshold of 1.8\,MeV) with well known cross section, followed by neutron capture. Experimentally, the clear signature of the coincidence signal, formed by the prompt positron signal followed by the delayed neutron capture in spatial correlation can be used for powerful reduction of accidental background. The visible energy $E_{\rm vis}$ of the prompt event is related to the energy of the incident neutrino $E_{\bar{\nu}_e}$ by $E_{\rm vis} \cong E_{\bar{\nu}_e} - m_n +m_p + m_e$.

In \cite{Petcov:2006gy}, the possibility of a high precision measurement of the solar mixing
parameters $\Delta m_{21}^{2}$ and $\sin\theta_{12}$ has been investigated,
assuming a LENA type liquid scintillator detector that is located at Fr\'ejus. Here, the reactor neutrino flux is highest compared to other sites under consideration. Fig.\,\ref{Fig:reactors} shows on the right the spectrum from inverse beta decays expected at Fr\'ejus. A dominant part of the total flux (67\,\%) is provided by the four nearest reactors within a distance of up to 160\,km in Switzerland and France. As the authors of \cite{Petcov:2006gy} point out, their distances are located between the first and the second survival probability minimum, and hence spectral information should provide a powerful tool to measure the oscillation parameters.

The large flux originating from French and Swiss nuclear power plants corresponds to a rate of of 1.7$\times$10$^4$ inverse beta decay events per year in a fiducial mass of 44\,kt, two orders of magnitude larger than the KamLAND event rate. A threshold of 2.6\,MeV on the visible prompt energy was applied to eliminate the signal from geoneutrinos for these numbers.

In this scenario, 3$\sigma$ uncertainties below 3\,\% on $\Delta m_{21}^{2}$ and of about 20\,\% on $\sin\theta_{12}$ could be obtained based on 1 year of exposure. After 7 years of data taking, the 3$\sigma$ uncertainties would diminish to 1\% in $\Delta m_{21}^{2}$ and 10\% in $\sin\theta_{12}$, respectively. While this would mean only a moderate improvement compared to present-day accuracies in the case of $\sin\theta_{12}$, the uncertainty in determining the
value of $\Delta m_{21}^{2}$ would decrease by almost an order of magnitude \cite{GonzalezGarcia:2010er}. 

In case of the Pyh\"asalmi site, the total event rate coming from currently operating reactors is lower by almost on order of magnitude. A third reactor core at the Olkiluoto plant is under construction since 2005, and the permission for a fourth was approved in 2010. Approval for an additional reactor in a site at 130\,km distance to Pyh\"asalmi is under discussion. The future situation with new reactors would correspond to a roughly doubled expected reactor neutrino flux. Therefore, the Pyh\"asalmi site will still be the preferred for the detection of geoneutrinos and the DSNB due to the lower reactor neutrino background. Nevertheless, one can  expect a useful total number of reactor neutrino events accumulated over the operation time of LENA, improving the determination of neutrino oscillation parameters.

%% file: oscillometry.tex

%

\noindent An extended liquid-scintillator detector LENA offers the opportunity for neutrino oscillometry \cite{Novikov:2011gp}. Based on a monoenergetic $\nu_e$ source, the characteristic spatial pattern of $\nu_e$ disappearance can be detected within the length of detector. Radioactive nuclides under-going electron capture produce monoenergetic neutrinos: Sufficiently strong sources of more than 1\,MCi activity are produced at nuclear reactors. In the three-flavor scenario, the investigation of the mixing parameters $\theta_{13}$ and $\Delta m_{13}^2$ are the most promising due to the short oscillation length $L_{13}$. Moreover, oscillometry is a unique tool to probe the existence of oscillations into a fourth sterile neutrino. LENA can be considered as a versatile tool for neutrino oscillation measurements at short baselines.

\subsubsection{Introduction}

The most precise and unambiguous way to detect neutrino oscillations is a determination of the oscillation pattern in the distance-dependent flux of the given neutrino flavor over the entire oscillation length. Since the oscillation length is proportional to the neutrino energy, neutrino oscillometry would require a detector hundreds or even thousands of kilometers long if used with the present or proposed neutrino beams that feature energies of several hundred MeV. As this is unrealistic, all beam experiments aiming at neutrino oscillations consider just a single or at most a two-point measurements instead of the full oscillometric approach. Also when using reactor neutrinos, the distance from the source to the first minimum is about 2\,km $-$ still beyond the technological and financial constraints for a detector. To be able to perform neutrino oscillometry using a realistic-size detector like LENA (100\,m long), one needs a strong source of monoenergetic neutrinos with an energy of a few hundred keV. Comparable sources have already been produced by neutron irradiation in nuclear reactors \cite{Hampel:1997fc,Abdurashitov:2006qr,Vergados:2010vp}. Neutrino oscillometry potentially provides a competitive and considerably less expensive alternative to long-baseline neutrino beams.

\subsubsection{Detection principle}

In liquid scintillator, $\nu_e$s at sub-MeV energies are detected by the recoil electrons from elastic neutrino-electron scattering. Any decrease in the detection rate along the detector that exceeds the geometric factor will give, for the first time, a continuous (oscillometric) measure of flavor disappearance. 
 
LENA is well suited for an oscillometric measurement due to its large height $h$ of 100\,m, the low detection threshold of $\sim$200\,keV, and the considerable fiducial volume of $\sim$35\,kt in this energy region (compare Sect.\,\ref{subsec::solar}). Assuming a light yield of 200\,pe/MeV for LENA, the expected position sensitivity is $\sim$25\,cm at 500\,keV electron recoil energy \cite{Alimonti:2008gc}. The energy resolution will be $\sim$10\% in this region \cite{Oberauer:2005kw}. 

As the cross-sections for $\nu$-e scattering are tiny, a very strong neutrino source has to be used to provide adequate statistics. Fortunately, there is a variety of radionuclei decaying via electron capture (EC\nomenclature{EC}{Electron Capture}). Since EC is a two-body process, the emitted electron neutrino is monoenergetic and carries most of the transition energy. Tab.\,\ref{Tab:OscMat} lists EC isotopes featuring suitable $Q$-values to produce monoenergetic neutrinos of a few hundreds of keV and with half-lives of a few months allowing for convenient handling. 

Sources of this kind have been produced in the past by neutron irradiation at nuclear reactors: Usually, a lighter ($A$-1), stable isotope of the same element is exposed to the intense neutron flux generated inside a reactor. For the calibration of the GALLEX\nomenclature{GALLEX}{GALLium EXperiment} experiment \cite{Hampel:1997fc}, a $^{51}$Cr source of an initial activity of 62\,PBq (1.7\,MCi) \cite{Cribier:1996cq} was produced by placing 36\,kg of metallic chromium, enriched in $^{50}$Cr, at the core of the Siloe reactor in Grenoble (35\,MW thermal power) for a period of 23.8 days. In principle, the 36\,kg batch of enriched chromium is still available and could be reused for LENA. Assuming an activity of 5\,MCi, about 1.9$\times$10$^{5}$ neutrino events would be expected. This already enormous statistics could be further increased either by repeating the cycle of neutron activation and measurement runs in LENA or by a further increase in source activity.

\begin{table*}
\begin{tabular}{ccccccc}
\toprule
Nuclide~~~& ~~~$T_{1/2}$\,[d]~~~ & ~~~$Q_\epsilon$\,[keV]~~~ & ~$E_\nu$\,[keV] (BR)~ & ~$E_{e,\mathrm{max}}$\,[keV]~ & ~~~Material~~~ & ~$\nu$ intensity\,[Bq]\\
\colrule
$^{37}$Ar	& 35 	& 814	& 811 (100\%)	& 617	& $^{40}$Ca,\,Ar	& 8.3$\times$10$^{15}$ \\
$^{51}$Cr	& 28 	& 753	& 747 (90\%)	& 560	& $^{50}$Cr	& 2.3$\times$10$^{16}$ \\
$^{75}$Se	& 120 	& 863	& 450 (96\%)	& 287	& Se	& 1.1$\times$10$^{14}$ \\
$^{113}$Sn 	& 116 	& 1037	& 617 (98\%)	& 436	& Sn	& 8$\times$10$^{11}$ \\
$^{145}$Sm 	& 340 	& 616	& 510 (91\%)	& 340	& Sm	& 2$\times$10$^{12}$ \\
$^{169}$Yb 	& 32 	& 910	& 470 (83\%)	& 304	& Yb	& 1.1$\times$10$^{15}$ \\
\botrule
\end{tabular}
\caption{Potential EC $\nu_e$ sources that can be produced by neutron irradiation in nuclear reactors. The half-life $T_{1/2}$, the $Q$-value of the reaction, the energy $E_\nu$ of the neutrino line and the corresponding branching ratio BR, as well as the maximum electron recoil energy $E_{e,\mathrm{max}}$ are shown. The achievable neutrino source intensities have been estimated for 1\,kg batches of the irradiated elements, assuming natural isotope abundances and a 10-day irradiation with a neutron flux of 5$\times$10$^{14}$\,n/cm$^2$s. Neutron-capture cross sections were taken from \cite{lbnl:web}.
}\label{Tab:OscMat}
\end{table*}

\subsubsection{Short baseline neutrino oscillations}

The survival probability of electron neutrinos in a short baseline experiment can be approximated as
\begin{eqnarray} \label{eq::pee}
P_{ee}(\ell) = 1-\sin^22\theta_{ij}\cdot\sin^2(\pi\cdot\ell/L_{ij}),
\end{eqnarray}
as long as the mixing effects are clearly disentangled due to different oscillation length $L_{ij}$. The length $L_{ij}$ can be written to
\begin{eqnarray} \label{eq::osclen}
L_{ij} = 2.48\,\mathrm m \cdot \frac{E_\nu}{\mathrm{MeV}}\,\frac{\mathrm{eV}^2}{\Delta m^2_{ji}}.
\end{eqnarray}
In the three-flavor scenario, the short baseline is $L_{23}\approx L_{13}$. Assuming the value $\Delta m^{2}_{32}=2.5\times10^{-3}$\,eV$^2$ for the mass squared difference that can be derived from the global oscillation analysis, the baseline (when expressed in meters) is approximately equal to the neutrino energy in keV:
\begin{eqnarray}\label{eq::osc0}
L_{13}\mathrm{[m]} \approx E_\nu\mathrm{[keV]},                                                                                                                                                                   \end{eqnarray}
about 3\,\% of the solar oscillation length $L_{12}$. However, if a four-flavor scenario including the sterile neutrino of the Reactor Antineutrino Anomaly (RAA\nomenclature{RAA}{Reactor Antineutrino Anomaly}) is considered, the shortest baseline is due to $\Delta m^{2}_{41}\geq 1.5$\,eV$^2$ \cite{Mention:2011rk}. In this case, the oscillation length $L_{14}$ for a EC source experiment is $\sim$1\,m.
The number of events in an differential volume d$V$ in the cylinder can be written in the following form \cite{Vergados:2010vp}:
\begin{eqnarray}\label{eq::osc1}
\mathrm d N(\ell) = \frac{N_\nu}{4\pi \ell^2} n_e \sigma(E_{\nu})p(E_{\nu},\ell,\sin^22\theta_{ij})\mathrm dV(\ell),                                                                      
\end{eqnarray}
\noindent where $N_{\nu}$ is the $\nu_e$ source intensity, $\ell$ is the distance of the detection region from the source, $n_e$ is the detector electron density ($n_e = 3\times10^{29}\,\mathrm m^{-3}$ for an LAB-based scintillator), and $\sigma(E_\nu)$ stands for the $\nu_e$-$e$ scattering cross-section at the neutrino energy $E_\nu$. The detection probability $p$ is a function of the distance to the source, the mixing angle $\theta_{ij}$ and the oscillation length $L_{ij}$:
\begin{eqnarray}
p(\ell) = 1-\chi(E_\nu) \sin^2 2\theta_{ij}\cdot\sin^2(\pi \ell/L_{ij}),
\end{eqnarray}
\noindent where $\chi(E_\nu)$ takes into account the effect of the other flavors. For sterile neutrinos, $\chi(E_\nu)=1$. The integral number of events $N_\mathrm{int}$ can be deduced from Eq.\,(\ref{eq::osc1}). It can be presented in the form:
\begin{eqnarray}\label{eq::osc2}
N_\mathrm{int} = N_0 \left( 1- g(L_{ij},h)\cdot\sin^22\theta_{ij}\right),                                                                                                        \end{eqnarray}
\noindent where $N_0$ is the expected event number without oscillations, while the fraction of ''disappearing'' neutrino events is a function of the oscillation probability and the geometric factor $g$ that depends on the fraction of the oscillation length contained inside the detector of height $h$. Eq.\,(\ref{eq::osc2}) can be used as base for a sensitivity estimate for the detection of neutrino mixing.

\subsubsection{Physics case for oscillometry}
\label{sss::oscsen}

Neutrino oscillometry offers an elegant way to address a number of questions related to neutrino oscillations: a precise determination of the mixing angle $\theta_{13}$ and of the oscillation length $L_{13}$, confirming the results of the ÒglobalÓ analysis. Beyond the standard oscillation picture, oscillometry will be very sensitive to $\nu_e$ oscillations into sterile neutrinos on the eV mass scale predicted by the RAA: Based on the observed rate deficit in reactor and radiochemical neutrino experiments, the existence of this fourth neutrino has been recently proposed in \cite{Mention:2011rk}. An oscillometric measurement in LENA will allow a precise determination of the associated mixing parameters $\theta_{14}$ and $L_{14}$. 
 
\medskip
\noindent\textbf{Mixing parameters $\theta_{13}$ and $L_{13}$.}
For a precise determination of $\theta_{13}$, an advantage of the short baseline oscillometry is the absence of matter effects. These effects cause a degeneracy in the determination of the oscillation parameters in long-baseline beam experiments (Sect.\,\ref{subsec::beam}).

\begin{figure*}
\begin{minipage}[d]{0.55\textwidth}
\includegraphics[height=0.26\textheight]{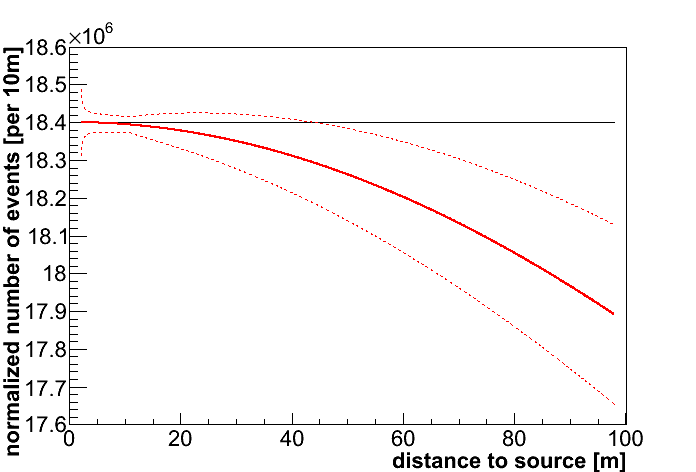}
\caption{Differential $\nu$-e scattering event rate for a measurement campaign based on a $^{51}$Cr source installed on top of LENA (5$\times$55\,d, 5\,MCi). The dashed lines indicate the statistical uncertainties (1$\sigma$) assuming a bin width of 10\,m. A correction for the solid angle has been applied, resulting in an increase of uncertainties with distance. The assumed oscillation amplitude is $\sin^22\theta_{13}$ = 0.17.}\label{Fig:OscRat}
\end{minipage}
\hfill
\begin{minipage}[d]{0.41\textwidth}
\includegraphics[height=0.26\textheight]{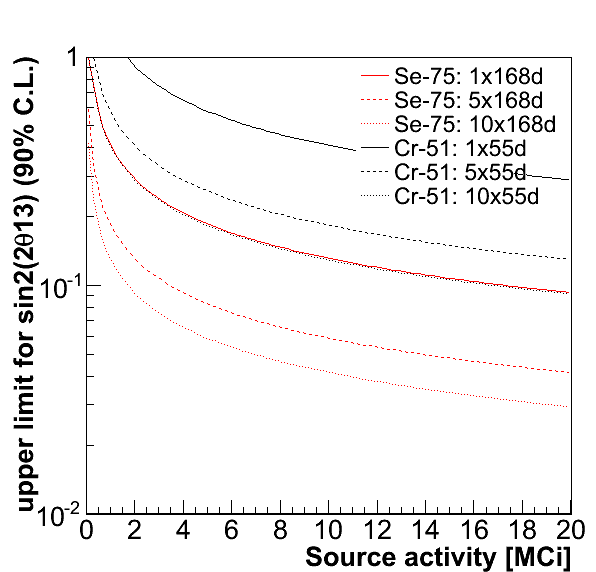}
\caption{ Upper limits for $\sin^22\theta_{13}$ (90\%\,C.L.) as a function of the initial source strength and measurement repetitions. Results for both $^{51}$Cr and $^{75}$Se are shown, considering statistical uncertainties only.\\}\label{Fig:T13Sen}
\end{minipage}
\end{figure*}

The oscillometric approach to determining $\theta_{13}$ is a measurement of the differential rate d$N$/d$\ell$ of $\nu$-e scattering events as a function of the distance $\ell$ from the neutrino source. Due to the still relatively large oscillation length of several hundred meters, such a measurement would require a strong EC source and multiple measurements: Fig.\,\ref{Fig:OscRat} shows d$N$/d$\ell$ for the case of a 5$\times$55\,days measurement campaign based on a $^{51}$Cr source of an initial activity of 5\,MCi. A detection threshold of 200\,keV is assumed (Sect.\,\ref{sss::oscbgs}). The rates have been normalized to the full solid angle. The dashed lines indicate the statistical 1$\sigma$ uncertainties on the differential rate, assuming a bin width of 10\,m: For large $\ell$, these uncertainties increase substantially due to the geometric decrease in the detected event rate with $\ell^2$.

Alternatively, the mixing angle can be determined by the integral number of events via Eq.\,(\ref{eq::osc2}): Monte-Carlo calculations return $N_{0}=3.8$$\times$10$^4$/MCi and $g=0.65\,\%$ for $^{51}$Cr (55\,days), or $N_{0}=2.9$$\times$10$^4$/MCi and $g=2.3\,\%$ for $^{75}$Se (160\,days). The achieved sensitivity is only a function of source strength and the number of measurement runs: Fig.\,\ref{Fig:T13Sen} shows 90\,\% exclusion limits for oscillations for both isotopes. $^{75}$Se reaches by far better limits due to the lower neutrino energy and therefore larger value of $g$. In this way, sensitivity to $\sin^22\theta_{13} \approx 0.1$ could be reached by five runs with a 3.5\,MCi Se-source. However, the required exposure will further increase when uncertainties introduced by the subtraction or suppression of backgrounds are considered (Sect.\,\ref{sss::oscbgs}).


In principle, also the oscillation length $L_{13}$ could be determined by oscillometry, although $-$ even for $^{75}$Se $-$ the spatial oscillation pattern is only partially contained within the extensions of LENA. The result could be compared with the neutrino energy that is usually well known  or that can be measured independently very precisely \cite{Blaum:2009eu}. For $^{51}$Cr, the neutrino energy is presently known with a precision of 0.03\,\%. Since Eq.\,(\ref{eq::osc0}) is valid if the global-analysis value of $\Delta m^2_{32} = 2.5\times10^{-3}\,\mathrm{eV}^2$ is used, this comparison will be helpful for assessment of the global analysis itself.


\medskip
\noindent\textbf{Sterile neutrinos, $\theta_{14}$ and $L_{14}$.} 
The LENA detector provides unique sensitivity for the possible fourth (sterile) neutrino that is introduced by the RAA. Since the new neutrino $\nu_s$ is sterile, its existence will manifest in a disappearance of $\nu_{e}$s into $\nu_{s}$, the amplitude being governed by the mixing angle $\theta_{14}$. The $\nu_e$ survival probability $P_{ee}(\ell,L_{14})$ is given by Eq.\,(\ref{eq::pee}). The best fit values for the mixing parameters are $\sin^22\theta_{14}=0.16$ and $\Delta m_{42}^2 \geq 1.5$\,eV$^2$ \cite{Mention:2011rk} (see also \cite{Giunti:2010jt}). 

According to Eq.\,(\ref{eq::osclen}), the oscillation length $L_{14}$ should be rather short, $L_{14}\leq 1.24$\,m for the case of $^{51}$Cr. 
Therefore, the oscillation $\nu_e\leftrightarrow\nu_s$ could be observed several times within the first 10\,m from the source. This opens an excellent possibility for direct oscillometry. It is worthwhile to note here that oscillation lengths for active and sterile neutrinos,  $L_{13} = 742$\,m  and $L_{14}=1.24$\,m (both for $^{51}$Cr) are fully disentangled and can be derived independently.

\begin{figure*}
\begin{minipage}[d]{0.55\textwidth}
\includegraphics[height=0.26\textheight]{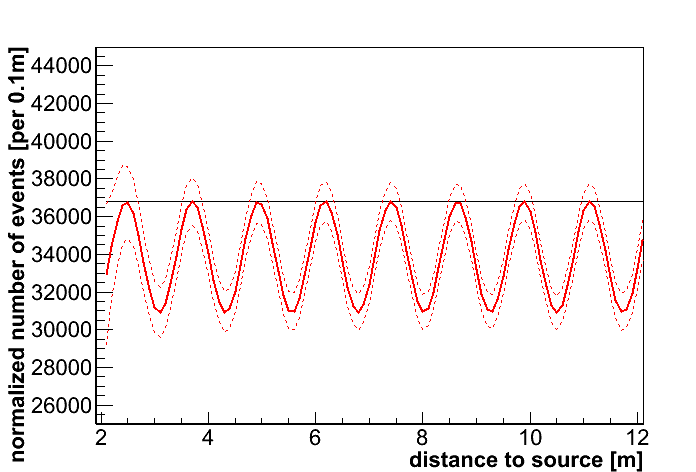}
\caption{Differential $\nu$-e scattering event rate for a 55-day run with a 5\,MCi $^{51}$Cr-source installed on top of LENA, including a correction for the solid angle. The first 10\,m are shown. The dashed lines indicate the statistical uncertainties (1$\sigma$) assuming a bin width of 0.1\,m. RAA best-fit mixing parameters are used.}\label{fig::oscratste}
\end{minipage}
\hfill
\begin{minipage}[d]{0.41\textwidth}
\centering
\includegraphics[height=0.26\textheight]{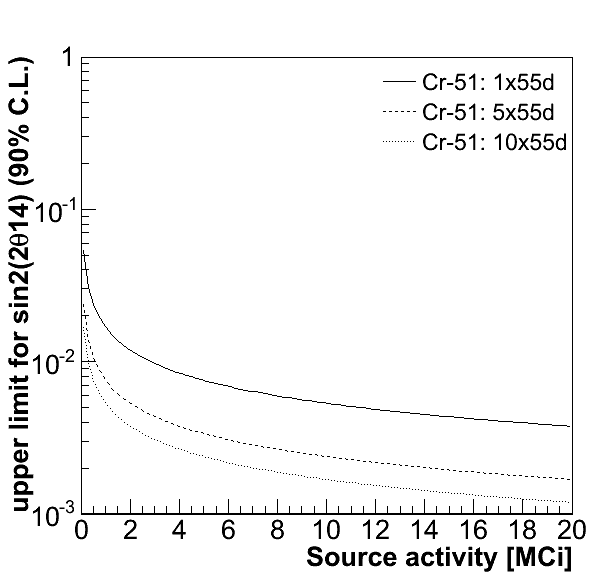}
\caption{ Upper limits for $\sin^22\theta_{14}$ (90\%\,C.L.) as a function of the initial source strength and measurement repetitions. Results for $^{51}$Cr are shown, considering statistical uncertainties only.\\}\label{Fig::T14Sen}
\end{minipage}
\end{figure*}

The differential event number d$N$/d$\ell$ as obtained from Eq.\,(\ref{eq::osc1}) is depicted in Fig.\,\ref{fig::oscratste}, assuming the best-fit RAA mixing parameters and a single 55-days run with a $^{51}$Cr source. Statistical uncertainties for a bin width of 1\,m are far smaller than the disappearance amplitude.

Like for $\theta_{13}$, we determine the sensitivity to the amplitude $\sin^22\theta_{14}$ by the integral event number in LENA, using Eq.\,(\ref{eq::osc2}). As the oscillation is fully contained within the detector, $g$ reaches the maximum value of 50\,\%. The resulting sensitivity as a function of source strength and runs is shown in Fig.\,\ref{Fig::T14Sen} for $^{51}$Cr: The high sensitivity of LENA is clearly demonstrated: a single run with a 5\,MCi source would be sufficient to exclude the best-fit value of the RAA. Similar results are expected for $^{75}$Se. 

The RAA analysis only gives a lower limit for $\Delta m_{14}^2\geq 1.5$\,eV$^2$, and therefore an upper limit to the oscillation length $L_{14}\leq1.24$\,m (for $^{51}$Cr). As long as $L_{14}$ is large compared to the spatial resolution of about 25\,cm, a precise determination of this parameter can be expected. Therefore, the sensitivity for $L_{14}$ will vanish for $\Delta m_{14}^2 \approx 7.5$\,eV$^2$. Due to the lower energy and therefore lower spatial resolution of $^{75}$Se-$\nu$ recoils, $^{51}$Cr seems the better candidate for this search.
                      


\subsubsection{Experimental uncertainties}
\label{sss::oscbgs}

While Sect.\,\ref{sss::oscsen} describes the optimum results for oscillometry achievable in LENA, an actual experiment will suffer from a number of uncertainties reducing the sensitivity. In the following, we discuss two main aspects, the uncertainty of the initial source strength and the subtraction of background events from the signal rate.  

\medskip
\noindent\textbf{Source activity.} In the calibration campaigns of GALLEX/GNO and SAGE experiments that used strong sources based on $^{51}$Cr ($\sim$2\,MCi) and $^{37}$Ar (0.4\,MCi), great care was given to an exact determination of the source activity \cite{Hampel:1997fc,Abdurashitov:2006qr}. Various methods were used, ranging from precision measurements of source weight and heat emission to direct counting of decays in aliquots of the sources. The greatest accuracy reached for $^{51}$Cr was 0.9\,\% \cite{Hampel:1997fc}, 0.4\,\% in case of $^{37}$Ar \cite{Abdurashitov:2006qr}.
While this uncertainty will not play a dominant role for the detection of large oscillation amplitudes as in the case of $\theta_{14}$, it has considerable influence if the expected effect is of the same order of magnitude, i.e. for $\theta_{13}$: However, since $L_{13}$ is known from global analysis, a precise measurement of the neutrino rate in the first 10$-$20\,m of LENA can be exploited as a normalization for the search in the remaining volume, provided the number of events in the near volume exceeds $\sim$10$^4$.

\medskip
\noindent\textbf{Background subtraction.} Independent of the used isotope, solar neutrinos pose an irreducible background for all oscillometric measurements. $^7$Be-$\nu$s will be detected at a rate of $\sim$0.5 counts per day and ton, featuring a maximum recoil energy of 665\,keV, only slightly above the spectral maximum of $^{51}$Cr. In addition, radioactive impurities inside the scintillation volume have to be considered: $^{14}$C sets the energy threshold of detection to $\sim$200\,keV, while traces of the isotopes $^{85}$Kr, $^{210}$Po and $^{210}$Bi dissolved in the scintillator will cause background contributions over the whole energy range of the source signals \cite{Arpesella:2008mt}.

In the $^7$Be analysis of Borexino \cite{Arpesella:2008mt}, most of the background contributions are eliminated by pulse-shape discrimination and a spectral fit to the signal region, separating the neutrino recoil shoulder from background spectra. A similar technique could be applied in LENA, the efficiency depending on the achieved photoelectron yield and pulse shaping properties. This analysis will be aided by the fact that the EC source can be removed from the detector, providing an exact measurement of the background rates. Nevertheless, the $\ell^2$ decrease of the signal rate will mean that the background rate will dominate in the far-region of the detector, considerably enhancing the signal rate uncertainties. This does not assail the search for sterile neutrinos that mainly concentrates on the parts of the detector closest to the source. However, it is a serious issue for $\theta_{13}$-experiments in which the oscillation signature is limited to the far-region. The feasibility of an oscillometric search will depend on the availability of strong sources, background conditions and the efficiency of spectral separation.

\subsubsection{Conclusions}

Thanks to its low energy detection threshold ($\sim$200\,keV) and considerable length ($\sim$100\,m), LENA is exceptionally well suited to perform determination of neutrino oscillation parameters. The needed electron-capture source emitting high-intensity monoenergetic and low-energy neutrinos can be produced by neutron irradiation in a nuclear reactor: currently, MCi-sources of $^{51}$Cr and $^{75}$Se seem the most promising candidates. The disappearance of electron neutrinos can be monitored over the full length of the detector by the neutrino-electron scattering event rate. However, radioactive background and the signal of solar $^7$Be neutrinos will reduce the accuracy in the far region of the detector. The resulting oscillometric curve as well as the integral event number potentially allow for an accurate determination of the mixing angles $\theta_{13}$ and $\theta_{14}$ as well as the associated oscillation lengths $L_{13}$ and $L_{14}$.  

LENA will achieve great sensitivity for the sterile neutrinos predicted by the RAA \cite{Mention:2011rk}: The best-fit mixing parameters could be conclusively tested by a single run with a 5-MCi $^{51}$Cr-source. Also $L_{14}$ can be determined precisely, provided $\Delta m_{14}^2$ is not too large. However, a search for oscillations driven by $\theta_{13}$ will be far more demanding: Multiple runs with strong sources as well as excellent detector performance and background conditions would be required to reach a sensitivity in $\sin^22\theta_{13}$ that is competitive to current reactor $\bar\nu_e$ and long-baseline experiments. 

Although not discussed here in detail, a similar experiment can be conducted with a strong antineutrino source, using e.g.~$^{90}$Sr($^{90}$Y) source with a spectral endpoint of 2.3\,MeV. While the range of $\bar\nu_e$ energies will be limited to the range from 1.8 to 2.3\,MeV due to the threshold of the inverse beta decay, this will require an analysis resolved both in space and energy to compensate the slight differences in oscillation length. However, exploiting the final-state coincidence will correspond to a strong reduction of background. This will allow to perform the search for sterile neutrinos also in the $\bar\nu$ sector \cite{Cribier:2011fv}. For $\theta_{13}$ and $L_{13}$, the oscillation baseline will be too long to show significant disappearance within the detector.


%% file: daedalus.tex

\begin{figure}[htp]
\includegraphics[width=0.48\textwidth]{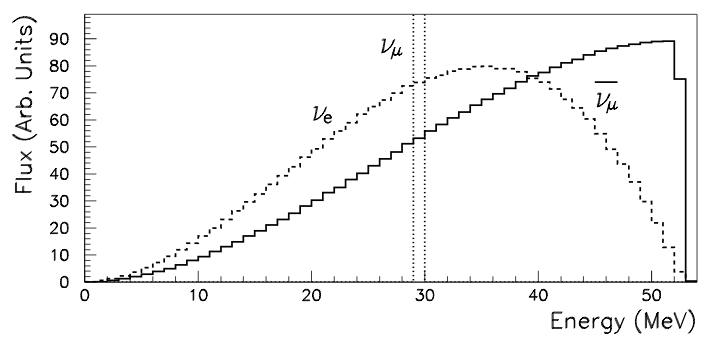}
\caption{Energy distribution of neutrinos in a $\pi$ decay at-rest beam \cite{Alonso:2010fs}.}
\label{fig::daedalusspectrum}
\end{figure}

The DAE$\delta$ALUS (Decay At-rest Experiment for $\delta_\mathrm{CP}$ studies At the Laboratory for Underground Science) concept \cite{Alonso:2010fs} proposes a neutrino oscillation experiment on three comparatively short baselines of 1.5, 8 and 20\,km. Neutrinos are created by charged pions decaying at rest, which are in turn produced by high-power synchrotrons. Via the decays
\begin{eqnarray}
\pi^+ & \to & \mu^+\nu_\mu \nonumber\\
\mu^+ & \to & e^+\nu_e\bar\nu_\mu, \nonumber 
\end{eqnarray} 
monoenergetic $\nu_\mu$ as well as spectra of $\nu_e$ and $\bar\nu_\mu$ are generated (Fig.\,\ref{fig::daedalusspectrum}). The $\bar\nu_\mu$ energies range to a kinematic maximum of $\sim$50\,MeV, matching the relatively short oscillation baselines. The neutrinos propagate from three locations at different distances to a single large-volume detector. The sought-for signal is the appearance of $\bar\nu_e$ from $\bar\nu_\mu\to\bar\nu_e$ oscillations driven by the small mixing angle $\theta_{13}$. Most importantly, the relative rates observed for the medium and far baselines depend on the size of the CP-violating phase $\delta_\mathrm{CP}$.

The original proposal foresees the LBNE Water Cherenkov detector(s) for the $\bar\nu_e$ appearance measurement. However, LENA features an intrinsic capability for the identification of inverse beta decay events, offering excellent background discrimination for this channel. The necessary cross-calibration of the neutrino intensities from the three $\pi$-decay sources will be possible via neutrino-electron scattering and charged-current reactions on $^{12}$C. Therefore, a combination of the DAE$\delta$ALUS and LENA programs seems a promising alternative (or extension) to the discussed long-baseline neutrino beam scenarios (Sect.\,\ref{subsec::beam}). However, detailed calculations on the event and background rates as well as the expected sensitivity are needed.

%% file: doublebeta.tex

%

\noindent The huge amount of instrumented mass provided by LENA might open the possibility of a large neutrino-less double-beta ($0\nu2\beta$\nomenclature{$0\nu2\beta$}{Neutrino-less Double Beta decay}) decay experiment, based on $^{136}$Xe dissolved in the liquid scintillator. The solubility in organic liquid scintillators of $^{136}$Xe at atmospheric pressure is about 2\,\% in weight, allowing potentially an experiment with 200 tons of active mass or more. The energy resolution is a very crucial parameter in $0\nu2\beta$ experiments. Therefore, a more dense PMT\nomenclature{PMT}{PhotoMultiplier Tube} coverage might be required, at least in the central region of the detector. Further studies are needed to asses the real sensitivity of such an experiment. Nevertheless, this seems the only realistic way for $0\nu2\beta$ experiments at the 100\,t scale, which arguably would be able to attack the normal hierarchy region of neutrino masses.

%% file: rareproc.tex

%

\noindent The unique expected characteristics of the detector such as ultra low background and large target mass provide the possibility of search for others low energy rare processes \cite{Derbin:2005nc}. They are searches for nucleon decay into invisible channels $N\rightarrow 2\nu$ and $NN\rightarrow 2\nu$ \cite{Back:2003wj}, solar neutrino decay $\nu_H\rightarrow \nu_L + \gamma$ \cite{Derbin:2002mf}, heavy neutrino emission in $^8\rm{B}$-beta decay \cite{Back:2003ae} and high energy solar axions \cite{Bellini:2008zza,Bellini:2012}. Pauli Exclusion Principle in nuclei
\cite{Bellini:2009jr} and electron stability \cite{Back:2002xz} can be tested at more high level. The precise measurements of solar neutrino-electron elastic scattering and excitation of nuclear levels by neutral currents allow to study non standard neutrino interactions \cite{Berezhiani:2001rt,Barranco:2011wx,Pospelov:2011ha}.

%% file: pdecay.tex

%

\noindent Due to the large target mass and the intended long measurement time, LENA offers the opportunity to search for nucleon decays. Currently, the best limits on proton lifetime are hold by Super-Kamiokande \cite{Nishino:2009gd,Kobayashi:2005pe}, and it seems not likely that LENA will substantially improve the limit for $p\rightarrow\pi^0e^+$. However, the sensitivity for the decay mode $p\rightarrow K^+\bar\nu$ is an order of magnitude larger than in water Cherenkov detectors. Moreover, the search in LENA is expected to be background-free for about 10 years, allowing to set a limit of $\tau_{p}>4$$\times$10$^{34}$\,yrs (90\,\% C.L.)  if no event is observed after this period \cite{MarrodanUndagoitia:2006qn}. This already probes a significant fraction of the proton lifetime range predicted by SUSY theories \cite{Pati:2003qia,Lucas:1996bc}.

\subsubsection{Theoretical predictions}

In the standard model of particle physics, protons are stable. This is a consequence of the baryon number (B) conservation which has actually been introduced empirically into the model. It is interesting to realize that there is no fundamental gauge symmetry which generates the conservation of B. For this reason, the validity of B-conservation can be considered as an experimental question. However, several theories beyond the standard model actually predict an instability of the proton:

\medskip
\noindent\textbf{GUT SU(5).} In the minimal Grand Unified Theory\nomenclature{GUT}{Grand Unified Theories} SU(5), $M_x \sim 10^{15}$\,GeV/c$^2$, the predicted lifetime is $\tau_{p \to \pi^0e^+}= 10^{29}$\,years. The first generation of large water-Cherenkov detectors motivated by this prediction observed no evidence of proton decay in the $p \to \pi^0 e^+$ mode and therefore ruled out the model. The lifetime of the proton largely depends on the mass scale of the super-heavy particles mediating the decay process ($X$ and $Y$ bosons). Further extensions of the SU(5) model predict a longer proton-decay lifetime with a larger uncertainty, typically from 10$^{30}$ to 10$^{36}$~years \cite{Perez:2008ry}.

\medskip
\noindent\textbf{GUT SO(10).} The proton lifetime predicted by the SO(10) extension of the SU(5) model is around 10$^{32\pm1}$\,yrs for non-supersymmetric models and 10$^{34\pm1.5}$\,yrs if Supersymmetry is included \cite{Shaban:1992vv}.

\medskip
\noindent\textbf{SUSY SU(5).} In the minimal supersymmetric\nomenclature{SUSY}{SUper SYmmetry} SU(5) model, the dominant decay modes of the proton involve pseudo-scalar bosons and anti-leptons \cite{Nath:2006ut}:
\begin{equation}
K^+\bar\nu ,\ \pi^+\bar\nu,\ K^0e^+,\ K^0\mu^+,\ \pi^0e^+ ...
\end{equation}
where the relative strengths depend on the specific exchange of the SUSY particles involved. However, in most of the models the proton-decay channel $p \to K^+\bar\nu$ is favored \cite{Nath:2006ut,Arnowitt:1985iy,Pati:2003qia,Lucas:1996bc}. The predictions concerning the lifetime of the proton are in the order of $10^{33}$ to 10$^{34}$ years \cite{Pati:2003qia,Lucas:1996bc}.

\subsubsection{Detection mechanism}
\label{sec::pddetection}

Within the target volume of LENA, about 1.6$\times$10$^{34}$ protons, both from carbon and hydrogen nuclei, are candidates for the decay. This number has been calculated for PXE\nomenclature{PXE}{Phenyl-Xylyl-Ethane, organic solvent} (Sect.\,\ref{subsec::scintillator}), which is assumed as reference in the following. As all decay particles must be contained inside the active volume, the fiducial volume is about 5\,\% smaller. In the case of LAB, the proton number will be further reduced by about 6\,\% due to its lower density (see also Tab.\,\ref{tab::solventproperties}). 

In the case of protons from hydrogen nuclei ($\sim$0.25$\times$10$^{34}$ protons in the fiducial volume of LENA), the proton can be assumed at rest. Therefore, the proton decay $p\to K^+\bar\nu$ can be considered as a two-body decay problem, where $K^+$ and $\bar\nu$ always receive the same energy. The energy corresponding to the mass of the proton, $m_{p}=938.3$\,MeV is thereby given to the decay products. Using relativistic kinematics, it can be calculated that the particles receive fixed kinetic energies, the antineutrino 339\,MeV and the kaon 105\,MeV. 

The antineutrino escapes without producing any detectable signal. However, the large sensitivity of LENA for this decay channel arises from the visibility of the ionization signal generated by the (kinetic) energy deposition of the kaon. A water Cherenkov detector is blind to this signal as the kaon is produced below the Cherenkov threshold in water; only the secondary decay particles are visible, greatly reducing the detection sensitivity.

In LENA, the prompt signal of the decelerating kaon is followed by the signal arising from the decay particle(s):  After $\tau_{K^+}=12.8$\,ns, the kaon decays either by $K^+\to\mu^+\nu_\mu$ (63.43\,$\%$) or by $K^+\to\pi^+\pi^0$ (21.13\,\%). In 90\% of these cases, the kaon decays at rest\,\cite{Hayato:1999az}. If so, the second signal is again monoenergetic, either corresponding to the 152\,MeV kinetic energy of the $\mu^+$ or 246\,MeV from the kinetic energy of the $\pi^+$ and the rest mass of the $\pi^0$ (which decays into two gamma rays creating electromagnetic showers). A third signal arising from the decay of the muon will be observed with a large delay ($\tau_{\mu^+}=2.2$\,\textmu s). A more detailed discussion can be found in \cite{Undagoitia:2005uu}.

If the proton decays inside a carbon nucleus ($\sim$1.2$\times$10$^{34}$ protons in the fiducial volume), further nuclear effects have to be considered. First of all, since the protons are bound to the nucleus, their effective mass will be reduced by the nuclear binding energy $E_b$, 37\,MeV and 16\,MeV for protons in s-state and p-states, respectively. Secondly, decay kinematics will be altered compared to free protons due to the Fermi motion of the proton. The Fermi momenta in carbon have been measured by electron scattering on ${}^{12}$C \cite{Nakamura:1976mb}. The maximum momentum is about 250\,MeV/c. A range for the effective kinetic energy of the kaon has been derived by Monte Carlo simulations: $(25.1-198.8)$\,MeV for the s-state and $(30.0-207.2)$\,MeV for the p-state \cite{Undagoitia:2005uu}. 

In any case, the experimental signature of the proton decay in LENA is not substantially affected by nuclear effects or the kaon decay mode: A coincidence signal arising from the kinetic energy deposited by the kaon and from the delayed energy deposit of its decay particles will be observed.

\subsubsection{Background rejection}

The main background source in the energy range of the proton decay are atmospheric muon neutrinos $\nu_\mu$. Via weak charge-current interactions, these $\nu_\mu$ create $\mu$ inside the detector, a substantial fraction in the energy range relevant for the proton decay search. Moreover, additional kaons can be produced in deep inelastic scattering reactions at higher $\nu_\mu$. In the following, the arising background rates and possible rejection cuts are shortly outlined. For a thorough discussion, see \cite{Undagoitia:2005uu}.

\medskip
\noindent\textbf{Muon events.} The rate of muon events from atmospheric neutrinos in the relevant energy range can be derived from Super-Kamiokande measurements \cite{Hayato:1999az}. At Pyh\"asalmi, the rate corresponds to 1190.4 $\nu_\mu$-induced muons per year \cite{Undagoitia:2005uu}.

\begin{figure}[tbp]
\begin{center}
\includegraphics[width=0.46\textwidth]{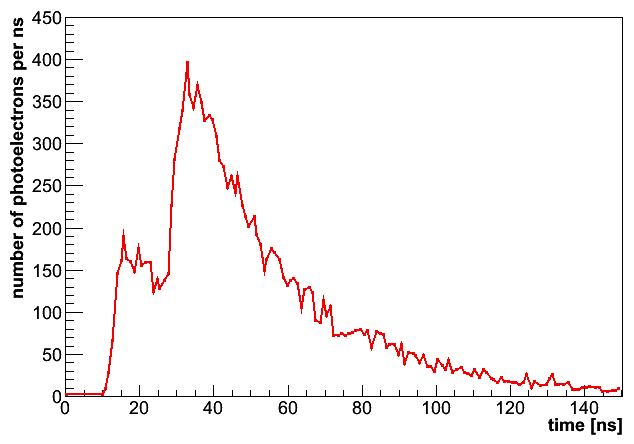}
\caption{Signature of the proton decay into kaon and antineutrino in LENA, based on the analog sum of all PMT channels. The prompt signal is generated by the deceleration of the kaon. After a delay of (in this case) 14\,ns, the second peak is caused by its decay particles \cite{Undagoitia:2005uu}.}
\label{pic::pdecay}
\end{center}
\end{figure}

In order to distinguish the real proton decay signals from muon background events, a pulse shape analysis can be applied. MC simulations show that the kaon deposits its energy within 1.2\,ns, leading to a fast but resolvable coincidence with the kaon decay products after $\tau_{K^+}$ (Sec.\,\ref{sec::pddetection}). A typical time profile is shown in Fig.\,\ref{pic::pdecay}. This double signature can be used to discriminate atmospheric $\nu_\mu$ events as long as the kaon decay is sufficiently delayed to produce a discernible double signal, i.e.\,the delay is large compared to the time resolution of the detector.

In the analysis presented in \cite{Undagoitia:2005uu}, signal and background events are discriminated via the signal rise time. No time-of-flight correction is applied. Based on 2$\times$10$^4$ proton decay and muon events in the relevant energy regime, an analysis cut can be defined that rejects all muons and retains a detection efficiency of $\varepsilon_{p}\approx 65$\,\% for proton decay. The sensitivity $\varepsilon_{p}$ is an order of magnitude larger than the one obtained in the Super-Kamiokande analysis, corresponding to a similar increase in the proton lifetime limit. 

The corresponding background rejection efficiency is at least $\varepsilon_\mu \geq 1-5$$\times$10$^{-5}$. This results in an upper limit of $\sim$0.05 muon events per year that are misidentified as proton decay events.

The prominence of the double-peak signature generated by the kaon and decay particles depends on the performance of the liquid scintillator. The two crucial parameters are the optical transparency, especially the scattering length, and the fast fluorescence time constant $\tau_1$ discussed in Secs.~\ref{subsec::scintillator} and \ref{subsec::baspar}. Moreover, there is also a dependence on the position of the event within the detection volume. The studies performed in \cite{Undagoitia:2005uu} used a fixed value of $\tau_1=3.5$\,ns (similar to PXE) and varied the light propagation parameters as well as the vertex position. Assuming reasonable values for absorption and scattering lengths, $\varepsilon_{p}$ is changing only on a level of a few percent. An LAB-based scintillator might feature a slightly lower efficiency due to the larger $\tau_1=5.2$\,ns. A study performed using a pessimistic value of $\tau_1=6$\,ns returned an efficiency of $\varepsilon_{p}= 58$\,\% \cite{Marrodan:2009}.

\medskip
\noindent\textbf{Hadronic event.} In case of charged current reactions of atmospheric $\nu_\mu$'s at larger energies, hadrons can be produced along with the final state muon. These events are dangerous if they are able to mimic the double signature of the proton decay. While this is not the case in pion and hyperon production, interaction modes creating an additional kaon in the final state may be mistaken as signal events \cite{Drakoulakos:2004gn}. In principle, these events can be discriminated by the additional decay electron of the muon created in the CC reaction. However, this signal is sometimes covered by the muon signal itself: Monte Carlo simulations return an upper limit of 0.06 irreducible background events per year for this channel.

\subsubsection{\label{ProtonDecaySensitivity}Proton decay sensitivity}

Based on the efficiencies of the rise time cut, the sensitivity of LENA for the proton decay search can be determined. The observed activity due to proton decays is given by the expression:
\begin{equation}
 A=\varepsilon_{p} N_{p}t_{m} / \tau_{p} 
\end{equation}
where $\varepsilon_{p}=0.65$ is the efficiency, $N_{p}=1.45$$\times$10$^{34}$ is the number of protons in the fiducial volume, $t_{m}$ is the measurement time and $\tau_{p}$ is the lifetime of the proton.

If the proton lifetime corresponds to the current best limit to this channel by Super-Kamiokande, ($\tau_{p}=2.3\cdot10^{33}$\,yrs)~\cite{Kobayashi:2005pe}, about 40.7 proton decay events will be observed in LENA in a measurement time of ten years.

For establishing a new lower limit on proton lifetime, the number of background events observed over the measurement time is the main issue. Combining the expected background rates from atmospheric neutrino-induced muon and kaon production, a rate of 0.11 background events per year or 1.1 events in 10 years can be obtained. This result is an upper limit on the expected background rate \cite{Undagoitia:2005uu}.

In case there is no signal observed in LENA within these ten years, the lower limit for the lifetime of the proton will be placed at $\tau_{p}>4$$\times$10$^{34}$\,yrs at 90\,\% C.L. If one candidate is detected, the lower limit will be reduced to $\tau_{p}>3$$\times$10$^{34}$\,yrs (90\,\% C.L.), featuring a 32\,\% probability that this event is due to background \cite{Undagoitia:2005uu}.

\subsubsection{\label{Conclusions}Conclusions}

The MC studies carried out in \cite{Undagoitia:2005uu} determine the efficiency for the proton decay search in LENA to $\sim$65\%. Based on this, a new lower limit for the proton lifetime of $\tau>4$$\times$10$^{34}$\,yrs (at $90\%$ C.L.) can be reached if no proton decay event is observed within ten years. The high efficiency is based on the  distinct pulse shape of the proton decay mode $p\to K^+\bar\nu$ in LENA. Since the values predicted by the favored theories for the proton decay in this channel are of the order of the value resulting from this analysis \cite{Pati:2003qia,Lucas:1996bc}, it is obvious that LENA measurements would have a deep impact on the proton decay research field.

LENA might also provide relevant sensitivity levels to other nucleon decay channels. While the analysis presented here is independent of the tracking capabilities of the detector (Sect.\,\ref{subsec::tracking}), in others (e.g.~$p\to\pi^0e^+$) the possibility of reconstructing the decay vertex might be necessary to discriminate background signals. However, these aspects require further studies.

%% file: beam.tex

%

\noindent Accelerator-based neutrino beam experiments might prove to be the only viable way to determine
the value of the CP-violating phase $\delta_\mathrm{CP}$ in the leptonic sector
and the neutrino mass hierarchy. The oscillation baselines discussed today range from hundreds to 
thousands of km, corresponding to GeV neutrino energies. 
In spite of its focus on low-energy neutrinos, LENA might serve as a far detector for such an experiment.
We review general properties of future beam experiments and discuss both a conventional neutrino beam to Pyh\"asalmi and a beta-beam to Fr\'ejus in the context of the tracking capabilities of LENA.

\subsubsection{Concept and goals}
During the last decade we witnessed drastic changes in our understanding of neutrinos. A number of experiments have shown that neutrinos violate lepton flavor through oscillations and therefore must have mass. But despite the large progress many fundamental questions remain unanswered. Some of these questions can be answered by sending an artificial neutrino beam over a long distance (several 100 km) to LENA. This is the subject of this chapter. The questions that can be answered with a neutrino-beam are:
\begin{itemize}
  \item What is the value of the last unknown mixing angle $\theta_{13}$\,\footnote{There are first 
        indications for a non-zero value of {$\theta_{13}$} from the T2K, Minos, DoubleChooz 
        experiments. Even if these indications turn out to be valid, LENA 
        will still provide one of the most precise measurements of {$\theta_{13}$}.}?
  \item What is the hierarchy of the neutrino masses (sign of $\theta_{23}$)?
  \item Is the mixing angle {$\theta_{23}$} maximal?
  \item Do neutrinos violate the CP symmetry?
\end{itemize}
A number of other experiments are trying to find the value of {$\theta_{13}$}. 
These are reactor neutrino experiments (DoubleChooz, DayaBay, Reno) and long-baseline experiments 
(Minos, T2K). First indications of a value between 5 and 10 degrees have been found. 
The value is important to see the pattern of neutrino mixing and to understand the sensitivity of most 
other measurements. 
Even if measurements are achieved in the near future, LENA will still provide one of the most precise 
measurement of {$\theta_{13}$} and similarly of {$\theta_{23}$}. 
To our current knowledge {$\theta_{23}$} is consistent with maximum mixing for 45$^{\circ}$. 
Here precision is important to understand whether {$\theta_{23}$} is exactly maximal or 'just' 
accidentally close to 45$^{\circ}$. 
Furthermore we can determine the sign of $\sin \theta_{23}$ and therefore the mass hierarchy.

But the most important goal from these measurements is the search for CP-violation in neutrino oscillations. Today we know that two of the three mixing angles are substantially larger in neutrino mixing compared to the quark mixing, allowing for much larger CP-violation in the lepton sector. The relevant quantity -- Jarlskog's determinant -- is approximately {$4 \cdot 10^{-5}$} for the quarks and {$0.028\sin \delta$}for the leptons, if we assume a value of 5$^{\circ}$ for {$\theta_{13}$}, not far below the current limits. The CP-violating phase {$\delta$} is unknown today. Its measurement will be the prime goal.

Moreover, matter effects modify the survival probabilities of $\nu$ and $\bar\nu$ for beams over very long distances. The sign of this change can in principle be exploited to determine the neutrino mass hierarchy. However, the combined effects of CP violation and matter might be hard to disentangle.

\medskip\noindent Artificial neutrino beams are produced from the decay of certain unstable particles, emitting neutrinos in their decay. High intensity beams of the mother particle are produced and directed towards LENA, creating a more or less collimated neutrino beam. The neutrino beam travels through the earth while all other beam particles are absorbed. The technologies to produce neutrino beams may be split into three classes.

\medskip\noindent In so-called \textbf{conventional neutrino beams} a high intensity proton beam is directed onto a target. In the induced hadronic interactions pions are produced which subsequently decay as {$\pi \to \mu \nu_\mu$}. Only the charged pions 
contribute. Magnetic horns focus pions of one charge in the direction of LENA and defocus the other charge. By switching the polarity of the horns one may choose between neutrinos $\nu_\mu$ and anti-neutrinos {$\bar\nu_\mu$}. The beam is broad in energy and has a small contamination from electron-neutrinos.\\

\medskip\noindent \textbf{Beta-beams} emerge from the decay of radioactive ions which are accelerated and stored in a storage ring with a straight section pointing towards the detector. Accelerating {$\beta^+$} emitters produces a beam of $\nu_e$s and 
{$\beta^-$} emitters $\bar\nu_e$s. Producing these ions in sufficient quantity is a technological challenge. An accelerator complex is needed to accelerate the ions to high energies. The beam is well focused, broad in energy and pure in flavor.\\

\medskip\noindent A \textbf{neutrino factory} produces pions in the technology of the conventional neutrino beam. But now the muons from the decay of the pions are captured, reformed into a beam and accelerated. From the decay of negative muons ($\mu^- \to e^- \bar\nu_e \nu_\mu$) a neutrino beam with two flavors is created. Accelerating positive muons creates the opposite flavors. The beam is higher in energy, somewhat less focused and pure in its two flavors.

\medskip\noindent The first two technologies are described in more details in the following sections. The neutrino factory is not pursued further as it needs a magnetized detector to distinguish $\nu$ from $\bar\nu$ interactions. It is unlikely that LENA will be magnetized because of the negative effect on the PMT performance. There might be the possibility to use recoil neutrons and protons to identify $\nu$ and $\bar\nu$. However, further MC studies are needed on this aspect.

\subsubsection{Conventional neutrino beam}

For the beam source, the most evident candidate is SPS\nomenclature{SPS}{Super Proton Synchrotron} at CERN, producing a beam of 400 GeV protons. The current maximum power is 300\,kW, a 400\,kW update is intended. Using the planned PS2 with SPS will also permit to increase the proton power to 1.2\,MW (by 2016). In case of other future upgrades in the proton production chain, other options may be available. To 
produce neutrinos of 3--5 GeV by pion (and kaon) decay we need at least 20 GeV protons. Larger energies may produce more flux, but on the other hand the high-energy tail may induce more neutral current background.
 
\begin{figure}[tbp]
\includegraphics[width=0.43\textwidth]{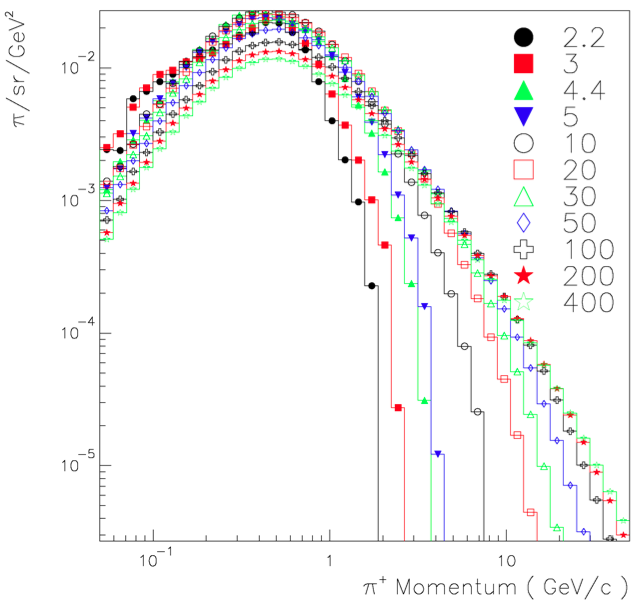}\hspace{12pt}
\caption{Simulated pion spectra (flux/proton energy), for different proton energies from \cite{Ferrari:2002yj}. To get 4 GeV neutrinos we need at least 10 GeV pions and consequently 15 GeV protons, though at least 20 GeV proton beam would be preferred. 5 GeV proton beam will give neutrinos of 1--2 GeV and below.}
\label{fig::pionspectra}
\end{figure}

\begin{figure*}[tbp]
\begin{center}
\includegraphics[angle=0, width=0.6\textwidth]{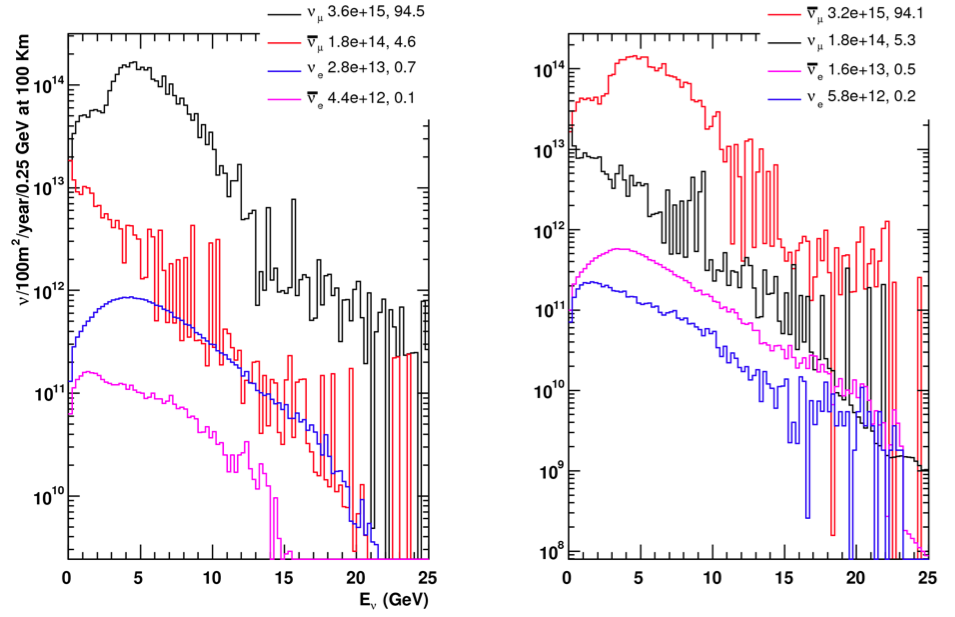}
\includegraphics[angle=0, width=0.6\textwidth]{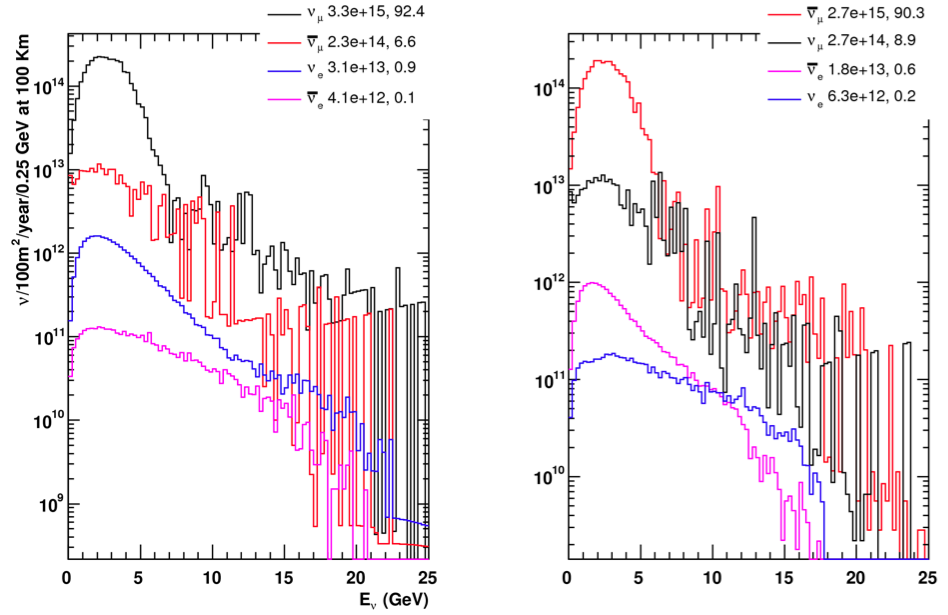}
\caption{Recent neutrino spectra from a conventional beam by Longhin. Left column: neutrino mode; right coulumn: anti-neutrino mode. The configuration in the upper plots is optimized for long-distances 
($>2000$ km) and the lower plots for medium-distance (1000 km). The flux of each flavour is indicated on
the plot in units of neutrinos per {$100 m^2$} and year at a distance of 100 km from the target and as
a relative fraction of the total flux in percent.}
\label{fig::beamspectra}
\end{center}
\end{figure*}

Neutrinos are produced via decays of pions and kaons whose fractions are typically 10 pions for one kaon. The decay modes are:
\begin{equation}
\begin{array}{rcll}
\pi^+ & \to & \mu^+ \nu_\mu & (99.98 \%)   		\nonumber\\
      &     & e^+ \nu_e     & (0.01 \%) 			\nonumber\\
      \nonumber \\
K^+  & \to & \mu^+ \nu_\mu       & (63.4\,\%)   \nonumber\\
     &     & e^+ \nu_e           & (0.0015\,\%) \nonumber\\      
     &     & \pi^0 e^+ \nu_e     & (5\,\%)      \nonumber\\
     &     & \pi^0 \mu^+ \nu_\mu & (3\,\%)      \nonumber\\      
\end{array}
\end{equation}
Neutrinos are also produced by the muon decay, though this is mainly background:
\begin{eqnarray}
\mu^+ &\to& e^+ \nu_e \overline{\nu}_\mu \nonumber
\end{eqnarray}
Independent of the initial proton energy, the $\pi$ spectra are always peaked at just below 500 MeV (Fig.\,\ref{fig::pionspectra}). The flux depends quite linearly on both proton energy and luminosity. The resulting $\nu$ spectra feature a maximum energy of $E_{\rm max}= 0.43 E_\pi$ or $E_{\rm max} = 0.95 E_K$, respectively. The shape of the $\nu$ spectrum in forward direction depends also on the magnetic focusing system for $\pi$/K's. It determines the number and spectrum of the pions entering the decay pipe. Optimizations and simulations for the beam are being made elsewhere \cite{Longhin:2009zz,Bernabeu:2010rz}. Fig.\,\ref{fig::beamspectra} depicts some sample spectra from those simulations.

Using the CNGS beam as an example, the $\nu_\mu$ beam is 97\,\% pure, with small admixtures of $\nu_e$ (1\,\%), 
$\bar\nu_e$ (0.1\,\%) and $\bar\nu_\mu$ (2\,\%). The $\bar\nu_\mu$ would feature equal but opposite contamination. The beam will be run in two phases, one with positive focus, (CP$^+$) and other with negative focus (CP$^-$). Asymmetrical running times might be used, e.g.\,2\,yrs ($\nu$) and 6\,yrs ($\bar\nu$), because of the lower cross section of $\bar\nu$s.

\begin{table}
\centering
\begin{tabular}{lcccc}
\toprule
accelerator & \multicolumn{3}{c}{SPS} &PS2\\
& old & upgrade & with PS2 & \\
\colrule
beam enegy [GeV] & 400 & 400& 400 & 50\\
pot\nomenclature{pot}{Protons On Target} [10$^{19}$/y] &7.6 & 11 & 33 & \\
$E_p N_{pot}$ [10$^{22}$\,GeV\,pot\,yr] & 3 & 4.4 & 13.2 & \\
Beam power [MW] & 0.3 &  0.4  & 1.2 & 0.4 \\
\botrule
\end{tabular}
\caption{Assumed properties of potential accelerators to be used as the neutrino source: The currently running SPS (old), a possible upgrade of SPS, SPS combined with PS2 (planned for 2016), and the PS2 on its own.}
\end{table}

\subsubsection{Beta-beams}

For beta-beams, the choice of isotopes depends on the baseline. Potential ions are listed in Tab.\,\ref{beta-beam-isotopes}. {$^6$He} and {$^{18}$Ne} are considered in case of the short baseline from CERN to Fr\'ejus. In contrast, {$^8$Li} and {$^8$B} are the best choice for a large baselines such as CERN to Pyh\"asalmi as both isotopes feature relatively high $Q$-values of about 13\,MeV. The maximum neutrino energy is given by the relativistic {$\gamma$} factor at production times $Q$. For example, with the SPS (450 GeV {$p^+$} energy), the baseline for a detector at the first oscillation maximum is 1100\,km and 2100\,km for the two isotopes. The baseline from CERN to Pyh\"asalmi is 2300\,km. It is obvious that a beta beam to Pyh\"asalmi would be technically very demanding, and we primarily consider the CERN-Fr\'ejus baseline in the following.

\begin{table}
\centering
\begin{tabular}{ccccccccc}
\toprule
type & isotope & Z & A & A/Z &{$T_{1/2}$}&{$Q_\beta$}&{$\langle E^\ast_\nu \rangle$}&{$\langle E^{lab}_\nu\rangle$}\\
        & & &   &     &   \small{[sec]}     &  \small{[MeV]}      &   \small{[MeV]}         &   \small{[GeV]}                      \\
\colrule
($\beta^+$) & {$^8$B} & 5 & 8 & 1.6 & 0.77      & 13.9      &  7.37                        & 4.15 \\
& {$^{18}$Ne} & 10&18 & 1.8 & 1.67      & 3.4       &  1.86                        & 0.93 \\
\colrule
($\beta^-$) & {$^6$He} & 2 & 6 & 3.0 & 0.81      & 3.5       & 1.94                         & 0.58 \\
 &{$^8$Li} & 3 & 8 & 2.7 & 0.84      & 13.0      & 6.72                         & 2.27 \\        
\botrule
\end{tabular}
\caption{Potential isotopes for the creation of a beta-beam to LENA. {$\langle E^\ast_\nu \rangle$} is measured in the rest frame of the 
decaying isotope.}
\label{beta-beam-isotopes}
\end{table}

\begin{figure*}[tbp]
\begin{center}
\includegraphics[width=0.85\textwidth]{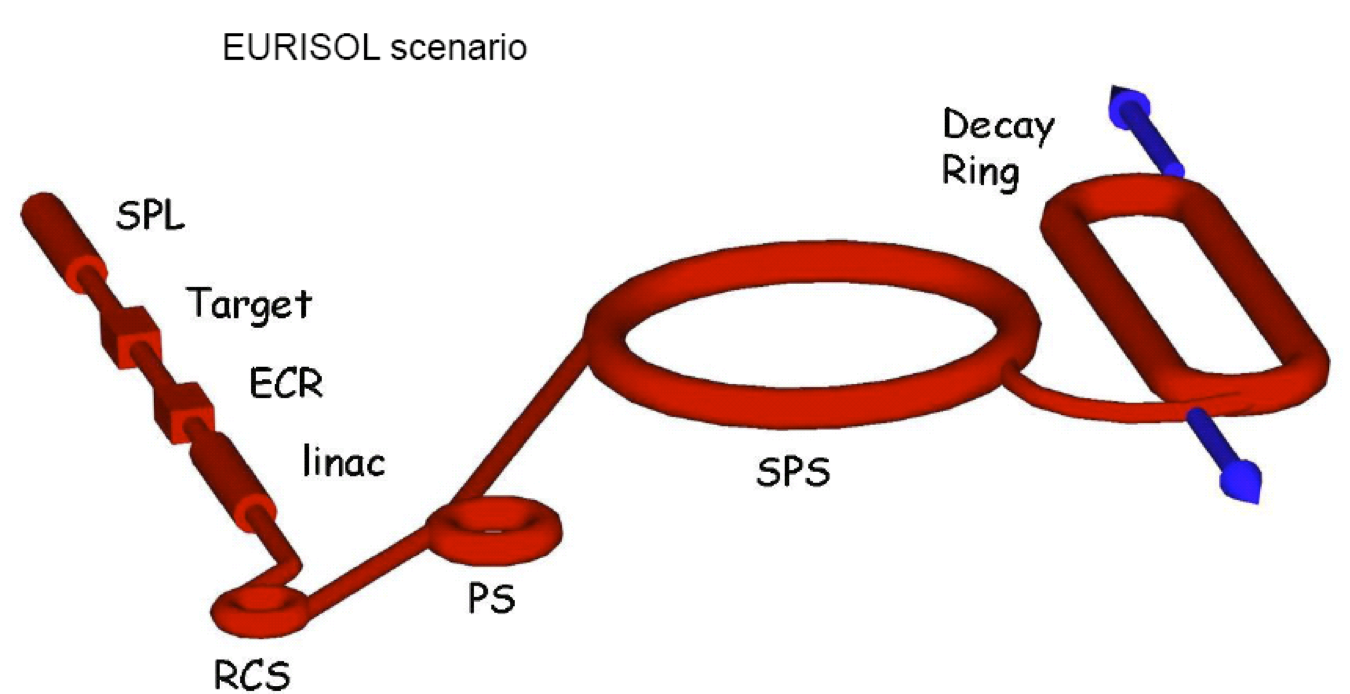}
\caption{Conceptual layout of a beta-beam facility at CERN from the EURISOL design study.}
\label{beta-beam-cern}
\end{center}
\end{figure*}

Fig.\,\ref{beta-beam-cern} shows the layout of a beta-beam facility at CERN \cite{Bernabeu:2010rz}. Radioactive ions are produced as neutral gas, ionized in an ECR\nomenclature{ECR}{Electron Cyclotron Resonance souce} source and accelerated. The acceleration starts with a LINAC\nomenclature{LINAC}{LINear ACcelerator} and a rapid cycling synchrotron and continues with the existing PS and SPS machines. Finally, the ions are injected into a decay ring with a straight section to the detector. Ions are continuously injected into the decay ring which is running at fixed energy. The intensity goal is to direct in the order of $10^{18}$ neutrinos to the detector per year. 

Even for short baselines, the beta-beam concept poses technological challenges. The biggest is the production, collection
and ionization of a sufficient number of isotopes. A number of different concepts are discussed. The ISOL method is considered in the EURISOL\nomenclature{EURISOL}{European Isotope Separation On-Line radioactive ion beam facility} study, production in a ring was proposed in \cite{Rubbia:2006pi}, and for some isotopes direct production with a deuteron beam is possible \cite{Hass:2008zza}. Other challenges are the injection and storage of such a 
large number of ions in the decay ring especially for rings with high {$\gamma$} and the collimation of the decay losses in the accelerators.

\subsubsection{Synergies and perspectives}

At the time LENA will start data taking, the value of $\theta_{13}$ might be already established. If it is not too small, LENA might stand a good chance to discover CP-violation by determining the phase $\delta_\mathrm{CP}$ in the PMNS matrix. In a long-baseline oscillation experiment, $\delta_\mathrm{CP}$ exhibits itself by different oscillation probabilities for $\nu$ and $\bar\nu$. 

\medskip\noindent\textbf{Conventional Beam to Pyh\"asalmi.} In a conventional $\nu_\mu/\bar\nu_\mu$ beam, {$\mathcal{P}\left(\nu_\mu \to \nu_e \right)$} will be different from {$\mathcal{P}\left(\bar\nu_\mu \to \bar\nu_e \right)$}. To discover this difference, the beam is operated for a certain amount of time with $\nu$s and then a matching amount with $\bar\nu$s. For Pyh\"asalmi, the beam will have to travel along a relatively long baseline ($>$1000\,km). In this case, oscillation probabilities in the far detector are changed by the Earth matter effect: In fact, $\nu_e$ and $\bar\nu_e$ are affected differently due to their different interaction cross sections with the electrons contained in terrestrial matter. The sign of the changes depends on the neutrino mass hierarchy. This is a disadvantage in view of a clear determination of $\delta_\mathrm{CP}$, but also offers the opportunity to discover both CP violation and mass hierarchy at the same time. The possible degeneracy might be resolved by using two different baselines or a common analysis with other experiments.

The conventional beam is an appearance experiment $\nu_\mu\to\nu_e$ at the far detector. As presented in Sect.\,\ref{subsec::tracking}, LENA features excellent flavor identification and better than 5\,\% energy reconstruction for energies above 1\,GeV. However, backgrounds due to NC $\pi^0$ production play an important role in determining the sensitivity for $\delta_\mathrm{CP}$ and the mass hierarchy. Further studies on the discrimination of this background are necessary.

\medskip\noindent\textbf{Beta-Beam to Fr\'ejus.} Alternatively, if one is mainly interested in 
$\delta_\mathrm{CP}$ or $\theta_{13}$ turns out to be small, a beta-beam over the short distance 
from CERN to Fr\'ejus might prove the better option. For a beta-beam, $\theta_{13}$ is found by 
the appearance signal of $\nu_\mu$s at the far detector, and $\delta_\mathrm{CP}$ by the comparison 
of the oscillation probabilities  {$\mathcal{P}\left(\nu_e \to \nu_\mu \right)$} and 
{$\mathcal{P}\left(\bar\nu_e \to \bar\nu_\mu \right)$}. Therefore, it is necessary to reliably 
isolate the weak $\nu_\mu$ signal from the large number of $\nu_e$ events. The reconstruction 
studies presented in Sect.\,\ref{subsec::tracking} indicate an excellent rejection efficiency 
for quasi-elastic $\nu_e$ events and a reliable $\nu_\mu$ vertex reconstruction if the energy of 
the final state muon exceeds 200\,MeV. 
However, the discrimination efficiencies for NC/CC backgrounds producing a charged pion in the 
final state has not been evaluated yet.

In case of a combination of conventional and beta beam, the availability of beams of $\nu_e/\bar\nu_e$ and $\nu_\mu\bar\nu_\mu$ allow to test even more fundamental symmetries: T- or even CPT-invariance. T-invariance can be tested by comparing for example {$\mathcal{P} \left(\nu_\mu \to \nu_e \right)$} with {$\mathcal{P} \left(\nu_e \to \nu_\mu \right)$}. If CPT is conserved, a discovery of CP-violation must be accompanied by break-down of T-invariance. The CPT-symmetry can be tested in the comparison of {$\mathcal{P} \left(\nu_\mu \to \nu_e \right)$} with {$\mathcal{P} \left(\bar\nu_e \to \bar\nu_\mu \right)$}. It might be reasonable to start with either conventional or beta-beam to search for CP-violation, and to add the second beam if CP-violation is observed.

Extensive studies on the discovery range of CP-violation with LENA are still missing. A few beam configurations have been simulated with GLoBES\nomenclature{GLoBES}{General LOng-Baseline Experiment Simulator} \cite{Huber:2004ka,Huber:2007ji}. See for example \cite{Peltoniemi:2009zk,Peltoniemi:2009hv}. The existing simulations show that for not too small values of {$\theta_{13}$} a substantial fraction of the parameter space 
(typically 60 to 80\,\%) of the CP-violating phase {$\delta$} can be covered. Typically the sensitivity is good for values of {$\sin^2\,2 \theta_{13}$} down to 0.01 and then starts to diminish below.

%% file: atmospherics.tex

%
\noindent Based on the current status of the tracking studies presented in Sect.\,\ref{subsec::tracking}, LENA also offers the opportunity to investigate atmospheric neutrinos. The large volume substantially extends the sensitivity of previous large-volume liquid-scintillator detectors to the multi-GeV region, filling the energy-gap between previous underground experiments and high energy neutrino telescopes which become sensitive above $10$\,GeV.

\begin{figure*}[htp!]
\includegraphics[width=0.49\textwidth]{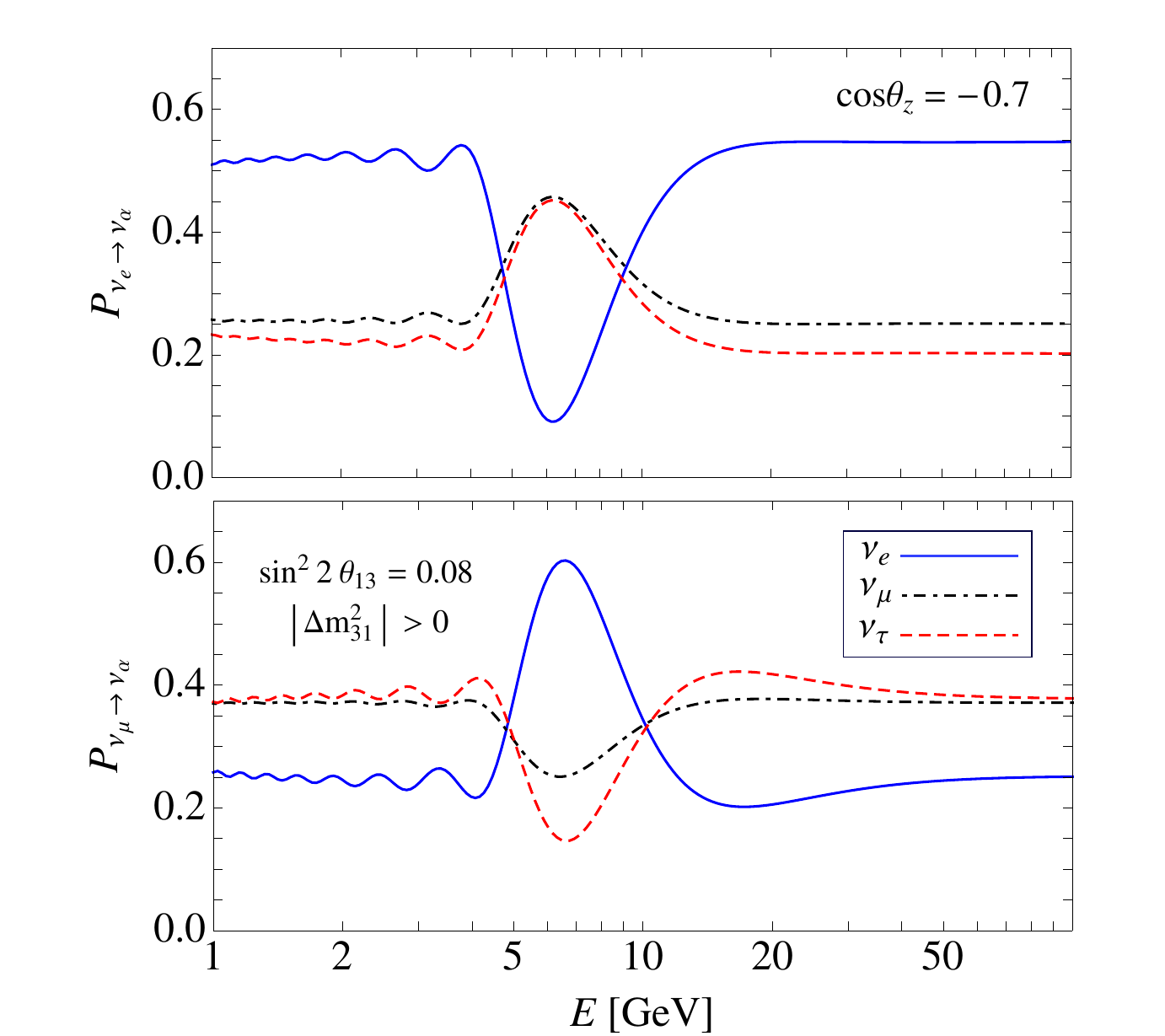}
\hfill
\includegraphics[width=0.49\textwidth]{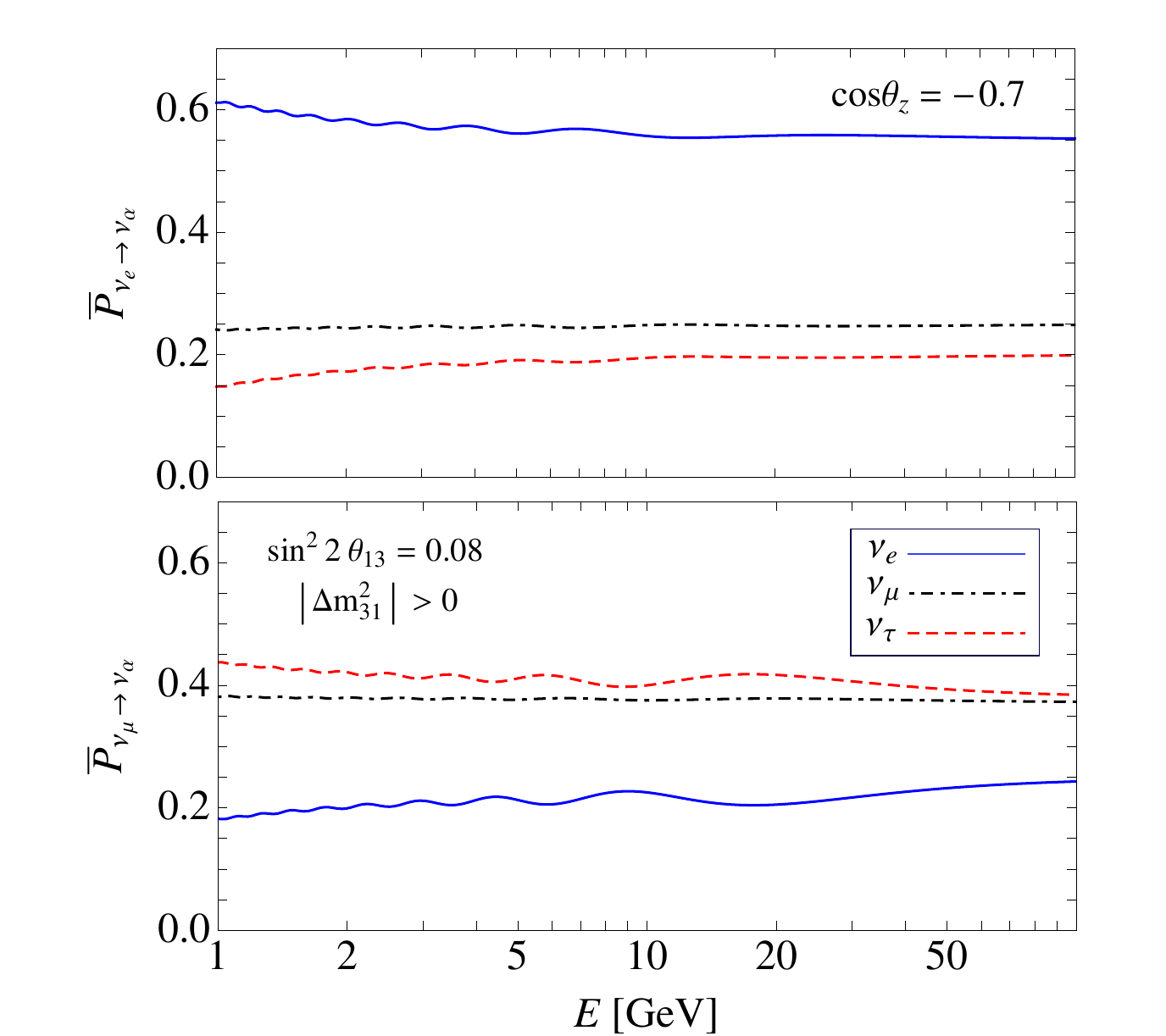}
\caption{Flavor conversion probabilities for neutrinos (left panel) and antineutrinos (right panel) for $\cos \theta=-0.7$ versus the (anti) neutrino energy. Both the $\nu_e \to \nu_\alpha$ ($\bar{\nu}_e \to \bar{\nu}_\alpha$) and $\nu_\mu\to \nu_\alpha$ ($\bar{\nu}_\mu \to \bar{\nu}_\alpha$) probabilities are plotted. We use normal hierarchy of $\nu$ masses and a vanishing CP phase along with the best-fit oscillation parameters (see text for details).}
\label{fig:fignadir}
\end{figure*}

Compared to Water-Cherenkov detectors a good energy resolution is expected up to $20 $\,GeV and even above. We therefore expect a good measurement of the flux and angular spectrum of 
atmospheric neutrinos up to a few tens of GeV.

These measurements will depend on the ability to identify the neutrino flavor in charged current interactions and to identify simultaneously the neutral current interaction rate for the total flux normalization. The separation of neutral currents from $\nu_e$ is based on the separation of $\pi^0$ from electrons, while the discrimination of charged pions from muons must rely on the identification of the $\pi^\pm$ decay. The performance of LENA regarding these issues is still under study (Sect.\,\ref{subsec::tracking}). 

The phenomenology of atmospheric neutrinos is rich in the multi-GeV region in particular with respect to oscillations \cite{Mena:2008rh,Akhmedov:2008qt}. The measurement of the direction and energy of the incoming neutrino will establish a long baseline experiment with variable baselines from a few tens of kilometers for neutrinos from above to more than 12\,000\,km for vertically up-going neutrinos originating from air showers on the other side of the Earth.

For vertical muon neutrinos the survival probability of oscillates with a broad minimum at about 20--25\,GeV and increases back to unity above that \cite{Mena:2008rh}. Towards the horizon this $1^\mathrm{st}$ oscillation shifts to smaller energy down to $1$\,GeV and the higher order minima at lower energy do likewise. The good energy resolution and statistics of LENA may allow to measure the zebra-shaped patterns of alternating oscillation minima and maxima with unprecedented resolution. This will allow to measure  $\theta_{23}$ and $\Delta m_{23}^2 $ to a high precision and hence, to probe the oscillation hypothesis in a 
previously not tested parameter region.

Correlated to the disappearance of muon neutrinos, we expect the appearance of tau neutrinos. Again, this would provide a unique tool to study the parameter space and verify the oscillations of  neutrinos. The ability to utilize the $\nu_\tau $ detection channel strongly depends on the ability of LENA to identify and separate tau neutrinos, which do, however, appear at a substantially higher rate than e.g. backgrounds from atmospheric $\nu_e$ or neutral current interactions: Above a few GeV, the ratio $R_{\mu e}$ of $\nu_\mu$ to $\nu_e$ fluxes increases from $R_{\mu e}\approx2$ at 1\,GeV to $R_{\mu e}\approx5-10$ at a few tens of GeV.

Very interesting structures in the zenith dependent oscillation probabilities appear, if also matter oscillations are taken into account \cite{Akhmedov:2008qt, Akhmedov:2006hb}. For nadir angles $\theta<33^\circ$ the  neutrinos have travelled through the core of the Earth and a strong resonance pattern appears, e.g. with maximum disappearance of $\nu_e $ at about $3$\,GeV. For neutrinos not crossing the core the effect abruptly shifts to larger energies. The amplitude of the above structures  depend on $\theta_{13} $ and differs for neutrinos and anti-neutrinos. The dependence on the CP phase and $\delta$ is small.  A change in hierarchy switches the resonance pattern from neutrinos to antineutrinos resulting in different event rates between the normal and inverted hierarchies.  The difference arises because of a smaller flux and cross-section for antineutrinos than neutrinos \cite{Mena:2008rh}. A complete study of the performance of LENA regarding neutrino mass hierarchy measurements is under development.

Fig.\,\ref{fig:fignadir}, extracted from \cite{Razzaque:2010kp}, depicts the conversion probabilities for neutrinos (left panel) and antineutrinos (right panel). The Preliminary Reference Earth Model \cite{Dziewonski:1981xy} is used for the density profile inside the Earth. The assumed input $\nu$ mixing parameters are: 
\begin{eqnarray}
\Delta m_{31}^2 & = & 2.4\times10^{-3} {\rm eV}^2, \nonumber\\
\Delta m_{21}^2 & = & 8\times10^{-5} {\rm eV}^2, \nonumber\\
\sin^2\theta_{12} & = & 0.31, \nonumber\\ 
\theta_{23} & = & \pi/4, \nonumber\\
\sin^2\theta_{13} & = & 0.02. \nonumber
\end{eqnarray} 
The CP violating phase is set to $\delta = 0$. We consider normal $\nu$ mass hierarchy only and $\cos \theta=-0.7$.  The upper and lower plots correspond to $\nu_e \to \nu_\alpha$ ($\bar{\nu}_e \to \bar{\nu}_\alpha$) and $\nu_\mu \to \nu_\alpha$ ($\bar{\nu}_\mu \to \bar{\nu}_\alpha$) conversions, respectively.  The anti-neutrino conversion probabilities are not affected by matter in case of normal $\nu$ mass hierarchy and vacuum conversion formalism apply.  For this value of the nadir angle, $\nu$'s do not pass through the Earth's core.  Conversions mostly take place in the mantle with an average density of $\langle\rho\rangle\sim 5\,$g\,cm$^{-3}$. The dip at $\sim$6\,GeV for $P_{\nu_e \to \nu_e}$ in Fig.\,\ref{fig:fignadir} (left panel, upper plot) corresponds to the 1-3 or high MSW resonance energy $E_H = \Delta m_{13}^2\cos 2\theta_{13}/(2\sqrt{2} G_{\rm F} \langle\rho\rangle) \approx 6$\,GeV. The width of the dip is $2\tan 2\theta_{13}E_H \approx 3.6$\,GeV. At energies $\gg E_H$, the conversion probablities are dominated by vacuum oscillation. A complete study of the performance of LENA regarding neutrino mass hierarchy measurements is under development.

A high statistics measurement with good energy- and angular resolution and flavor identification as it is anticipated by LENA has the opportunity to use atmospheric neutrinos as a new tool for science ranging from precision neutrino physics to an improved understanding of the Earth's interior.

%% file: conclusions.tex

%

\noindent Liquid scintillator is a very attractive detection target for the next generation of large-volume neutrino observatories: 

\begin{itemize}
\item\textbf{Availability.} The organic liquids that serve as the primary materials of the scintillator are used in very large quantities in industry. Therefore, they are both economically produced and easily available. Due to the large market, their industrial re-use is easy. 

\item\textbf{Performance.} The light yield of organic liquid scintillators is roughly 50 times higher than the light emission in water by the Cherenkov effect. It was demonstrated in Borexino that extremely high levels of radiopurity can be reached in liquid scintillators. This provides the opportunity to search for very rare events at the energy scale of natural radioactivity (down to 200\,keV), in particular for low energy neutrino astronomy and neutrino geology. 

\item\textbf{Detection channels.} Besides electrons and protons, organic liquids offer $^{12}$C (and $^{13}$C) as target material for neutrino interactions. The multitude of interaction channels is the base for a spectroscopic measurement differentiating between the flavors of neutrinos and anti-neutrinos. This might be crucial for investigating the complex neutrino signature of a Supernova explosion. In such an event, neutrino-matter as well as neutrino-neutrino interactions might give rise to extensive swaps and distortions of the initial flavor spectra.

\item\textbf{GeV tracking.} Recent investigations of the photon arrival times for GeV particles in a homogenous large-volume scintillation detector indicate an unexpected accuracy of directional information and particle identification. These parameters are very well determined for track lengths that exceed several tens of centimeters, corresponding to particle energies of several hundred MeV.  

\item\textbf{Versatility.} The detection sensitivity in a large-volume scintillation detector covers an energy range reaching from sub-MeV energies to the scale of several GeV, providing access to a large range of topics in neutrino physics, geology and astronomy.
\end{itemize}
In this work, we have presented a broad variety of scientific research areas that can be addressed by a liquid-scintillator detector. In the context of a European Large Infrastructure for astro particle physics, LENA is put forward as a viable and cost effective alternative for a next-generation neutrino detector. The fields of research enclose low energy neutrino astronomy as well as elementary particle physics, which can be accessed by the investigation of natural neutrino sources and includes also nucleon decay search. If LENA is used as far detector in a next-generation neutrino beam experiment, this will allow for a unique investigation of neutrino oscillation parameters as well as CP-violation in the lepton sector. 

Profound expertise has been obtained in construction and operation of the presently running liquid-scintillator detectors KamLAND and Borexino. Their successes in neutrino physics and astronomy reflect the technological maturity. The results of the LAGUNA design study as well as of the specific design studies investigating a possible realization of LENA in the Finnish Pyh\"asalmi mine indicate a time frame of 8 to 10 years for an executive design and the detector construction.